%% file: TrussTop.tex
\documentclass[acmtog,review=false]{acmart}
\makeatletter
\renewcommand\@formatdoi[1]{\ignorespaces}
\makeatother
\renewcommand\footnotetextcopyrightpermission[1]{} 
\usepackage{booktabs} 

\usepackage{amssymb}
\usepackage{amsmath}
\usepackage{algorithm}
\usepackage[noend]{algpseudocode}

\usepackage{overpic}
\usepackage{contour}\contourlength{0.2ex}

\def\veccc#1/#2/#3;{\begin{pmatrix}#1\\#2\\#3\end{pmatrix}}
\def\cput(#1,#2,#3){\put(#1,#2){\contour*{white}{#3}}}%
\def\colorboxput(#1,#2,#3,#4){\put(#1,#2){\textcolor{#3}{\vrule width.6em height.6em depth0pt}~#4}}

\def\uw{{\mathbf u}}

\def\ww{{\mathbf w}}
  
\def\fw{{\mathbf f}} \def\dw{{\mathbf d}}
\def\pw{{\mathbf p}}   \def\fw{{\mathbf f}}

\def\sw{{\mathbf s}}
\def\be{\begin{equation}}   \def\ee{\end{equation}}
\def\cput(#1,#2)#3{\put(#1,#2){\hbox to 0pt{\hss#3\hss}}}

\acmPrice{15.00}

\setcitestyle{square}

   \newcommand{\AD}[1]{{\color{black} #1}}
  \newcommand{\CG}[1]{{\color{black} #1}}

\begin{document}
\title{Computational Design of
	Lightweight Trusses}

\author{Caigui Jiang}
\orcid{0000-0002-1342-4094}

\affiliation{%
	\institution{MPI for Informatics}
	\streetaddress{Campus E1 4}
	\city{Saarbrucken}
	\state{Saarland}
	\postcode{66123}
	\country{Germany}}
\affiliation{%
	\institution{KAUST}
	\city{Thuwal}
	\postcode{23955-6900}
	\country{Saudi Arabia}}
\author{Chengcheng Tang}
\affiliation{%
	\institution{Stanford University}
	\state{CA}
	\postcode{94305}
	\country{USA}
}
\author{Hans-Peter Seidel}
\affiliation{%
	\institution{MPI for Informatics}
	\streetaddress{Campus E1 4}
	\city{Saarbrucken}
	\state{Saarland}
	\postcode{66123}
	\country{Germany}}

\author{Renjie Chen}
\affiliation{%
	\institution{MPI for Informatics}
	\streetaddress{Campus E1 4}
	\city{Saarbrucken}
	\state{Saarland}
	\postcode{66123}
	\country{Germany}}

\author{Peter Wonka}
\affiliation{%
	\institution{KAUST}
	\city{Thuwal}
	\postcode{23955-6900}
	\country{Saudi Arabia}}

\makeatletter
\let\@authorsaddresses\@empty
\makeatother
\renewcommand{\shortauthors}{Jiang, Tang, Siedel, Chen and Wonka}

\begin{abstract}

Trusses are load-carrying light-weight structures consisting of bars connected at joints ubiquitously applied in a variety of engineering scenarios. Designing optimal trusses that satisfy functional specifications with a minimal amount of material has interested both theoreticians and practitioners for more than a century.
In this paper, we introduce two main ideas to improve upon the state of the art. First, we formulate an alternating linear programming problem for geometry optimization. Second, we introduce two sets of complementary topological operations, including a novel subdivision scheme for global topology refinement inspired by Michell's famed theoretical study. Based on these two ideas, we build an efficient computational framework for the design of lightweight trusses. \AD{We illustrate our framework with a variety of functional specifications and extensions. We show that our method achieves trusses with smaller volumes and is over two orders of magnitude faster compared with recent state-of-the-art approaches.}
\end{abstract}



\begin{teaserfigure}
  \centering{
  	\begin{overpic}			
  		[width=6.0in]{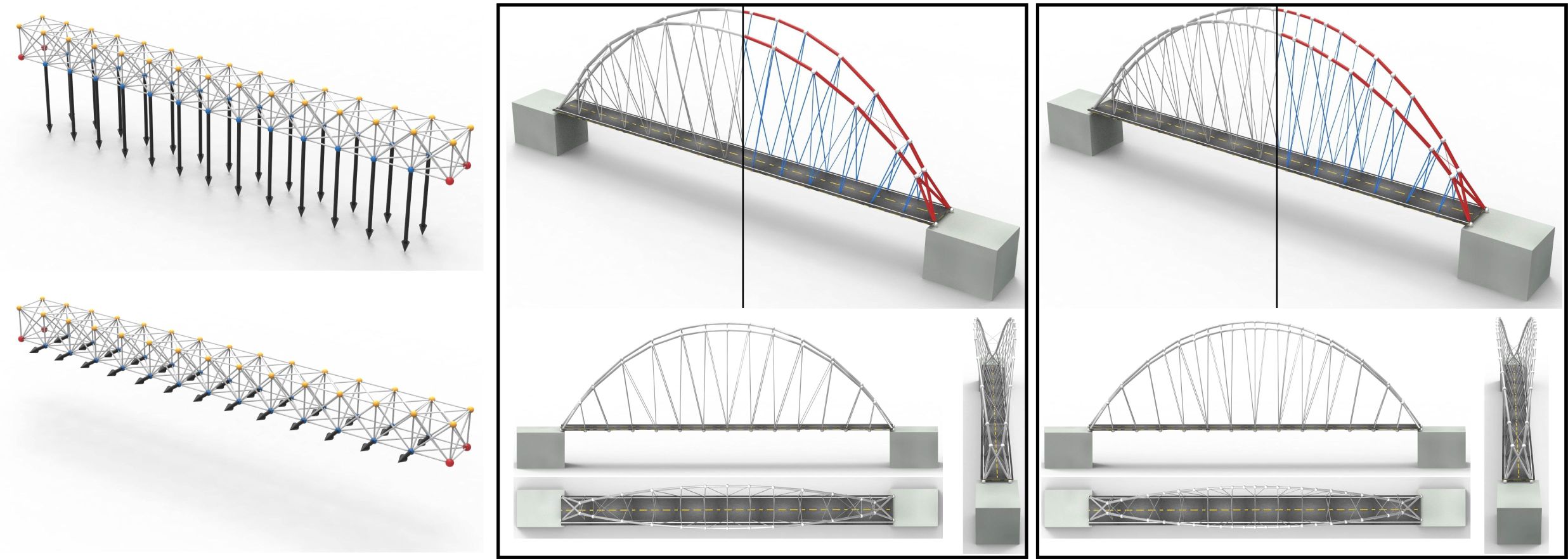}
  		\cput(10,20){\contour{white}{$(a)$}}
  		\cput(10,3){\contour{white}{$(b)$}}		
  		\cput(36,15){\contour{white}{$(c)$}}
  		\cput(70,15){\contour{white}{$(d)$}}		
  	\end{overpic}	
  }
  \caption{	\label{fig:Teaser} A bridge design problem from \cite{descamps2013lower}. 
  The input is an initial structure of 258-bars with supporting points (red) and two sets of forces: (a) downward loads of magnitude 1 and (b) horizontal loads of magnitude 0.2 perpendicular to the bridge's main direction.
  The result in \cite{descamps2013lower} has a total volume of 408.8 and took over 1000s to compute. (c) Our optimal geometry and topology viewed from different angles. The total volume is 333.4 and the running time is less than 10s. (d) Further structure refinement based on (c) to achieve a total volume of 331.9. }
\end{teaserfigure}

\maketitle
\thispagestyle{empty}
\input{1_Introduction}
\input{2_PreviousWork}
\input{3_Overview}

\input{3_Ourmethod}

\input{4_Algorithm}
\input{6_Results}
\input{7_Conclusion}


\bibliographystyle{ACM-Reference-Format}
\bibliography{truss}

\input{9_Appendix2}

\end{document}

%% file: 1_Introduction.tex
\section{Introduction}

Trusses are crucial and fundamental structures in multiple modern engineering domains. They consist of bar elements that are connected by pin joints.
Because of their efficiency and lightweight nature, trusses see considerable amount of usage in industrial design and architectural construction, e.g., for support structures of buildings, bridges, transmission towers, or even domes in playgrounds.

\begin{figure}[H]	\centering{
		\begin{overpic}			
			[width=.45\columnwidth,trim={2cm 9cm 2cm 9cm},clip]{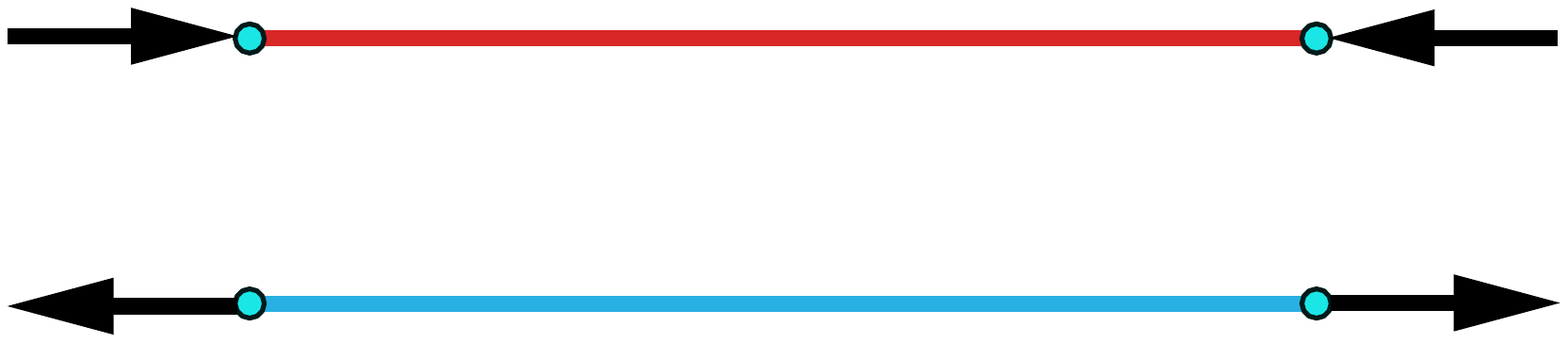}	
			\cput(50,2){\contour{white}{$(a)$}}		
		\end{overpic}
		\begin{overpic}			
			[width=.45\columnwidth,trim={2cm 9cm 2cm 9cm},clip]{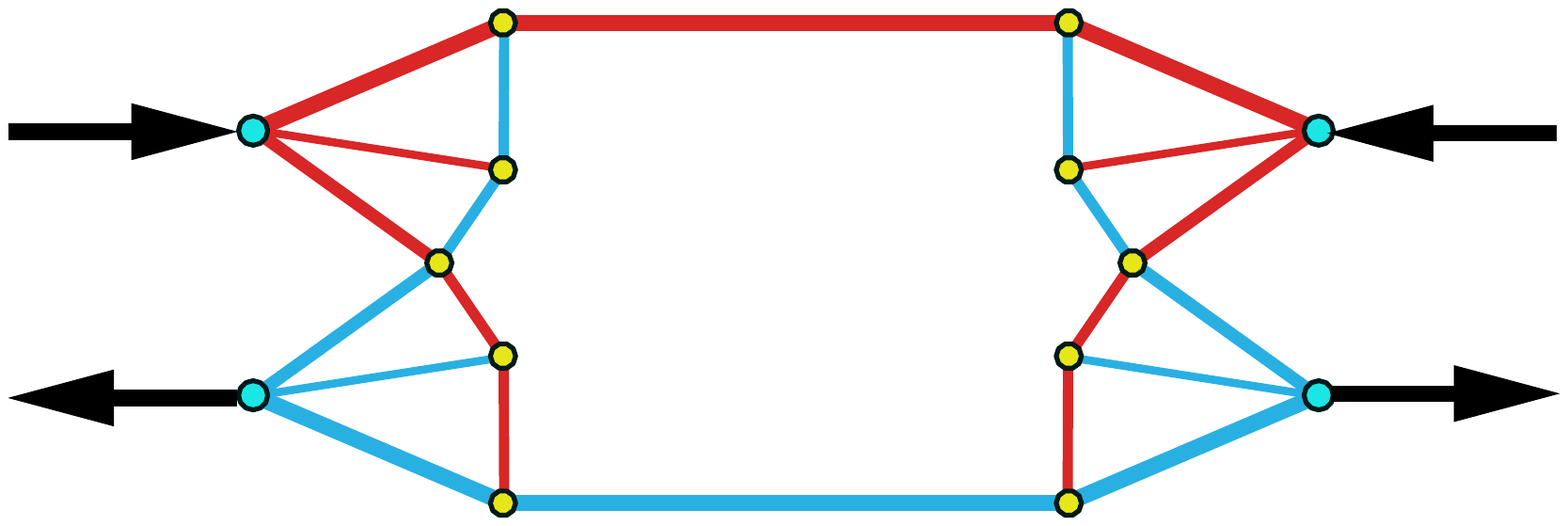}	
			\cput(50,2){\contour{white}{$(b)$}}	
		\end{overpic}

	}	
	\caption{
		\label{fig:2pairs}
		Two truss designs for the same functional specification. Left: a straightforward design with two bars. Right: a more complex and less intuitive design with less material usage.}
\end{figure}

Designing a lightweight truss typically starts with a functional specification, e.g., in the form of external forces that the structure has to withstand.
The design problem can then be formulated as an optimization problem to determine the geometry, topology, and the cross-sections of the truss. In other words, we have to find answers to the following questions: Where to put the intermediate joints? How to connect the joints with bars? What are the cross-section areas of the bars? These tasks are notoriously challenging because the optimization of geometry, topology, and cross-sections is interrelated, and there exists an infinite number of possible topologies which are difficult to classify and quantify. Even for a simple case, the optimal topology is not intuitive. As shown in Figure \ref{fig:2pairs}, to support two pairs of opposing forces lying on two straight lines, the simplest truss on the left with two bars, one in tension (blue) and another in compression (red), may be intuitively considered as the lightest truss. However, a lighter design with more intermediate joints and connections can be found as shown in the Figure \ref{fig:2pairs} right.
	Another simple functional specification problem is called the three forces problem (3FP) \cite{chan1966minimum,sokol2010solution}. 3FP is formulated as
	follows: find the lightest fully stressed truss transmitting three self-equilibrated
	co-planar forces. Although there are only three forces, the problem is still unsolved analytically for general cases. 

\begin{figure}[h]	\centering{
		\begin{overpic}			
			[width=.45\columnwidth,trim={2cm 10cm 2cm 11cm},clip]{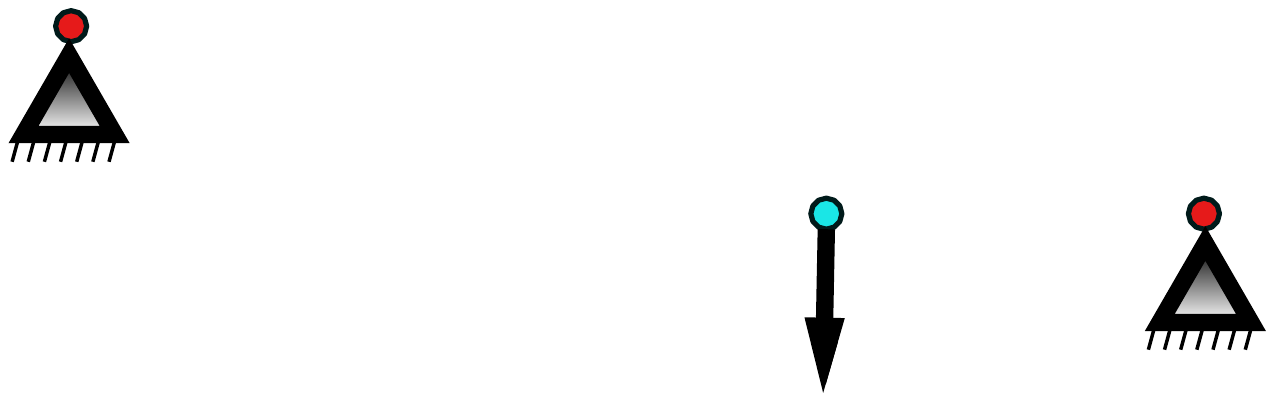}	
			\cput(50,2){\contour{white}{$input$}}		
		\end{overpic}
		\begin{overpic}			
			[width=.45\columnwidth,trim={2cm 9cm 2cm 10cm},clip]{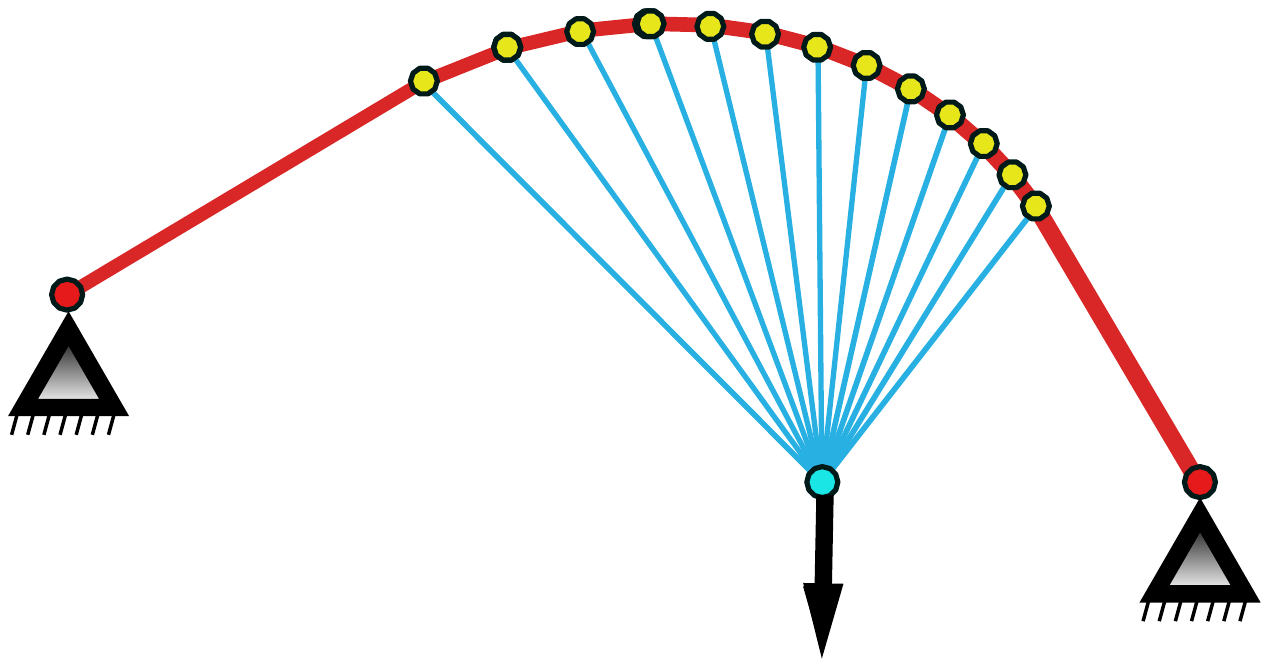}	
			\cput(50,2){\contour{white}{$output$}}	
		\end{overpic}
	}	
	\caption{
		\label{fig:inputoutput}
		Left: a functional specification including two supporting points and one external force. Right: an optimal truss.}
\end{figure}

In this paper, we mainly follow previous work and take functional specifications in the form of supporting points and applied forces as input. Our goal is then to construct a lightweight truss with optimal joint positions, topology, and cross-sections. As an example shown in Figure \ref{fig:inputoutput} left, two supporting points and one external force are given as inputs. Our computational method generates the topology automatically and optimizes the nodal positions and cross-section areas as shown in Figure \ref{fig:inputoutput} right.


There are two major strategies to tackle this problem in previous work. The first strategy is to start with a densely-connected structure and to subsequently identify which bars to remove (e.g., the ground structure method~\cite{dorn1964automatic} and its variations~\cite{gilbert2003layout,sokol2017numerical}). The main limitation of this strategy is its sub-optimality due to its heavy dependence on the initialization as it is extremely unlikely that the optimal structure is a sub-structure of the initial one.
The other strategy is to start with a sparse structure and to iteratively add new joints and bars. One of the most famous methods in~\cite{martinez2007growth} adds one joint at a time and can only deal with one single load 2D problem.

Even though topology optimization is such a longstanding and fundamental problem in structural engineering, one can identify large possible improvements to the current state of the art. First, the search space for truss topology is not properly explored by previous algorithms. We could observe that it is very difficult to optimize the topology in a single stage. Much better results can be achieved by proceeding in two stages of topology optimization: computing a coarse truss and truss subdvision. Our novel subdivision approach is inspired by Michell's pioneering theoretical treatment of optimal truss design in~\cite{michell1904lviii}. Second, the geometry optimization used in previous work is not efficient. To tackle this problem we decompose geometry optimization into alternating linear programming formulations to reduce the running time.

Our main contributions are as described in the following

	\begin{itemize}	
			\item{
			We propose two categories of complementary topology operations, local and global. While local operations have been used in previous work, our global operations based on subdivision are our original contribution.
			
		}
		\item{
			We introduce a novel algorithm for geometry optimization based on alternating linear programming (ALP) that jointly optimizes joint positions and bar cross sections.
		}

		\item{
			Based on these two technical contributions, we build a framework for lightweight truss design, a longstanding and important problem in structural engineering, architecture, graphics, and design. \AD{Compared with recent state-of-the-art approaches, our method creates trusses with smaller volumes, can handle more complex functional specifications, and is over two orders of magnitude faster.} 
			 }
	\end{itemize}


%% file: 2_PreviousWork.tex
\section{Previous Work}

In recent years, combining geometric modeling together with realistic engineering considerations, especially static equilibrium and manufacturability, have attracted the interests of many researchers in the graphics community.  Beyond applications in the virtual world \cite{smith2002creating}, those previous works enable novel and functional designs manufacturable with 3D printing ~\cite{wang2013cost,zpz_struct_ays_13}, laser cutting ~\cite{martinez2015structure}, masonry structure ~\cite{Block07,vouga-2012-sss,liu:2013,panozzo-2013,desbrun:2013,tang-2014-ff}, for toys ~\cite{prevost2013make,bacher2014spin}, furniture ~\cite{uim_guidedExploration_sigg12,Yao:2017:IDS:3068851.3054740}, and architecture ~\cite{jiang2015interactive,pietroni2015statics}. The most relevant works to ours are ~\cite{jiang2017design} and ~\cite{pellis-optimal-2017}. Jiang et al. propose a framework to design and optimize space structures where only a small set of cross-section areas are allowed. Therefore, Jiang et al.~\cite{jiang2017design} compute a specialized form of truss, but we focus on the classical problem of truss design without discrete restrictions on the cross sections. The main practical difference is that our proposed method can generate a truss from scratch, whereas Jiang et al. relies on a reasonable truss being given as input. In our results we also demonstrate that our proposed optimization technique ALP produces better results on our problem formulation than the geometric optimization technique used in~\cite{jiang2017design}.
Kilian et al.~\cite{pellis-optimal-2017} provide an interesting geometric understanding of "optimality" of surface-like lightweight structures. Compared with their work, we tackle a problem for common and general trusses in both 2D and 3D, instead of focusing on load-carrying surfaces.

The problem of designing a truss with a minimal volume of material that supports imposed external forces was first studied in ~\cite{michell1904lviii}.  In the milestone paper,  Michell proved that an optimal truss must follow orthogonal networks of lines of maximal and minimal strains in a constant-magnitude strain field. An optimal truss is usually called a Michell truss. Following his work, research on the topic of optimal truss can be divided into two categories: exact-analytical formulations and approximate-discretized formulations.

An exact-analytical formulation assumes that the truss is a continuum structure connected by an infinite number of bars with infinitesimally small cross-sections. In analytical formulations, the theoretical optimal design is determined exactly through the simultaneous solution of a system of equations expressing the conditions for optimality. The basic principles were establish in~\cite{maxwell1870reciprocal} and~\cite{michell1904lviii} and a more general treatment was outlined in~\cite{hemp1973michell} and ~\cite{prager1977optimization}. Recent works on deriving  exact solutions were presented in~\cite{rozvany1998exact},~\cite{lewinski2007exact},~\cite{lewinski2008exact}, and~\cite{lewinski2008analytical} for a series of benchmark problems. Basically, the analytical solutions are very hard to obtain and only available for some special boundary conditions. While they are less practical in most of the generic scenarios, these solutions could be used as references to verify the performance of numerical methods.

Discretized numerical formulations are more practical and efficient approaches for structural design tasks presented in the real world. The most influential method is the ground structure method (GSM) which was first proposed in~\cite{dorn1964automatic}. This method consists of generating a fixed grid of joints and
adding bars in some or all of the possible connections between the joints as potential structural or vanishing bars. 
The optimized structure for the imposed functional specification is found using the cross-section areas as
design variables, and the whole problem is formulated as a linear programming problem. Its optimal topology is achieved by eliminating the zero-area cross sections. The ground structure method has been recently improved in~\cite{gilbert2003layout} and~\cite{sokol2017numerical}.

Besides GSM, some other numerical methods are proposed recently, such as the method in~\cite{martinez2007growth}, carries out geometry optimization in conjunction with a heuristic `joint adding' algorithm, generating an increasingly complex truss structure from a relatively simple initial layout. However, this algorithm can only add one joint per time and only works for single load cases. An efficient algorithm proposed by He and Gibbert ~\shortcite{he2015rationalization} combines layout optimization with geometry optimization. Similar to GSM, its layout optimization starts from a densely connected truss and is formulated by a linear programming problem, and its geometry optimization is formulated by a non-linear optimization as a post-processing step.

%% file: 3_Overview.tex
\section{Overview}
Our framework has the following major components:

\begin{itemize}
	\item{{\textbf{Functional specification} (C1).}
		The input to our framework is the functional specifications including the external forces and supporting points together with a set of structural constraints, e.g., design regions, geometric obstacles, and material properties. (See Section \ref{sec:input})
	}
	\item{{\textbf{Initialization} (C2).} 
		To obtain an initial truss, we create a grid of intermediate joints and densely connect them. This grid is located inside the design region and its size is proportional to the bounding box of the points in the input specification. (See Section \ref{sec:initialization})
	}
	\item{{\textbf{Local topology operations} (C3).}
		 We locally manipulate the topology through some geometry operations such as removing bars with vanishing cross-section areas and joints without any connection, merging close joints, etc. (See Section \ref{sec:GeoOperations})
		}
	\item{{\textbf{Global topology refinement} using subdivision (C4).}
Use an optimized coarse truss as input, we further refine the truss through subdivision. (See Section \ref{sec:subdivision} )
		}
	\item{{\textbf{Geometric optimization} using ALP (C5).}
		Given a fixed topology, we propose an alternating linear programming algorithm (ALP) to reduce the total volume of the truss by adjusting the joint positions and cross-section areas of bars. This algorithm is an essential component and its details are introduced in Section \ref{ALP}.}
\end{itemize}

\paragraph{Pipeline Overview}The individual components work together as follows:
We start from an input specification to compute an initial truss.
Then, we proceed in two phases. In the first phase, coarse truss optimization, we interleave geometric optimization (C5) with local topology operations (C3).
In the second phase, structure refinement through subdivision, we interleave global topology refinement (C4) with geometric optimization (C5).
See Figure \ref{fig:framework} for an overview of our framework. We discuss each individual component and the overall framework in detail in the next sections.

%% file: 3_Ourmethod.tex
\section{Design Framework}
We provide a framework for the computational design of lightweight trusses.
In this section, we describe the input specification, the initialization, local topology operations, and global topology refinement.


\subsection{Functional Specification} \label{sec:input}
The input to our framework is the functional specification including the external forces and supporting points together with a set of structural constraints, e.g., design regions, geometric obstacles, and material properties. Throughout the paper, we visualize supporting joints as red dots, joints with active forces as blue dots, and intermediate joints as yellow dots. We also visualize bars in tension in blue and bars in compression in red. In addition, the thickness of the bars is visualized in proportional to the computed cross-section areas. Note that when the external forces are in self-equilibrium, the input specification may have no supporting points. For example, the three forces problem (3FP) in 2D. 

\begin{figure*}[tb]	\centering{
		\begin{overpic}
			[width=2.1\columnwidth,trim={0cm 0cm 0cm 0cm},clip]{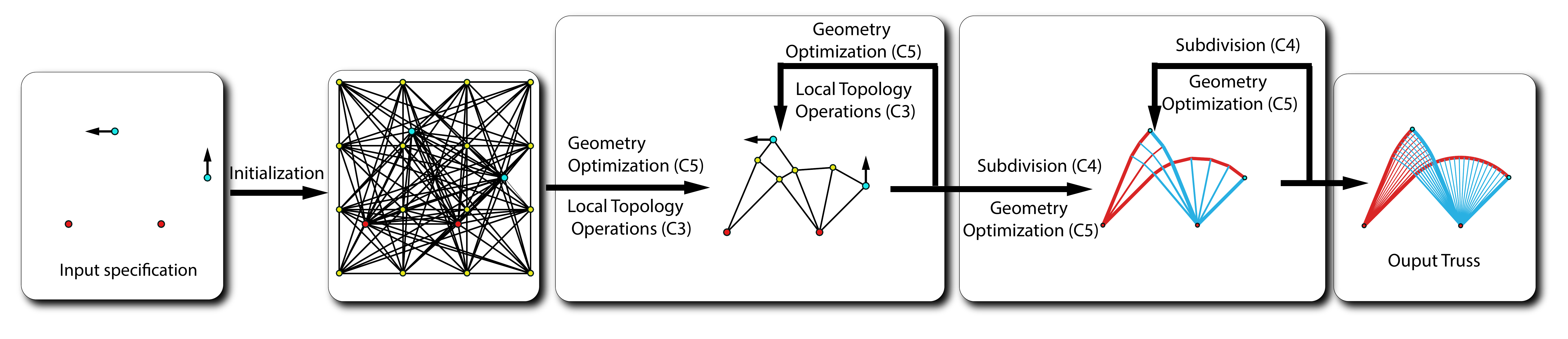}
				\cput(7,0){\contour{white}{$(a)$}}	
				\cput(48,0){\contour{white}{$(c)$}}	
				\cput(74,0){\contour{white}{$(d)$}}	
					\cput(28,0){\contour{white}{$(b)$}}	
					\cput(93,0){\contour{white}{$(e)$}}					
		\end{overpic}
	}	
	\caption{
		\label{fig:framework}
		Framework overview: Based on an input functional specification (a), our system creates an initial truss (b). Then, we proceed in two phases, coarse truss optimization (c) and structure refinement through subdivision (d). In the first phase, we interleave geometric optimization using ALP with local topology operations. In phase 2, we interleave subdivision with geometry operation using ALP. The output truss is shown in (e).}
\end{figure*}



\subsection{Truss Initialization} \label{sec:initialization}
\begin{figure}[H]	\centering{
		\begin{overpic}
			[width=.49\columnwidth,trim={37cm 11cm 32cm 10cm},clip]{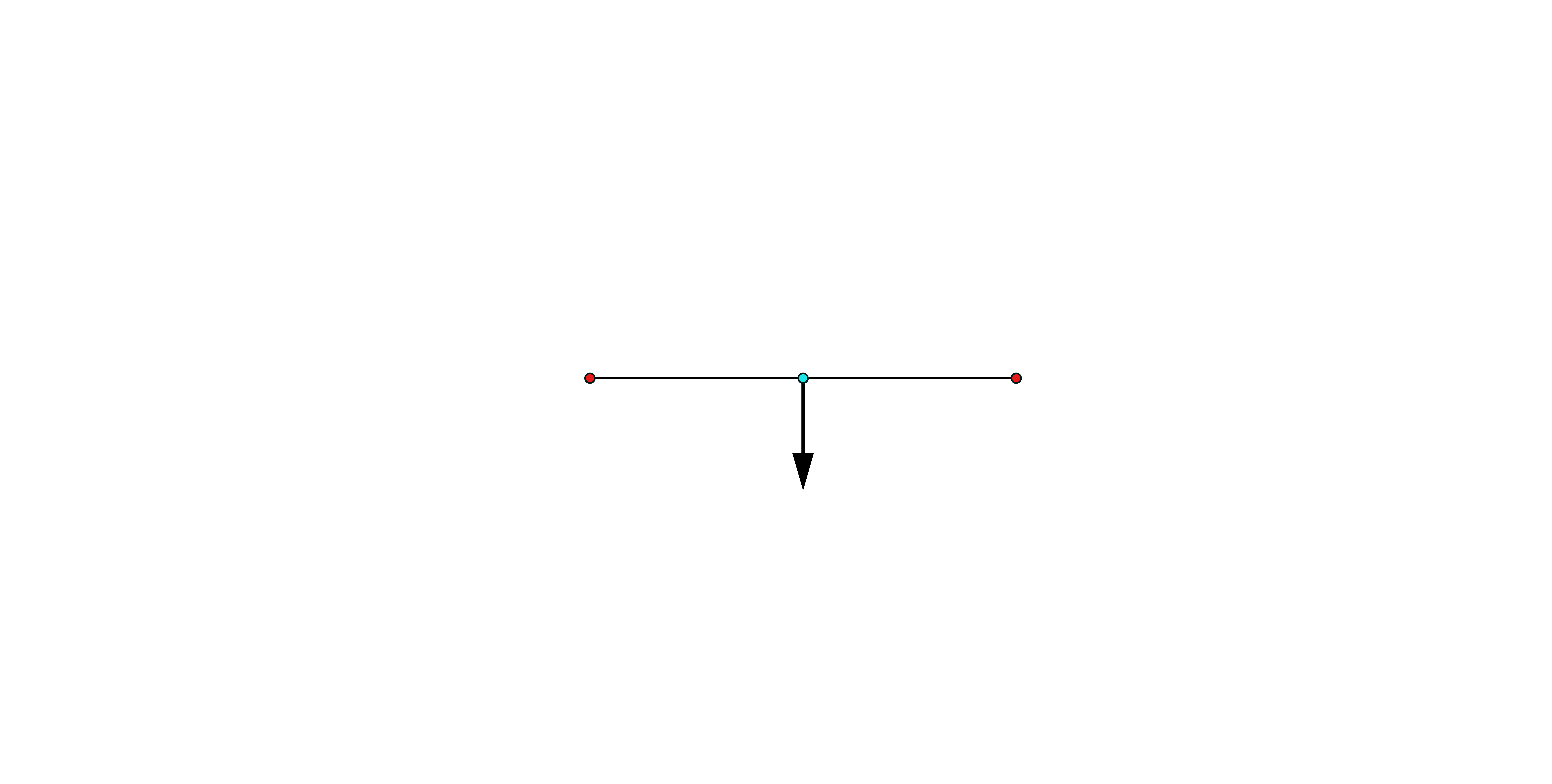}
		\end{overpic}
		\begin{overpic}
			[width=.49\columnwidth,trim={37cm 11cm 32cm 10cm},clip]{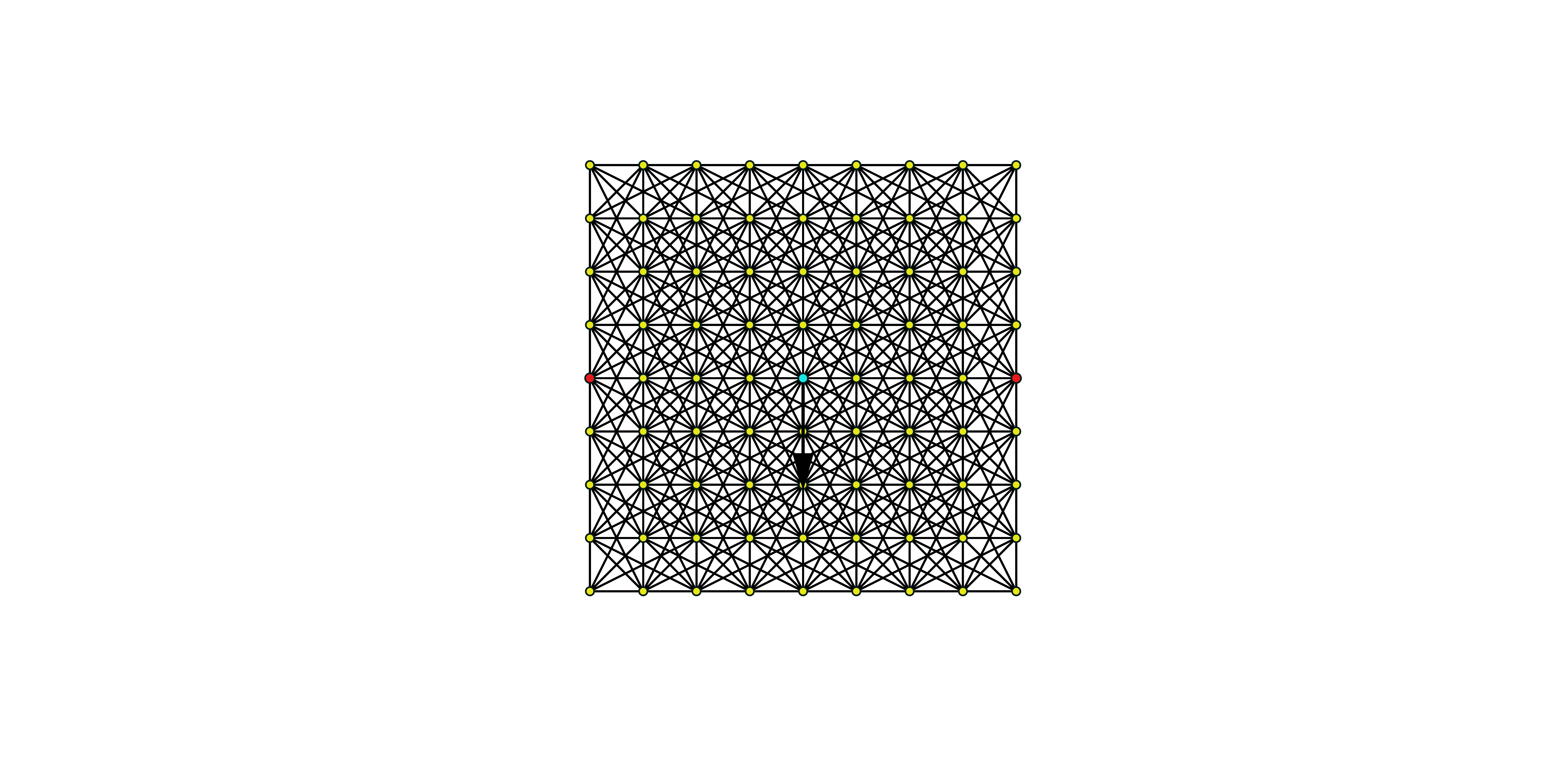}
		\end{overpic}
	}	
	\caption{
		\label{fig:badIni}
		Two kinds of initializations. Left: connect force application points and supporting points. Right: a densely connected initial structure for GSM.}
\end{figure}

We build on previous work to compute an initial truss.
There are two approaches to tackle this problem. One simply adds connections between provided joints in the functional specification
 (supporting joints and joints with active forces). For example, as shown in Figure \ref{fig:badIni} left, two bars connecting the joints with active forces (blue) and the supporting joints (red) are set as an initial truss. In some cases, this initialization is too simple to construct an equilibrium force system. Another method adds a grid of intermediate points over the design region and densely connects them as shown in Figure \ref{fig:badIni} right. The increasing method such as the work in ~\cite{martinez2007growth} used the first initialization. The GSM  usually uses the latter one with a large number of intermediate joints. 

In our framework, we first add some intermediate joints and connections. For instance, a size of $n\times n$ 2D grid points or $n\times n\times n$ 3D grid points and their dense connections, where $n$ is a user specified parameter. The default value of $n$ is the number of joints specified in the functional specification.
This is quite similar to the GSM, the difference is that the number of new joints that we add is usually much less.


\subsection{Local Topology Operations} \label{sec:GeoOperations}
\begin{figure}[H]	\centering{
		\begin{overpic}
			[width=.99\columnwidth]{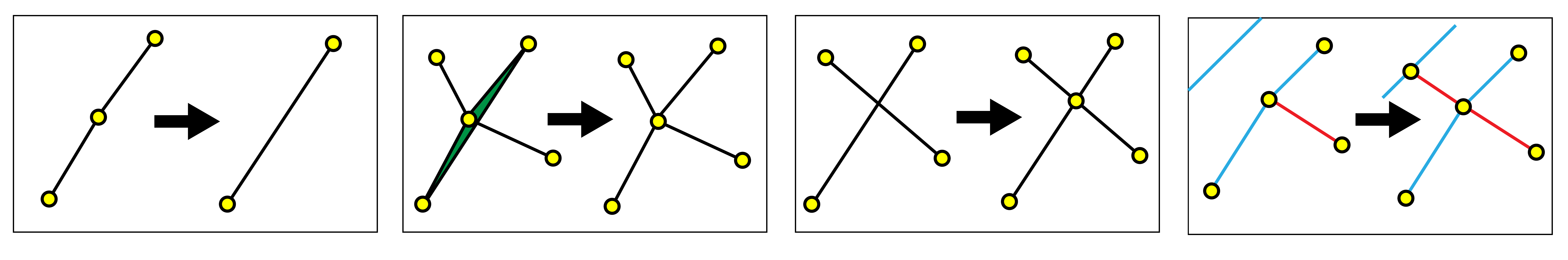}
			\cput(12.5,-2){\contour{white}{$(a)$}}	
			\cput(37.5,-2){\contour{white}{$(b)$}}	
			\cput(62.5,-2){\contour{white}{$(c)$}}
			\cput(87.5,-2){\contour{white}{$(d)$}}	
		\end{overpic}
	}	
	\caption{
		\label{fig:operation}
		 Four types of local topology operations: (a) Delete a joint of valence two. (b) Remove an ill-shaped narrow triangle. (c) Add an additional joint for a pair of intersecting bars.(d) Fix non-boundary T-junction.}
\end{figure}

We use the following local topology operations:

\begin{itemize}
	\item{
	Removing the bars with cross-section areas less than a small threshold $\epsilon_1$.
	}
	
	\item{
	Removing joints without any attached bars.
	}
	\item{
	Merging joints that are closer to each other than a small threshold $\epsilon_2$.
	}
	\item{Removing intermediate joints with valence two, as shown in Figure \ref{fig:operation}(a).}
	\item{Deleting the longest bar of a long narrow triangle as shown in Figure \ref{fig:operation}(b).}
	\item{Adding a new joint for each pair of intersecting bars. Split this pair of bars into four new bars and connect them at the new joint as shown in Figure \ref{fig:operation}(c). }
	\item{Fixing non-boundary T-junctions by adding a bar and a new joint connecting the new bar and the original truss as shown in Figure \ref{fig:operation}(d). The new joint is created at the point closest to the extension of the existing bar creating the T-junction.}

	\end{itemize}
	

These operations change the local topology and update joint positions. 



\subsection{Global Topology Operation --- Subdivision}\label{sec:subdivision}
The main idea of global topology refinement is to add joints and bars to the truss to be able to reduce its volume after geometry optimization. While previous work, e.g., \cite{martinez2007growth}, also proposes to add joints and bars, they add only one joint at a time by testing a large number of candidate locations. This results in a very expensive algorithm. By contrast, we propose to add new joints and bars based in a systematic manner. Our algorithm is inspired by two observations. First, Michell's theory~\cite{michell1904lviii} concludes that the minimum-weight truss should follow two families of continues curves which are orthogonal to each other, one in tension and one in compression. Second, an interesting aspect of truss design is that trusses with more bars can often be lighter than trusses with fewer bars, as more degrees of freedom are provided to approximate an analytical limit.
Our algorithm refines the discrete equivalent of such families of curves by subdivision in an efficient and coordinated manner. Most importantly, we insert multiple bars in one step.

\begin{figure}[H]	\centering{
		\begin{overpic}
			[width=.20\columnwidth]{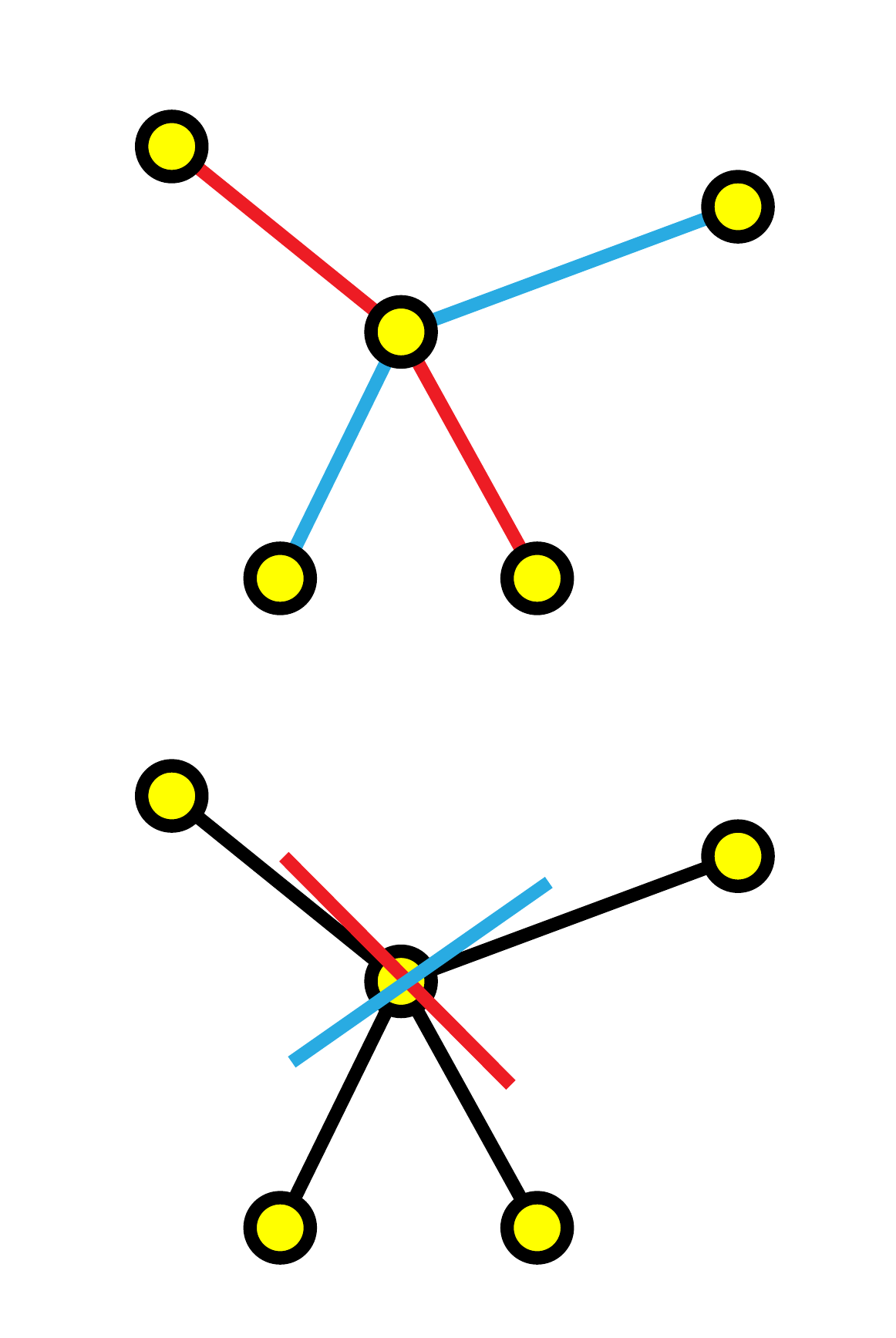}
			\cput(30,-2){\contour{white}{$(a)$}}	
			
		\end{overpic}
		\begin{overpic}
			[width=.49\columnwidth,trim={2cm 8cm 1cm 7.5cm},clip]{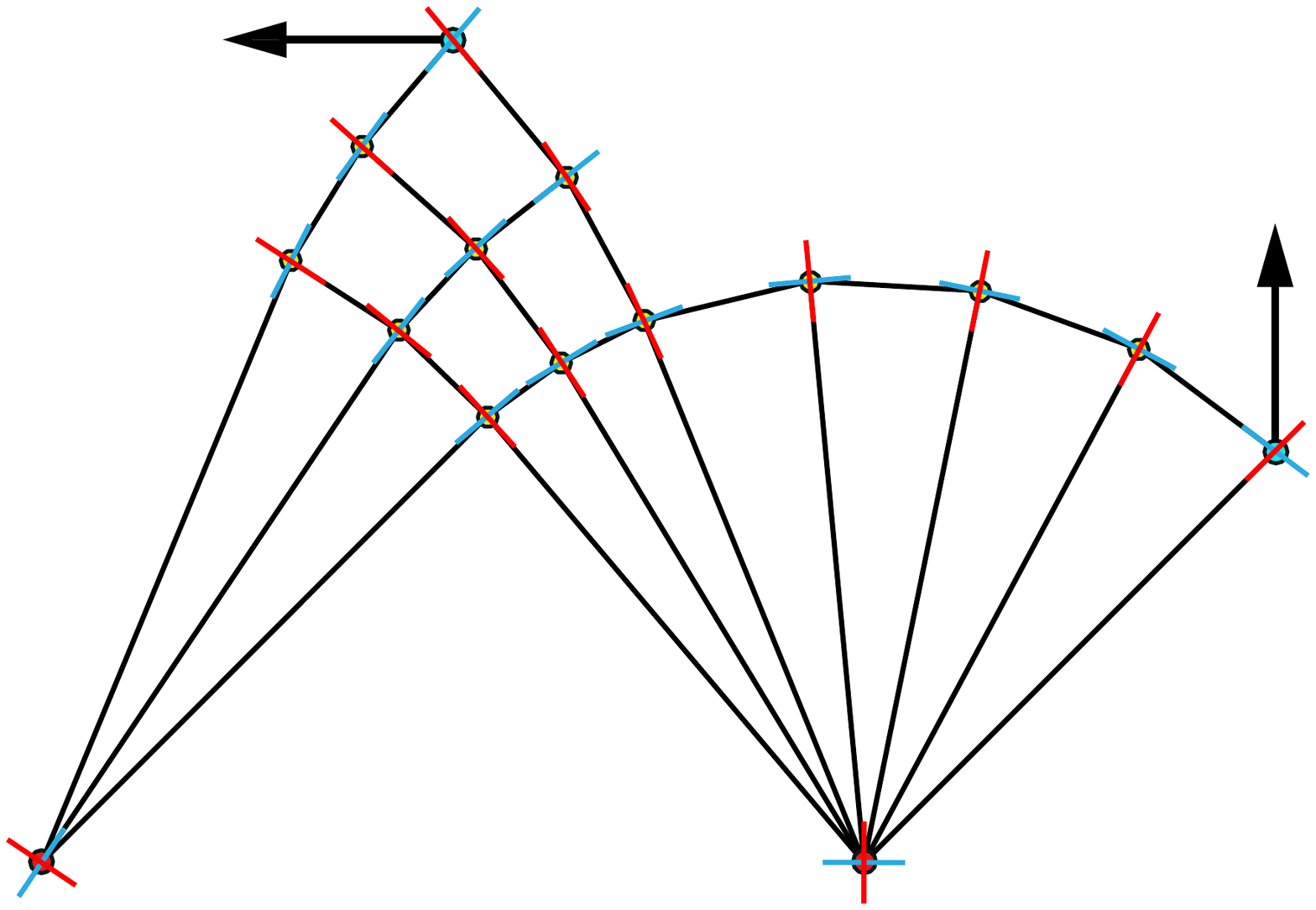}
			\cput(50,-2){\contour{white}{$(b)$}}				
		\end{overpic}
		\begin{overpic}
			[width=.25\columnwidth]{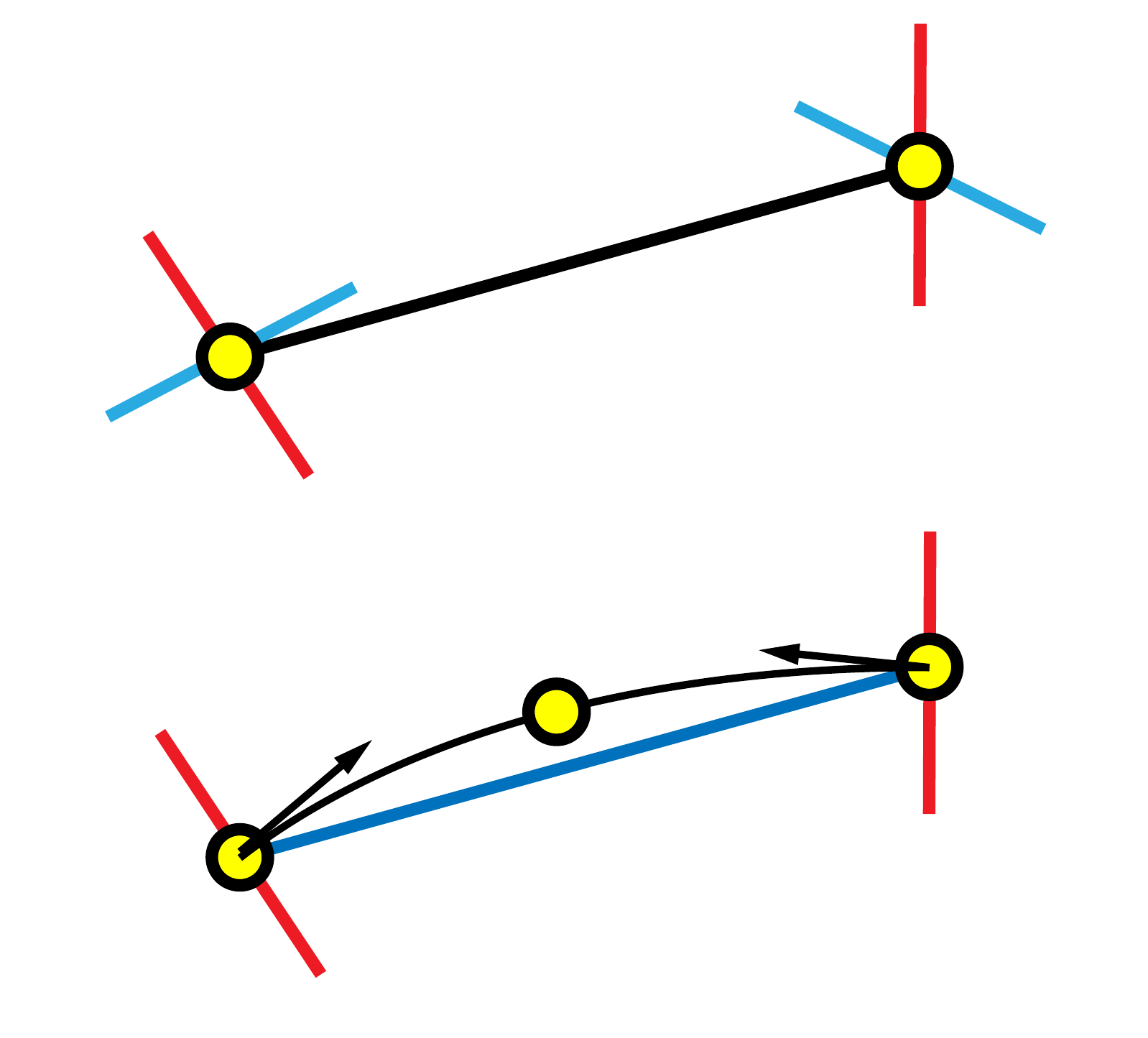}
			\cput(50,-2){\contour{white}{$(c)$}}
			\cput(15,5){\contour{white}{$p_i$}}	
			\cput(92,30){\contour{white}{$p_j$}}
					\cput(30,30){\contour{white}{$v_i$}}	
					\cput(65,40){\contour{white}{$v_j$}}				
		\end{overpic}
	}	
	\caption{
		\label{fig:subdivion1}
		Subdivision of a truss: (a) construction of compression-tension directions at each joint; (b) a compression-tension field for a truss; (c) the strategy to calculated an edge mid-point.}
\end{figure}

 We first calculate a pair of tension-compression directions at each joint. As shown in Figure \ref{fig:subdivion1}(a), for each joint, we separately average the bars connected with this joint according to their force signs (+ for compression (red) and - for tension (blue)) with their force magnitudes as weights. Figure \ref{fig:subdivion1}(b) shows the calculated nearly-orthogonal directions on a truss. Then we calculate the new joints for the bars which are estimated to be split. Take the bar in Figure \ref{fig:subdivion1}(c) for example, we know the coordinates of its two ends, $p_i$ and $p_j$, and the tension-compression directions at its two ends. For a bar in tension, we calculate the two directions, $v_i$ and $v_j$, which are orthogonal to the compression directions (red) at its two ends. Using $(p_i,p_j,v_i,v_j)$, we calculate a B\'{e}zier curve and set the mid-point of the curve as the new joint. For a bar in compression, we follow a similar procedure.

\begin{figure}[H]	\centering{
		\begin{overpic}
			[width=.89\columnwidth]{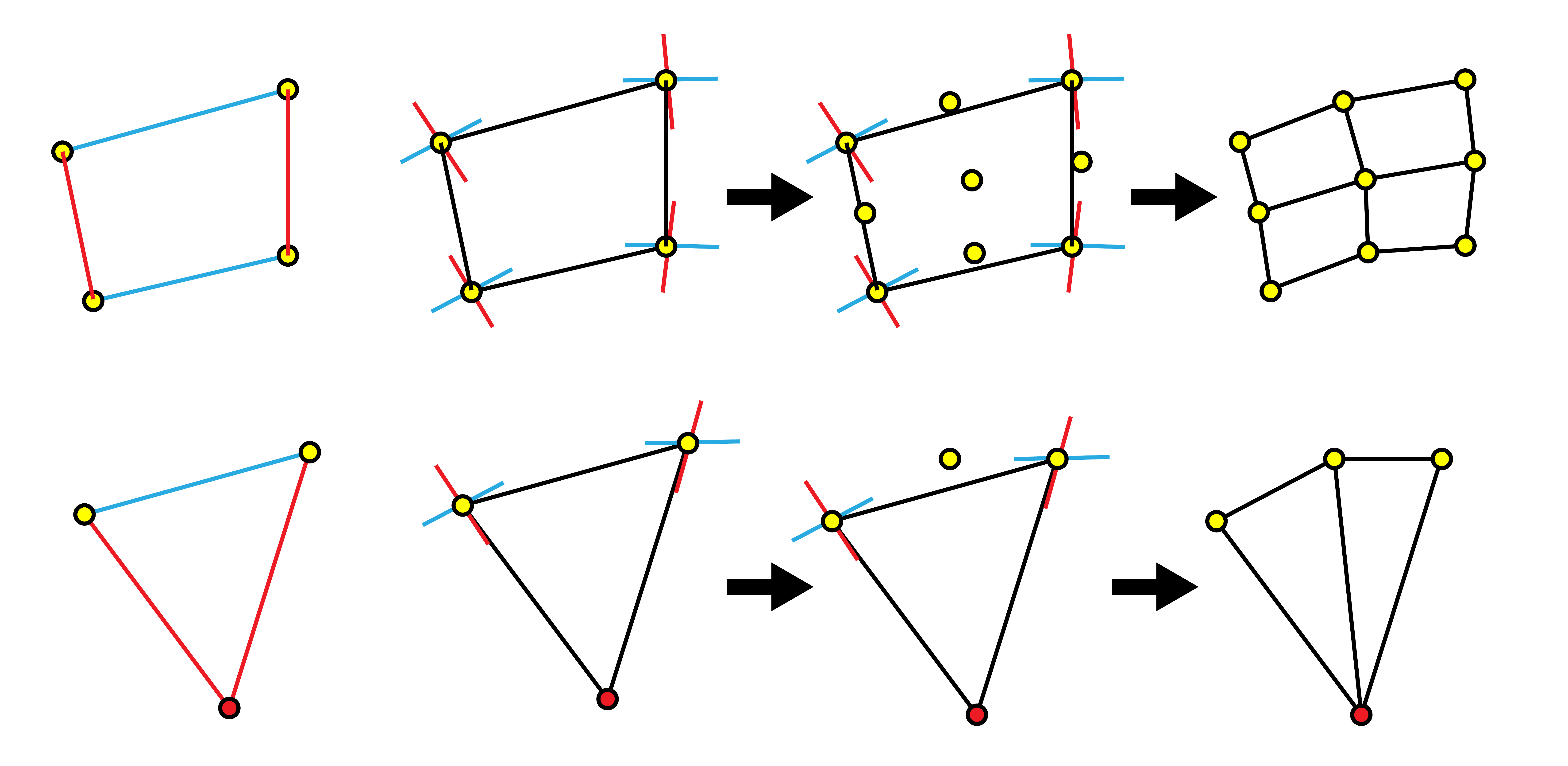}
		\end{overpic}
	}	
	\caption{
		\label{fig:subdivion2}
		Truss subdivision strategies for quadrilaterals (top) and triangles (bottom).}
\end{figure}

\begin{figure}[H]	\centering{
		\begin{overpic}			
			[width=.32\columnwidth,trim={5cm 9cm 4.5cm 9cm},clip]{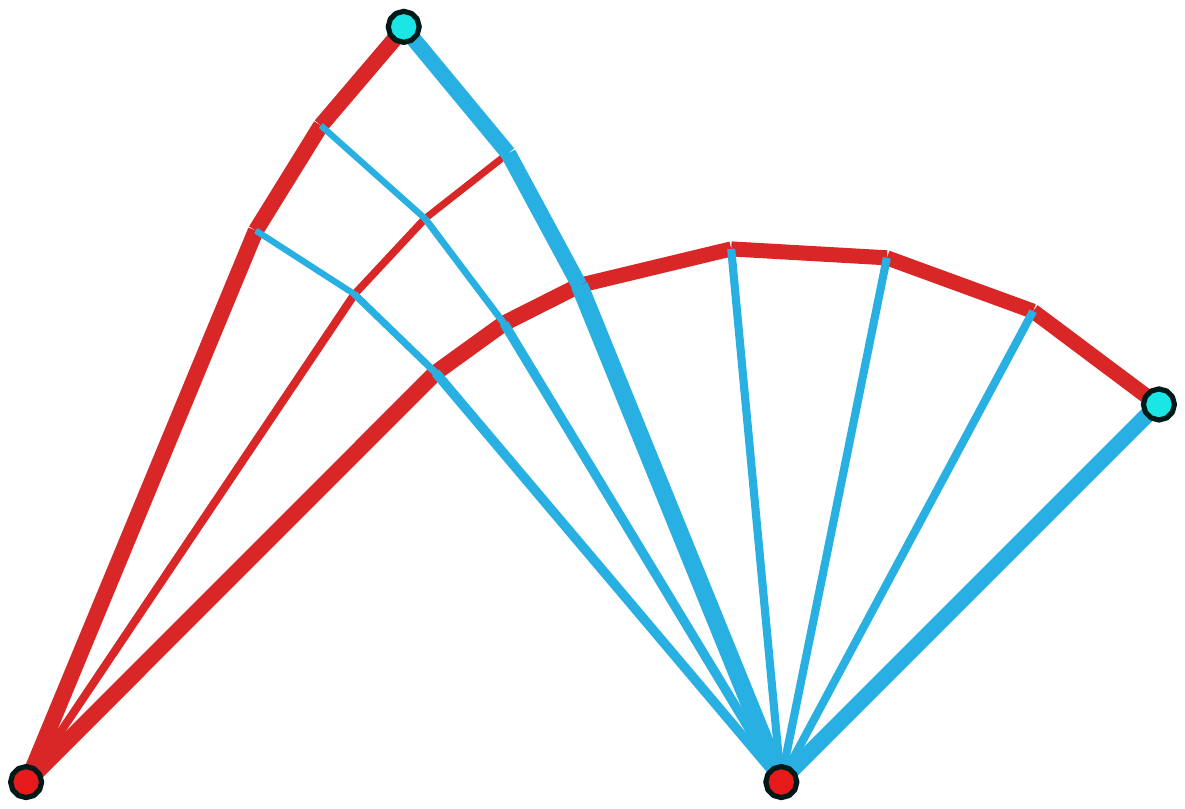}	
			\cput(50,2){\contour{white}{$(a)$}}		
		\end{overpic}
		\begin{overpic}			
			[width=.32\columnwidth,trim={5cm 9cm 4.5cm 9cm},clip]{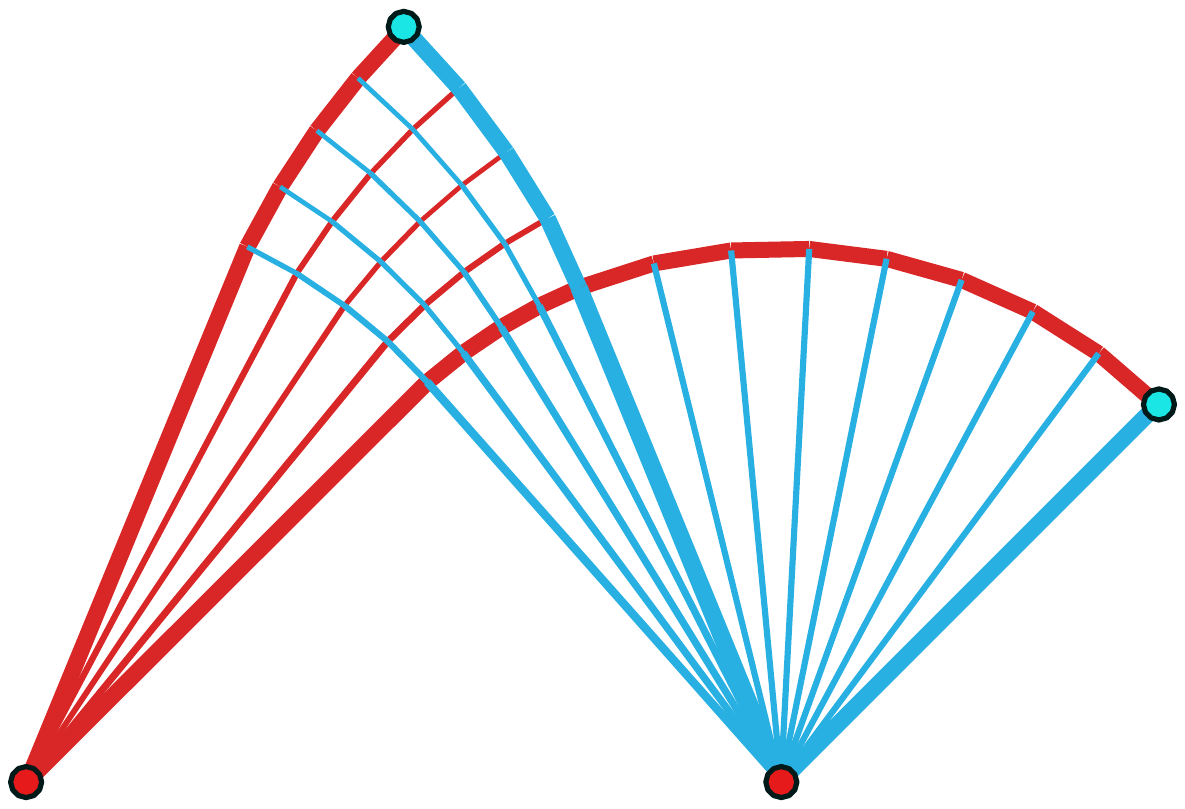}	
			\cput(50,2){\contour{white}{$(b)$}}	
		\end{overpic}
		\begin{overpic}			
			[width=.32\columnwidth,trim={5cm 9cm 4.5cm 9cm},clip]{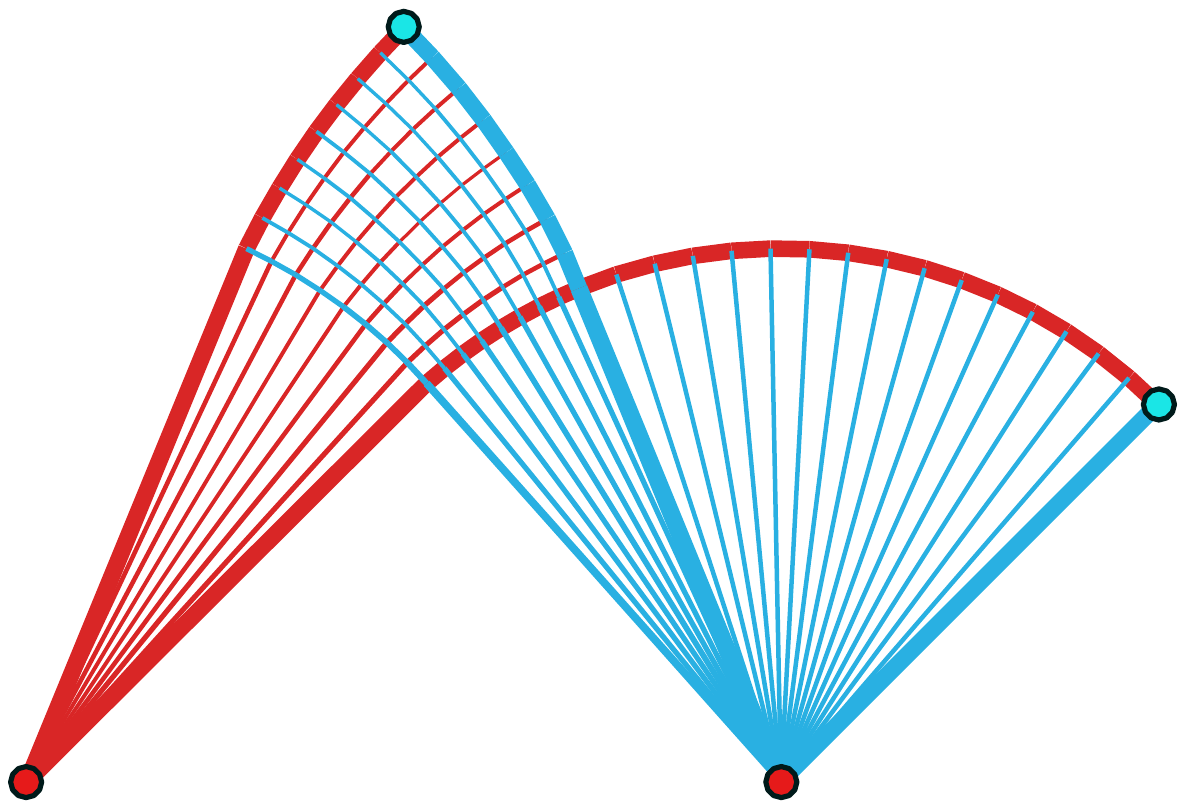}	
			\cput(50,0){\contour{white}{$(c)$}}		
		\end{overpic}
	}	
	\caption{
		\label{fig:subdivion3}
		Optimal truss designs in different subdivision levels. The input functional specification is given in Figure \ref{fig:framework}.}
\end{figure}

The purpose of truss subdivision is to improve the orthogonality of the bars in tension and in compression. 
Given an initial coarse truss, we know its geometry, topology, and the axial force of each bar. Consider the truss as a graph, we extract triangles and quadrilaterals. The triangles are usually formed by bars connected to joints specified in the functional specification. 
For a triangle as shown in Figure \ref{fig:subdivion2}, lower row, we add a new joint for the bar whose force sign is different from the other two and connect the new joint with the opposite joint. A quadrilateral is subdivided if it has two non-adjacent bars in compression and the other two non-adjacent bars in tension. As shown in Figure \ref{fig:subdivion2} upper row, we add four new joints for its four bars and one more joint at its face center initialized as the average of the previous four, and connect the face-center joint with each edge-middle joint. In the subdivided truss, we remove each bar where a new joint is added, and connect its two ends with the new joint as shown in Figure \ref{fig:subdivion2}. 
Figure \ref{fig:subdivion3} show the results of different levels of subdivisions using the functional specification in Figure \ref{fig:framework}(a). 

 
Although the above illustrations are for 2D cases, we can use the same subdivision strategies for 3D trusses. We extract all triangles and quads and test if they should be subdivided. In 3D, we use the same conditions as in 2D (see Fig.~\ref{fig:subdivion2}).

%% file: 4_Algorithm.tex
\section{Alternating LP} \label{ALP}

In this section, we introduce the alternating linear programming (ALP) step which optimizes joint positions and cross-section areas of bars for a given topology. ALP serves as the backbone of the proposed approach. Both the local and global topology operations in the previous section are based on optimization of joint positions and cross-section areas. Directly optimizing joint positions and axial forces is a highly nonlinear problem. Therefore, we split the problem into two linear problems.
 In Algorithm a, we solve force densities alone without changing the joint positions by the ground structure method. In Algorithm b, we update joint positions and force densities jointly based on results from Algorithm a.
 
\subsection{Algorithm a: The Ground Structure Method} 
Let us first recall the basic plastic formulation of the ground structure method ~\cite{zegard2014grand}, which solves a continuous linear programming problem to minimize the total volume of material under the premise of force balance with feasible axial forces:
\begin{align}
& \underset{a_i, s_i}{\text{minimize}}
& & \sum_{i=1}^{|E|} l_i a_i, \label{eqn:lp} \\
& \text{subject to}
& & \mathbf{B}^T \sw = -\fw,                                   \tag{\theequation a} \label{eqn:lp_a} \\
& & & a_i + s_i \geq 0 , & & \  i = 1, \ldots, |E|  \tag{\theequation b} \label{eqn:lp_b} \\
& & & a_i - s_i \geq 0 , & & \  i = 1, \ldots, |E|  \tag{\theequation c} \label{eqn:lp_c} 
\end{align}
where each scalar $a_i$ is the cross-section area of the $i$-th bar. $\mathbf{B}^T$ is the nodal equilibrium matrix, built from the directional cosines of the bars. \AD{More details about this matrix are given in the additional materials.} $\sw$ is a vector with the internal (axial) force for all bars, and $|E|$ is the number of bars. $\fw$ is a vector of the external force for all joints. The internal force $s_i$ should be within the range of admissible axial forces $[-\sigma_Ta_i, \sigma_Ca_i ]$. Here, we assume that the maximal compressive and tensile strains are the same, $\sigma_C$=$\sigma_T$=$\sigma$, which is a constant value. 
We set $\sigma=1$ in the formulation.
The inequality constraints, \ref{eqn:lp_b} and \ref{eqn:lp_c} are equivalent to  $a_i\geq|s_i|$. As the length of each bar, $l_i$, is positive, $l_i>0$, the objective function requires the cross-section area of each bar, $a_i$, to be its smallest permissible value, just enough to support the actual axial force of that bar. Then, we have $a_i=|s_i|$ and the following formulation.
\begin{align}
& \underset{s_i}{\text{minimize}}
& & \sum_{i=1}^{|E|} l_i |s_i|, \label{eqn:lp1} \\
& \text{subject to}
& & \mathbf{B}^T \sw = -\fw.                                   \tag{\theequation a} \label{eqn:lp1_a} 
\end{align}
The above formulation is equivalent when we use force densities $w_i=s_i/l_i$ as variables instead of axis forces $s_i$. The new formulation is transformed to:
\begin{align}
& \underset{w_i}{\text{minimize}}
& & \sum_{i=1}^{|E|} l^{2}_i|w_i|, \label{eqn:lp2} \\
& \text{subject to}
& & \mathbf{C}^T \ww = -\fw.                                   \tag{\theequation a} \label{eqn:lp2_a} 
\end{align}
Here, in Equation \ref{eqn:lp2_a}, the matrix $\mathbf{C}$ is a simpler expression than $\mathbf{B}$ because its elements are linear combinations of joint positions. \AD{More details about the matrix $\mathbf{C}$ are given in the additional materials.}

\subsection{Algorithm b: Relocation of Joints}
In Algorithm a, the joint positions are assumed to be fixed and the axial forces are the only variables. To further reduce the total volume of material, we complement it with Algorithm b and calculate the displacements of joints to leverage more degrees of freedom. 
We assume the initial values of force densities, $\ww$, are known by solving an LP problem in Equation \ref{eqn:lp} and set the difference of joint positions, $\uw$, and the difference of force densities of bars, $\Delta\ww$, as variables. By directly rewriting Equation \ref{eqn:lp2}, we have 
\begin{align}
& \underset{u_i, \Delta w_i}{\text{minimize}}
& & \sum_{i=1}^{|E|} \mathbf{sgn}(w_i+\Delta w_i)(l_i+\Delta l_i)^2 (w_i+\Delta w_i),  \label{eqn:lp4} \\
& \text{subject to}
& & (\mathbf{C}+\Delta \mathbf{C})^T (\ww+\Delta\ww) = -\fw.                                   \tag{\theequation a} \label{eqn:lp4_a}  
\end{align}
Here, we assume that the change of force densities, $\Delta \ww$, is small and that the signs of force densities remain the same, $\mathbf{sgn}(w_i+\Delta w_i)=\mathbf{sgn}(w_i)$. As Algorithm b is applied after Algorithm a, the values, $l_i$, $w_i$, $\mathbf{sgn}(w_i)$, and $\mathbf{C}$, are all known. 

\begin{figure}[h]	\centering{
		\begin{overpic}
			[width=.90\columnwidth]{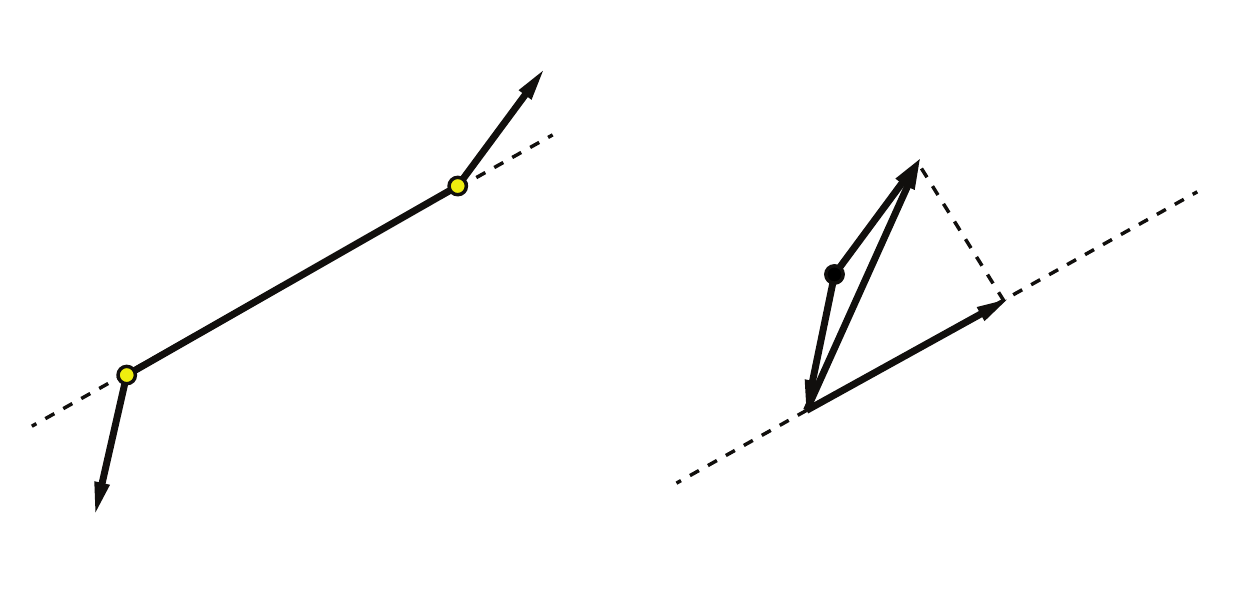}
			\cput(10,20){\contour{white}{$i1$}}
			\cput(25,21){\contour{white}{$i$}}
			\cput(37,36){\contour{white}{$i2$}}
			\cput(12,6){\contour{white}{$\uw_{i1}$}}		
			\cput(47,41){\contour{white}{$\uw_{i2}$}}			
			\cput(87,14){\contour{white}{$\Delta l_i=\dw_i\cdot(\uw_{i2}-\uw_{i1})$}}	
			\cput(25,4){\contour{white}{$(a)$}}	
			\cput(75,4){\contour{white}{$(b)$}}				
			\cput(62,20){\contour{white}{$\uw_{i1}$}}		
			\cput(67,30){\contour{white}{$\uw_{i2}$}}
		\end{overpic}
	}	
	\caption{
		\label{fig:barj}
		(a) The $i$-th bar connects the joints $i1$ and $i2$; $\uw_{i1}$ and $\uw_{i2}$ are joint displacements. (b) The length change along the bar direction, $\Delta l_i\approx\dw_i\cdot(\uw_{i2}-\uw_{i1})$.}
\end{figure}

To simplify the problem which has a cubic objective function and quadratic constraints, our goal is to approximate the above formulation with a linear programming problem and solve it in a sequential manner in conjunction with Algorithm a. 
By expanding the objective function, we have $(l_i+\Delta l_i)^2 (w_i+\Delta w_i)=(l_i^2w_i+2l_i\Delta l_iw_i+\Delta l_i^2w_i+l_i^2\Delta w_i+2l_i\Delta l_i\Delta w_i+\Delta l_i^2\Delta w_i)\approx(l_i^2w_i+2l_i\Delta l_iw_i+l_i^2\Delta w_i)$. Here, we remove the higher order terms, and use the fact that $l_i^2w_i$ is constant. The objective function is approximated by $\sum_{i=1}^{|E|} \mathbf{sgn}(w_i)(2l_i\Delta l_iw_i+l_i^2\Delta w_i)$. As shown in Figure \ref{fig:barj}, $\Delta l_i\approx\dw_i\cdot(\uw_{i2}-\uw_{i1})$, \CG{ where $\dw_i$ is the unit direction vector of the i-th bar connecting the joints $i1$ and $i2$}. The objective function is linear with $\uw_j$ and $\Delta w_i$ as variables, \CG{where  $j = 1, \ldots, |V|$ and $i = 1, \ldots, |E|$}. 

The force equilibrium constraint, $(\mathbf{C}+\Delta \mathbf{C})^T (\ww+\Delta\ww) = -\fw$, is equivalent to $\mathbf{C}^T\Delta\ww+\Delta \mathbf{C}^T\ww+\Delta \mathbf{C}^T\Delta\ww = \mathbf{0}$ because $\mathbf{C}^T \ww = -\fw$ is ensured by Algorithm a. Here we remove the small higher order term $\Delta \mathbf{C}^T\Delta\ww$. Then the force balance constraint is linearized as $\mathbf{C}^T\Delta\ww+\Delta \mathbf{C}^T\ww=\mathbf{0}$. The matrix $\mathbf{C}$ and the vector $\ww$ are known from Algorithm a and the elements in matrix $\Delta \mathbf{C}$ are linear combinations of nodal variations $\uw_j=(u_{jx},u_{jy},u_{jz})$. Then the constraint is also linear with respect to $\uw_j$ and $\Delta w_i$.

Finally, the formulation for Algorithm b is written as

\begin{align}
& \underset{u_i, \Delta w_i}{\text{minimize}}
& &  \sum_{
	i=1
	}^{|E|} \mathbf{sgn}(w_i)(2l_iw_i\dw_i\cdot(\uw_{i2}-\uw_{i1})+l_i^2\Delta w_i), \label{eqn:lp5} \\
& \text{subject to}
& & \mathbf{C}^T\Delta\ww+\Delta \mathbf{C}^T\ww= \mathbf{0},                                                                 \tag{\theequation a} \label{eqn:lp5_a} \\
& & & -\delta _i\leq\Delta w_i\leq \delta _i; \  i = 1, \ldots, |E|,  \tag{\theequation b} & &  \label{eqn:lp5_b} \\
& & & -\lambda  _j\leq u_{jx},u_{jy},u_{jz}\leq \lambda  _j ; \  j = 1, \ldots, |V|.& &   \tag{\theequation c} \label{eqn:lp5_c}  
\end{align}

where $\lambda _j$ and $ \delta _i $ are the bounds of the variables $u_i$ and $\Delta w_i$. In each iteration, we set small values for these bounds, e.g., $\delta _i=0.1|w_i|$ and $\lambda_i=0.1\bar{l}$, where $\bar{l}$ is the average length of all bars.
\subsection{Alternating Scheme}
\label{sec:53}
The above two algorithms are formulated as two LP problems in Equation \ref{eqn:lp} and Equation \ref{eqn:lp5}. The inputs to Algorithm a are the joint positions, $\pw$, and the functional specification such as the external forces, $\mathbf{LOAD}$, and the supporting points, $\mathbf{SUPP}$, and the outputs are the force densities of bars, $\ww$. The algorithm a is written as $[\ww,V]=\mathbf{ALGa}(\pw,\mathbf{LOAD},\mathbf{SUPP})$, where $V$ is the total volume of materials. The inputs of Algorithm b are the initial force densities, $\ww$, the initial joint positions, $\pw$, and the same functional specification. The outputs are the changing values of joint positions and force densities. Then, Algorithm b is written as $[\uw,\Delta \ww]=\mathbf{ALGb}(\pw,\ww,\mathbf{LOAD},\mathbf{SUPP})$.
In the whole algorithm, we organize them in an alternating way as shown in Algorithm 1. $N_\mathrm{max}$ is the maximum iteration number and $S_\mathrm{max}$ is the maximum line search step.
\begin{algorithm} \label{algorithm1}
	\caption{Alternating LP for truss geometry optimization}\label{euclid}
	\begin{algorithmic}[1]
		\Procedure{Alternating LP}{}
		\State Initial joint positions $\pw$;  $\mathbf{LOAD}$ and $\mathbf{SUPP}$;
		\State $[\ww,V]=\mathbf{ALGa}(\pw,\mathbf{LOAD},\mathbf{SUPP})$;
		\State $\mathbf{Flag} \gets True$; $N\gets 0$;
		\While{$\mathbf{Flag}$}
			\State	$[\uw,\Delta \ww]=\mathbf{ALGb}(\pw,\ww,\mathbf{LOAD},\mathbf{SUPP})$;
			\Procedure{Line search}{}
			\For{$j=0$ to $S_\mathrm{max}$}
			\State $s \gets 2^{-j}$; $\hat{\pw}\gets\pw+s\uw$;
			\State $[\hat{\ww},\hat{V}]=\mathbf{ALGa}(\hat{\pw},\mathbf{LOAD},\mathbf{SUPP})$;
					\If {$\hat{V}<V$} 
					\State $V \gets \hat{V}$; $\pw \gets \hat{\pw}$; $\mathbf{Break}$;
					\Else {\If{$j==S_\mathrm{max}$} $\mathbf{Flag}=False$; \EndIf}			
					\EndIf
			\EndFor
			\EndProcedure
			\State \textbf{endprocedure}
			\State $N\gets N+1$;\If{$N>N_\mathrm{max}$} {$\mathbf{Flag}=False$;} \EndIf				
		\EndWhile
		\State \textbf{endwhile}
		\EndProcedure
	\end{algorithmic}
\end{algorithm}

\begin{figure*}[h]	\centering{
\begin{overpic}			
	[width=1.8\columnwidth]{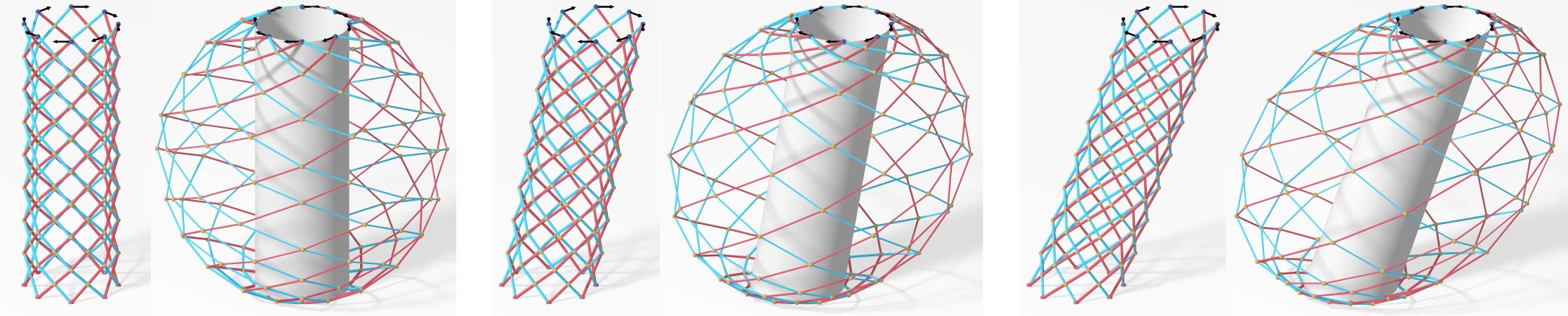}	
	\cput(10,0){\contour{white}{$(a1)$}}	
	\cput(26,0){\contour{white}{$(a2)$}}
	\cput(40,0){\contour{white}{$(b1)$}}
	\cput(60,0){\contour{white}{$(b2)$}}
	\cput(80,0){\contour{white}{$(c1)$}}
	\cput(97,0){\contour{white}{$(c2)$}}
\end{overpic}
	}

	\caption{\CG{Three trusses optimized through ALP for different input specifications, each with eight supporting joints and eight external loads along circles. From left to right, (a1), (b1), and (c1) are the initial trusses with a total volume of material consumption being 103.93, 105.34 and 109.58 respectively. (a2), (b2), and (c2) are the optimal trusses using ALP with a total volumes of 69.88, 68.53 and 69.21, respectively.}
		\label{fig:3Dsphere}
		}
\end{figure*}

\AD{In Figure \ref{fig:3Dsphere}, we show the effectiveness of ALP for truss optimization with three sets of load specifications involving torques.}

\CG{
\section{Extensions}
The generality of our formulation allows extensions of our approach for a broad range of scenarios and applications, such as tackling multiple load specifications, respecting stability analysis, and incorporating project-specific fabrication constraints.
\subsection{Multiple Load Specifications}
For the input specification with multiple sets of external forces, the ALP algorithm is adjusted accordingly. Static equilibrium is required for each set of external loads with generally different internal forces. Thus, assuming trusses are required to withstand $K$ sets of external forces ${\fw}^1$, ... , ${\fw}^K$, the formulation of Algorithm a is rewritten as:

\begin{align*}
& \underset{a_i, s_i^k}{\text{minimize}}
& & \sum_{i=1}^{|E|} l_i a_i,  \\
& \text{subject to}
& & \mathbf{B}^T {\sw}^1 = -{\fw}^1,                                    \\
& & & a_i + s_i^1 \geq 0 , & & \  i = 1, \ldots, |E|    \\
& & & a_i - s_i^1 \geq 0 , & & \  i = 1, \ldots, |E|    \\
& & & ... \\
& & & \mathbf{B}^T {\sw}^K = -{\fw}^K,                                    \\
& & & a_i + s_i^K \geq 0 , & & \  i = 1, \ldots, |E|   \\
& & & a_i - s_i^K \geq 0 , & & \  i = 1, \ldots, |E| 
\end{align*}
where $k = 1, \ldots, K$. Force equilibrium constraints similar to Equation \ref{eqn:lp} are required for each set of external forces. It is also worth noting that this set of equations is sufficient to guarantee force equilibrium in response to linear interpolation of the sets of specified external forces. As each bar needs to support the maximal axial forces from each set of the reaction forces, the cross-section of the $i$-th bar, $a_i=max\{|{s_i}^1|,...,|{s_i}^K|\}$. \AD{We denote $m_i\in\{1,...,k \}$ as the set index for which the $i$-th bar attains its maximal axial force, $|s_i^{m_i}|=max\{|{s_i}^1|,...,|{s_i}^K|\}$. As those indices, $m_i$, could be easily found from the result of Algorithm a, corresponding cross sections follow directly, $a_i=|s_i^{m_i}|$.} Similarly, by defining $w_i^{m_i}=s_i^{m_i}/l_i$, the formulation of Algorithm b for the multiple-load case can be written as:
\begin{align*}
& \underset{u_i, \Delta {w_i}^k}{\text{minimize}}
& &  \sum_{
	i=1
	}^{|E|} \mathbf{sgn}(w_i^{m_i})(2l_iw_i^{m_i}\dw_i\cdot(\uw_{i2}-\uw_{i1})+l_i^2\Delta w_i^{m_i}), \\
& \text{subject to}
& & \mathbf{C}^T\Delta\ww^1+\Delta \mathbf{C}^T\ww^1= \mathbf{0},    \\
& & & ...    \\
& & & \mathbf{C}^T\Delta\ww^K+\Delta \mathbf{C}^T\ww^K= \mathbf{0},    \\
& & & -\delta _i\leq\Delta w_i^k\leq \delta _i, \\
& & & -\lambda  _j\leq u_{jx},u_{jy},u_{jz}\leq \lambda  _j, 
& &    
\end{align*}
where $i = 1, \ldots, |E|$, $k = 1, \ldots, K$, and $j = 1, \ldots, |V|$. Here, only the nodal variations, $u_i$, and the change of force densities, $\Delta w_i^k$, are variables, others are known from Algorithm a. Using the same alternating scheme in Section \ref{sec:53} by adjusting the Algorithm a and b accordingly, our method can tackle cases of multiple-load input specifications. \AD{See Figure \ref{fig:multiload} for an example.}
\begin{figure}[h]\centering{
		\begin{overpic}			
			[width=0.99\columnwidth]{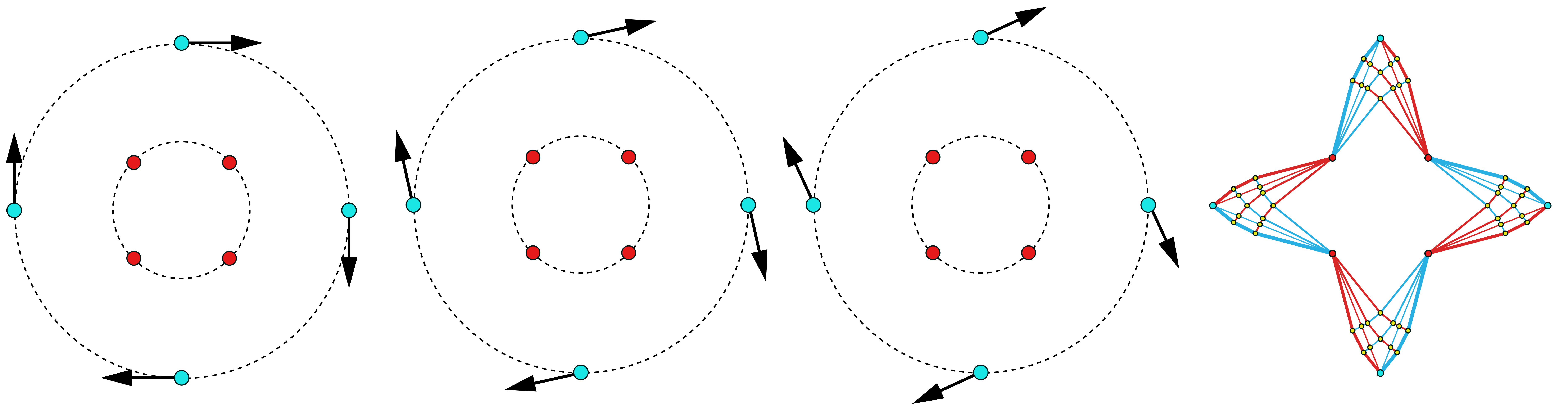}	
			\cput(19,1){\contour{white}{$(a)$}}	
			\cput(44,1){\contour{white}{$(b)$}}	
			\cput(70,1){\contour{white}{$(c)$}}	
			\cput(92,1){\contour{white}{$(d)$}}	
		\end{overpic}
	}	
	\caption{\CG{A truss designed to support three sets of different external loads. Here, \AD{red joints are fixed and blue joints have external loads applied}. With three sets of external loads shown in (a)-(c), our method creates an optimal truss that supports all of them, as shown in (d).} 
		\label{fig:multiload}}
\end{figure}
\subsection{Stability and Buckling}
Truss stability, in particular, the external stability, has been investigated extensively for its importance in keeping its rigid shape. By comparing the number of degrees of freedom against constraint counts, a condition for external stability is given in ~\cite{kassimali2011structural} as
\begin{align*}
|E|+r\geq d|V|.
\end{align*}
With $|E|$ being the number of bars and $r$ being the count of reaction components, the sum on the left-hand-side includes the number of both internal (e.g., fixed edge lengths) and external (e.g., supported vertices) constraints. The right-hand-side counts the total number of degrees of freedom, as a product of the number of joints, $|V|$, and the number of degrees of freedom for each joint, $d$, which is 2 for plane trusses and 3 for spatial trusses. A truss is externally stable if the condition is met and unstable otherwise. There are multiple approaches to integrate stability analysis to our framework, e.g., through a post-processing step checking the aforementioned condition and \AD{adding necessary} auxiliary bars with a minimum cross-section to increase the number of $|E|$ on the left side and thus achieving external stability. Figure \ref{fig:stability} provides an example. Similarly to stability analysis, buckling analysis through simulation packages like ANSYS are applicable as a feedback procedure to improve the truss structural properties of interest in an interactive manner.

\begin{figure}[h]\centering{
		\begin{overpic}			
			[height=0.18\columnwidth]{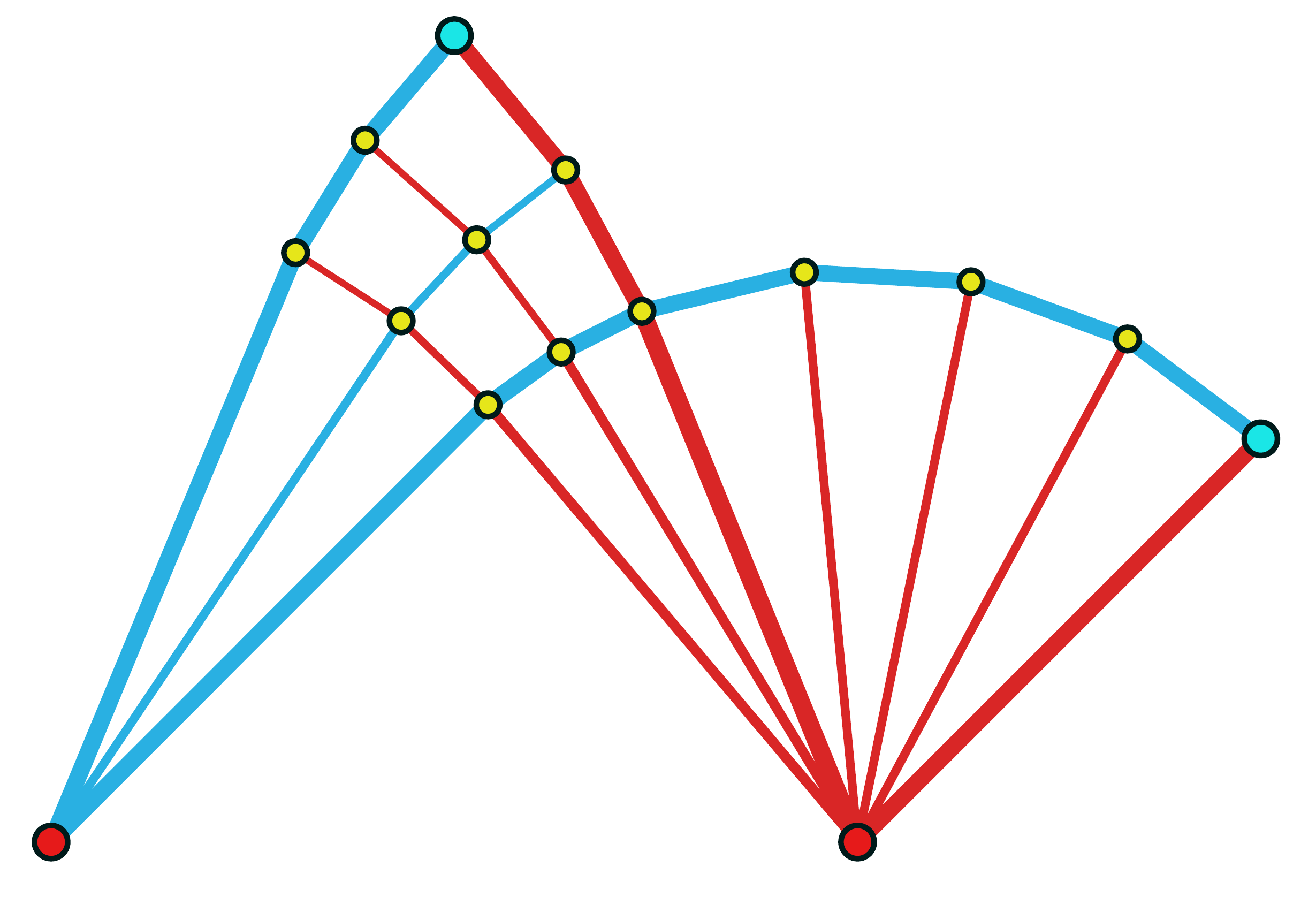}	
			\cput(29,1){\contour{white}{$(a)$}}	
		\end{overpic}
		\begin{overpic}			
			[height=0.18\columnwidth]{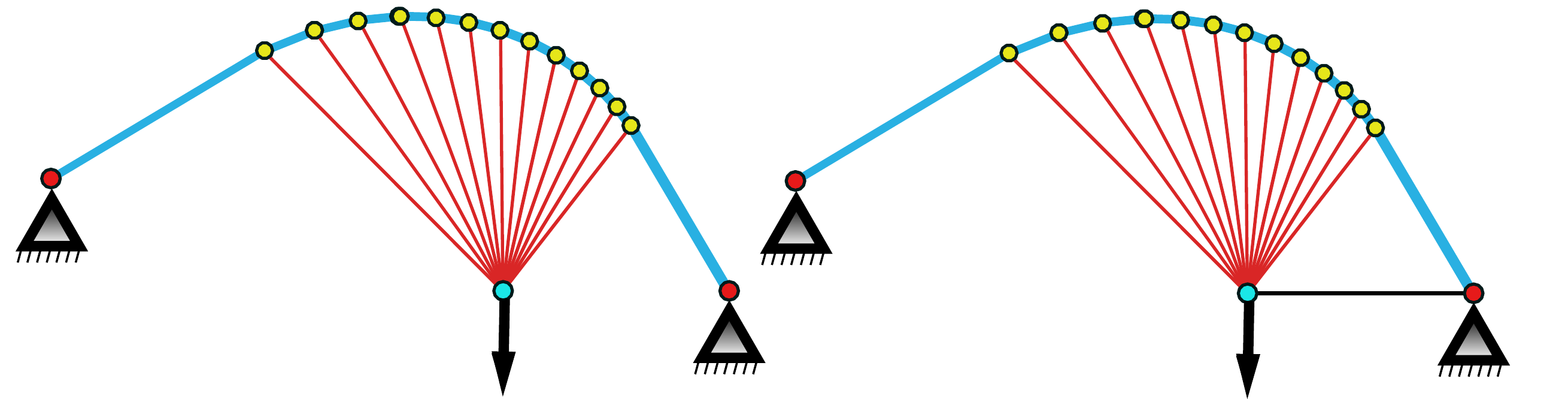}	
			\cput(15,1){\contour{white}{$(b1)$}}	
			\cput(65,1){\contour{white}{$(b2)$}}	
		\end{overpic}
	}	
	\caption{\CG{(a) With $|E|$=26, r=4, $|V|$=15, the truss is externally stable.  (b1) With $|E|$=27, r=4, $|V|$=16, the truss is is externally unstable. (b2) With one more bar added, $|E|$=28, r=4, $|V|$=16, the truss in (b1) becomes externally stable.} 
		\label{fig:stability}}
\end{figure}


\subsection{Fabrication Constraints}
Besides general engineering settings, our system is able to incorporate project-specific fabrication constraints through minor adjustments or complementary post-processing procedures. For projects requiring a limited number of parts, bounding the number of subdivision steps controls the amount of bars and joints. Meanwhile, while our algorithm reduces the total volume of material consumption based on a continuous selection of cross-sections in the default setting, for engineering practices which demand viable bars to be chosen from a set of predefined category or a discrete number of types, our method can be combined with other published algorithms. We show an example in limiting the types of cross sections by applying ~\cite{jiang2017design} directly as a post-processing procedure in Figure \ref{fig:discrete}.

\begin{figure}[h]\centering{
		\begin{overpic}			
			[height=0.16\columnwidth]{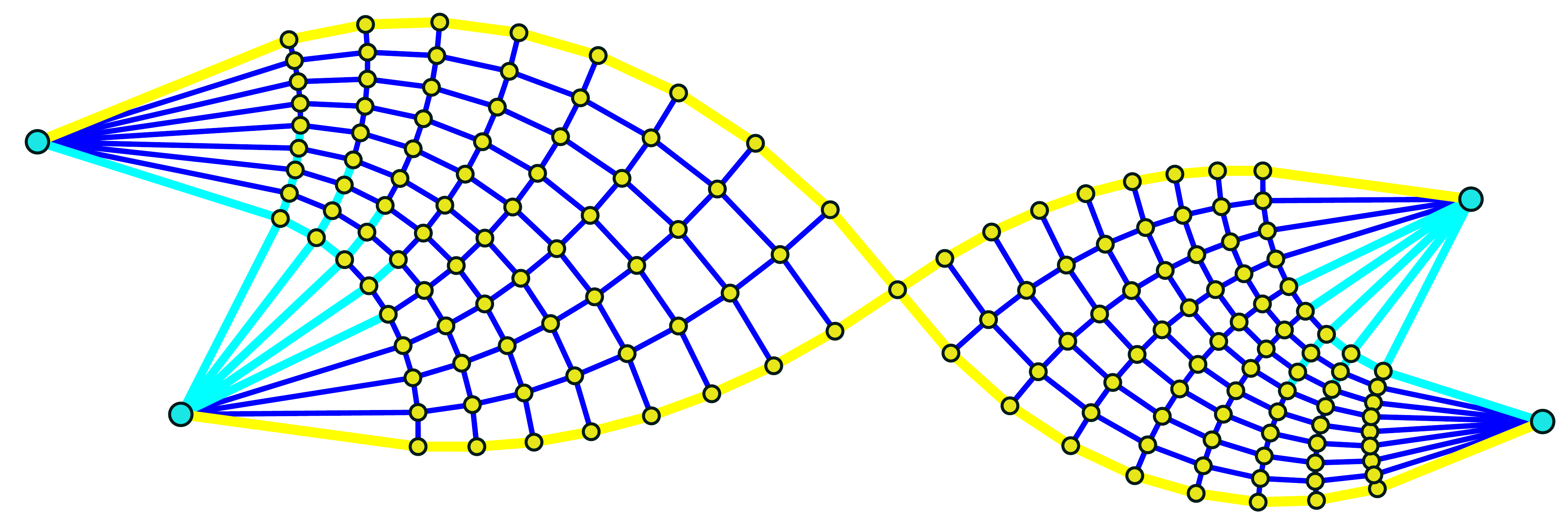}	
			\cput(50,1){\contour{white}{$(a)$}}	
		\end{overpic}
		\begin{overpic}			
			[height=0.16\columnwidth]{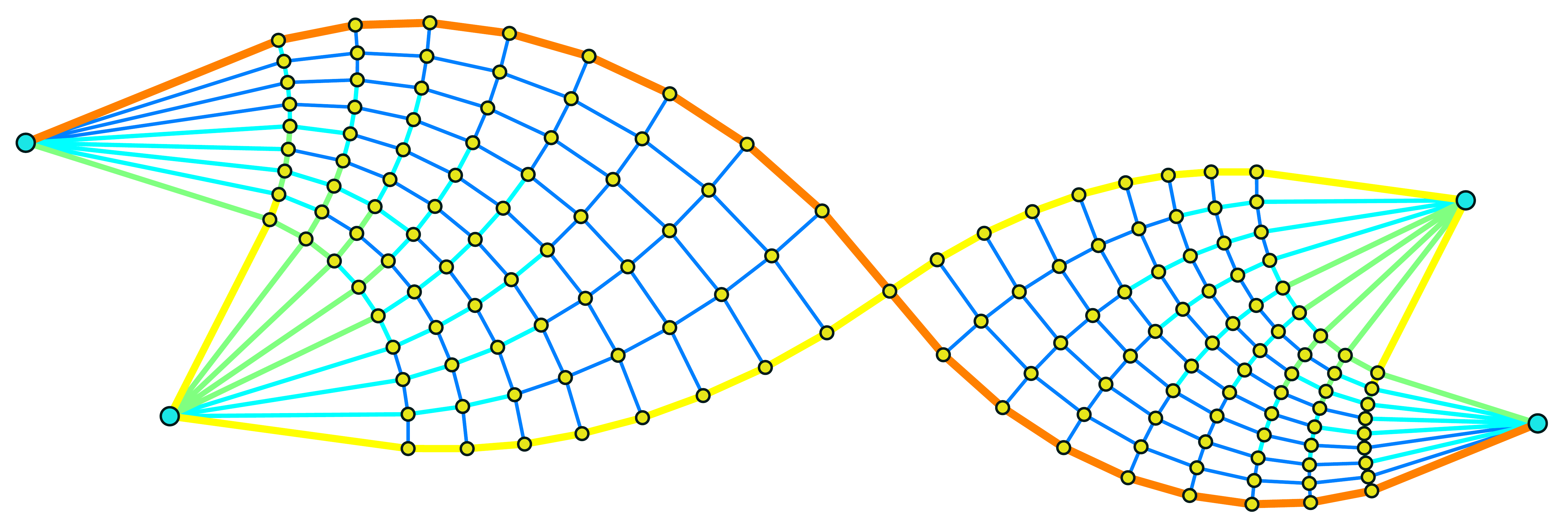}	
			\cput(50,1){\contour{white}{$(b)$}}	
		\end{overpic}
	}	
	\caption{\CG{A truss optimized by our method in Figure \ref{fig:nonconcurrentforce}, with a total volume of 3.188, is post-processed by a discrete optimization algorithm in ~\cite{jiang2017design} to obtain trusses constructed by (a) three types of cross-section areas with a total volume of 4.674, and (b) five types of cross-section areas with a total volume of 3.832.}
		\label{fig:discrete}}
\end{figure}



}

%% file: 6_Results.tex
\section{Results}

In this section, we illustrate truss designs using our framework for different types of input specifications and compare the results with state-of-the-art methods on selected benchmark design problems.

\subsection{Example Designs}
\begin{figure*}[tb]	\centering{
				\begin{overpic}			
					[width=.5\columnwidth,trim={4.5cm 9cm 1cm 9cm},clip]{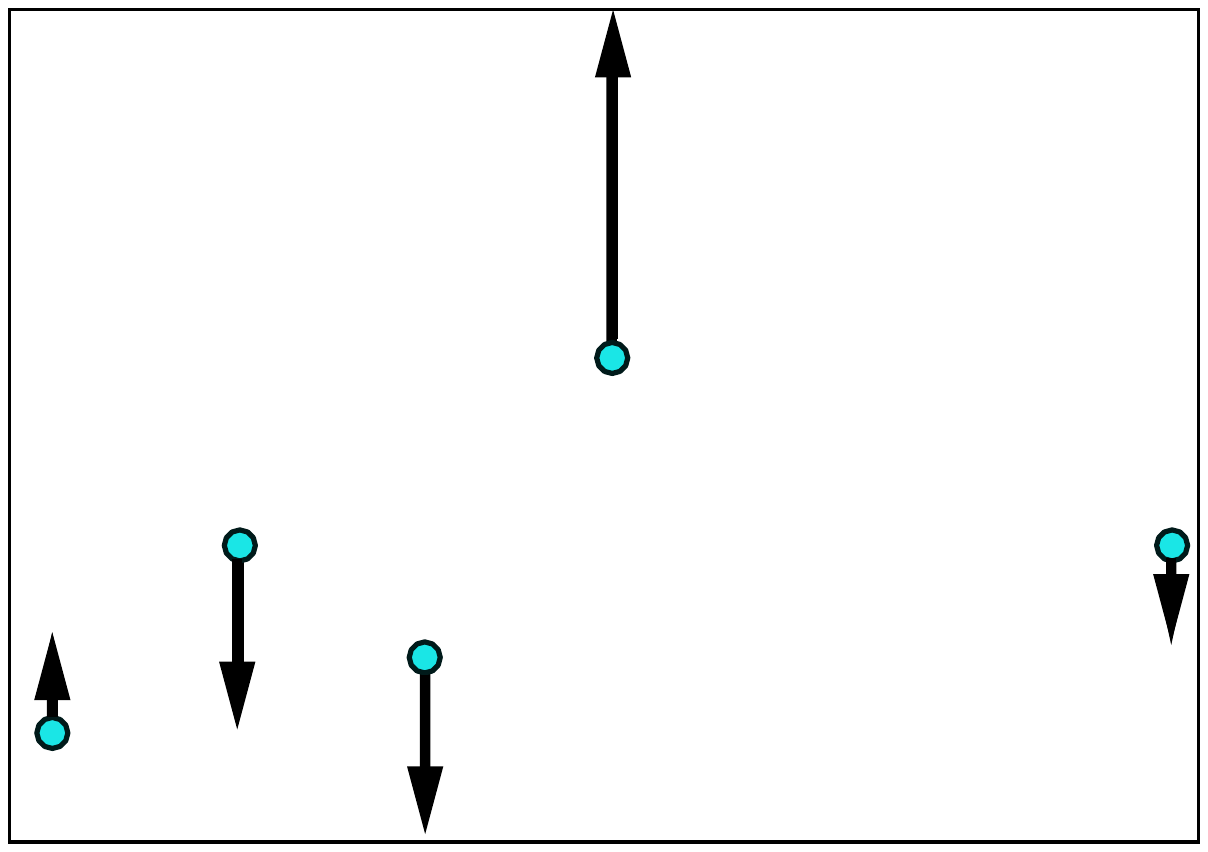}	
				\end{overpic}
		\begin{overpic}			
			[width=.4\columnwidth,trim={5cm 9cm 4cm 9cm},clip]{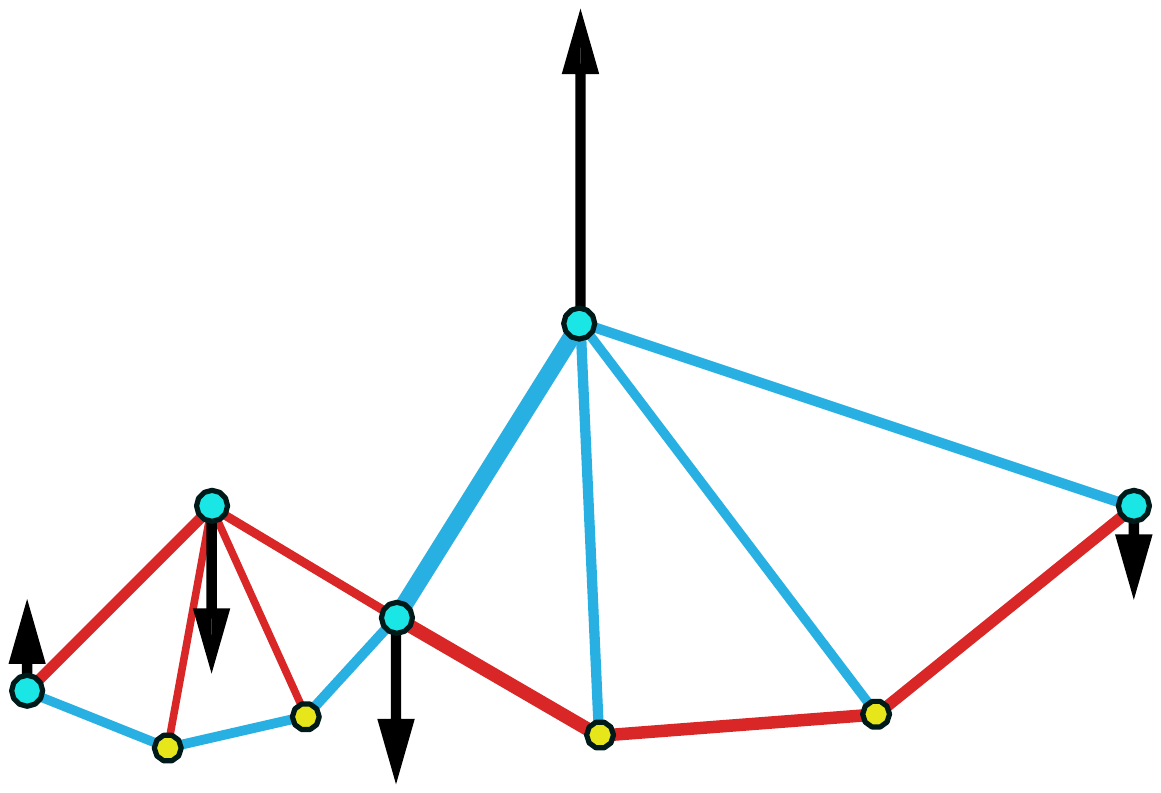}	
			\cput(30,2){\contour{white}{$(a)$}}		
		\end{overpic}
		\begin{overpic}			
			[width=.4\columnwidth,trim={5cm 9cm 4cm 9cm},clip]{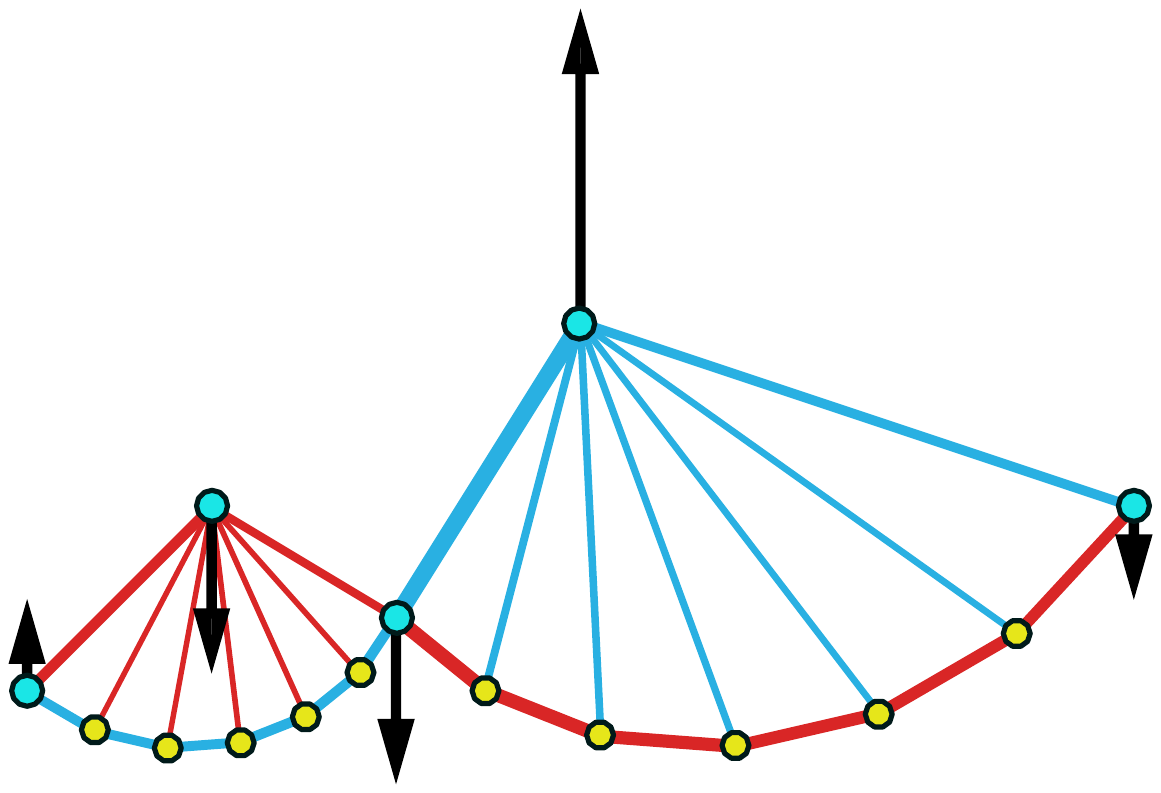}	
			\cput(30,2){\contour{white}{$(b)$}}	
		\end{overpic}
		\begin{overpic}			
			[width=.4\columnwidth,trim={5cm 9cm 4cm 9cm},clip]{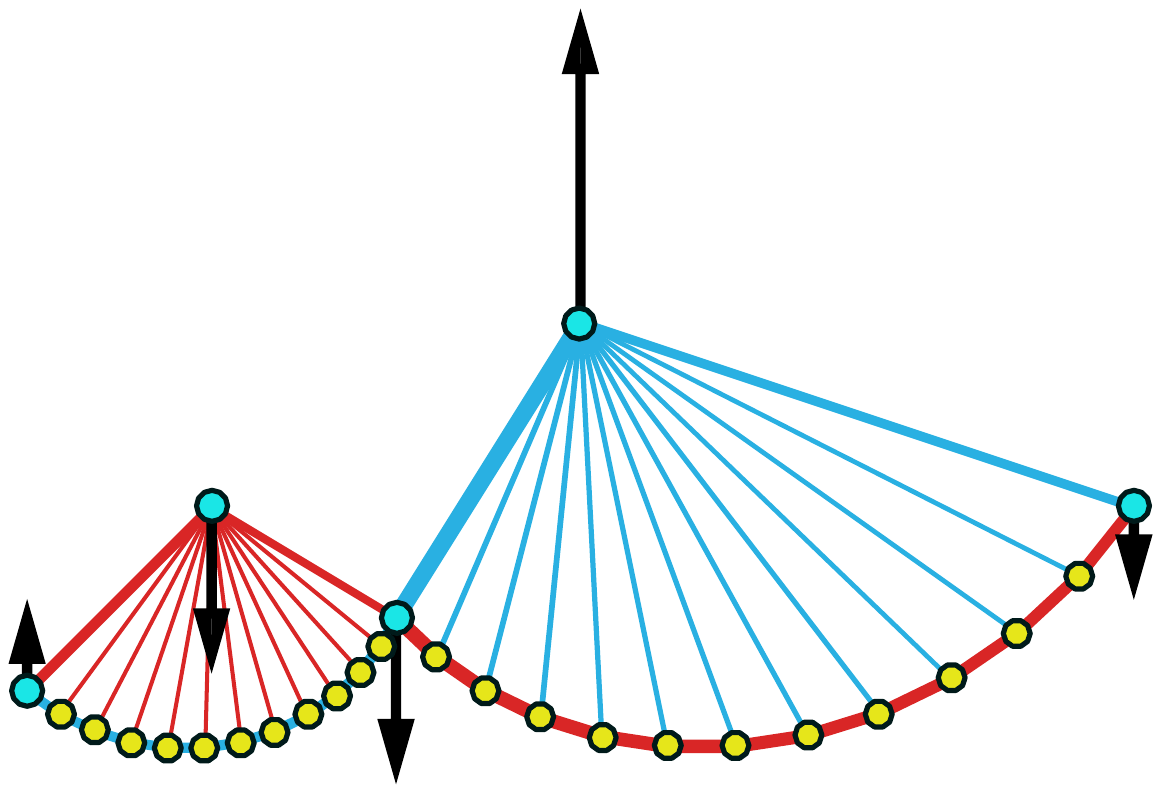}	
			\cput(30,0){\contour{white}{$(c)$}}		
		\end{overpic}
	}	
	\caption{
		\label{fig:parallelforce}
A truss design with the input being a 2D parallel force system. }
\end{figure*}

\begin{figure*}[tb]	\centering{
		\begin{overpic}			
			[width=.5\columnwidth,trim={3cm 8cm 0cm 10cm},clip]{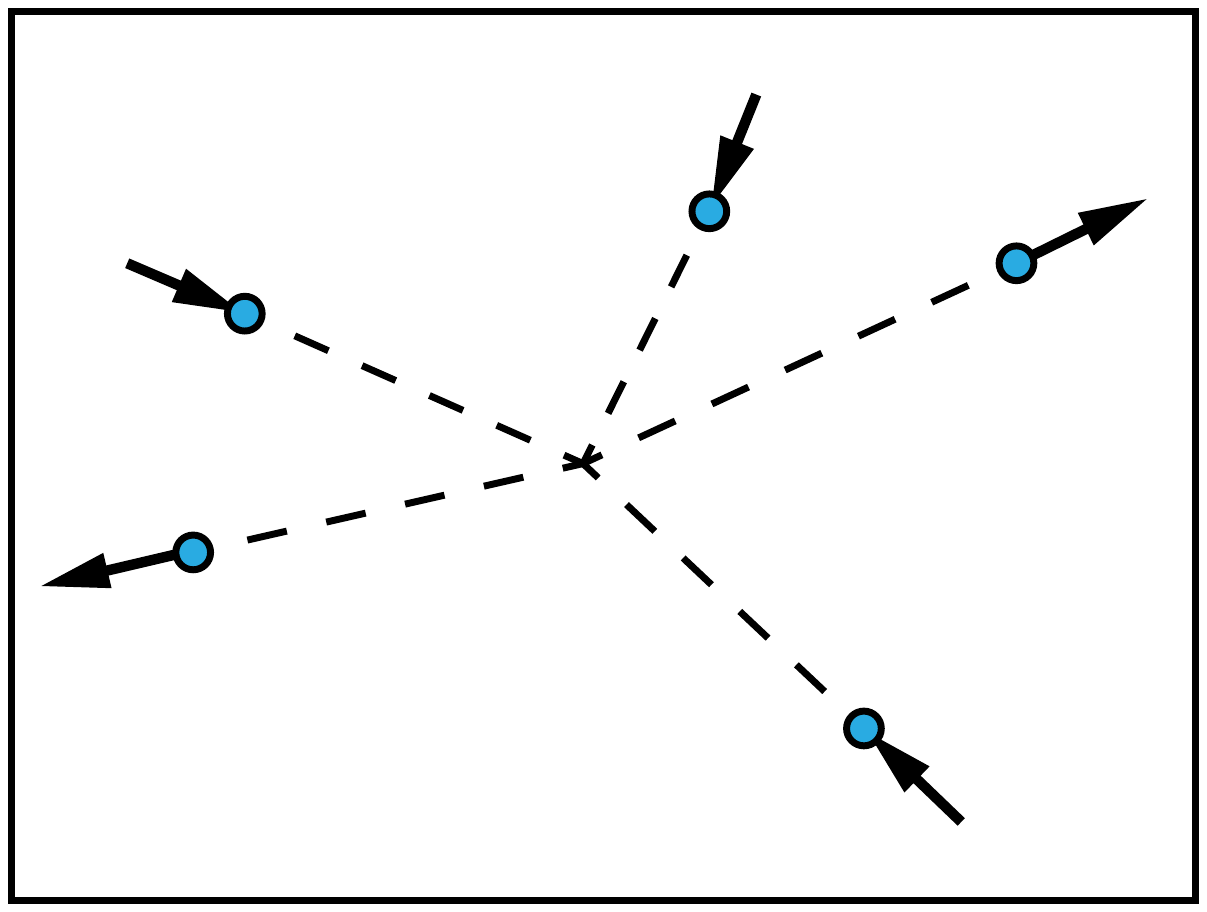}	
		\end{overpic}
		\begin{overpic}			
			[width=.40\columnwidth,trim={0cm 8cm -1cm 7cm},clip]{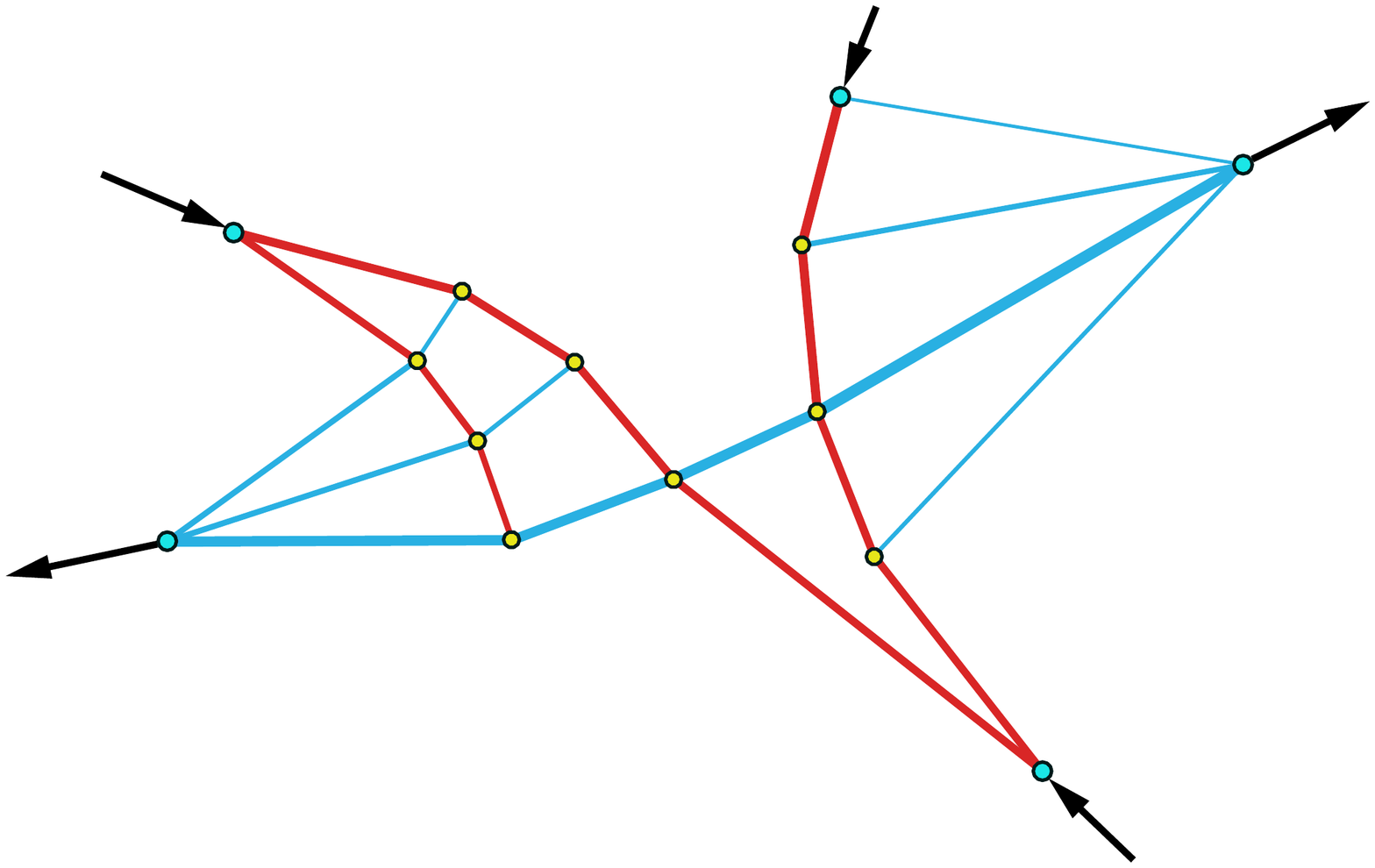}	
			\cput(30,2){\contour{white}{$(a)$}}		
		\end{overpic}
		\begin{overpic}			
			[width=.40\columnwidth,trim={0cm 8cm -1cm 7cm},clip]{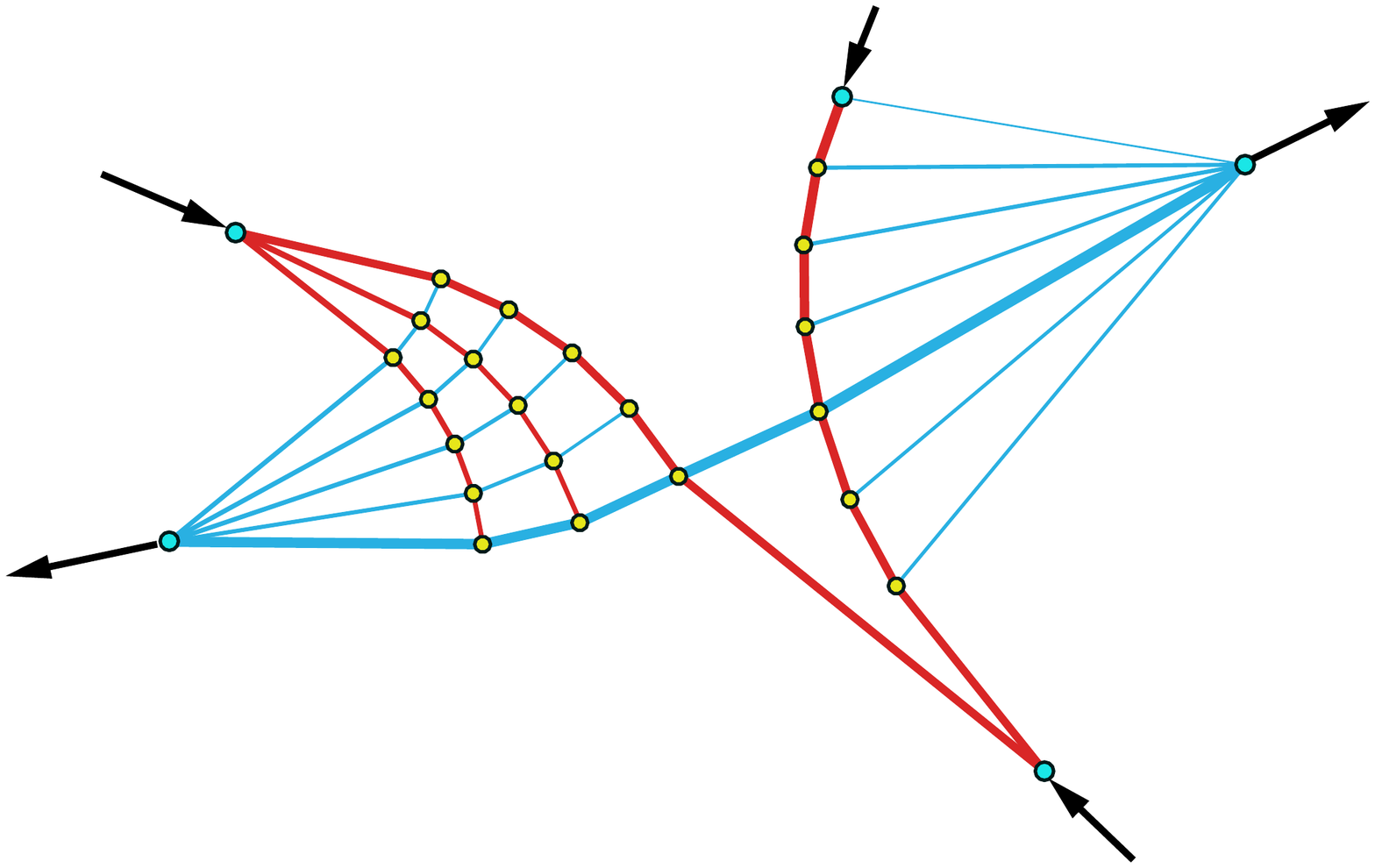}	
			\cput(30,2){\contour{white}{$(b)$}}	
		\end{overpic}
		\begin{overpic}			
			[width=.40\columnwidth,trim={0cm 8cm -1cm 7cm},clip]{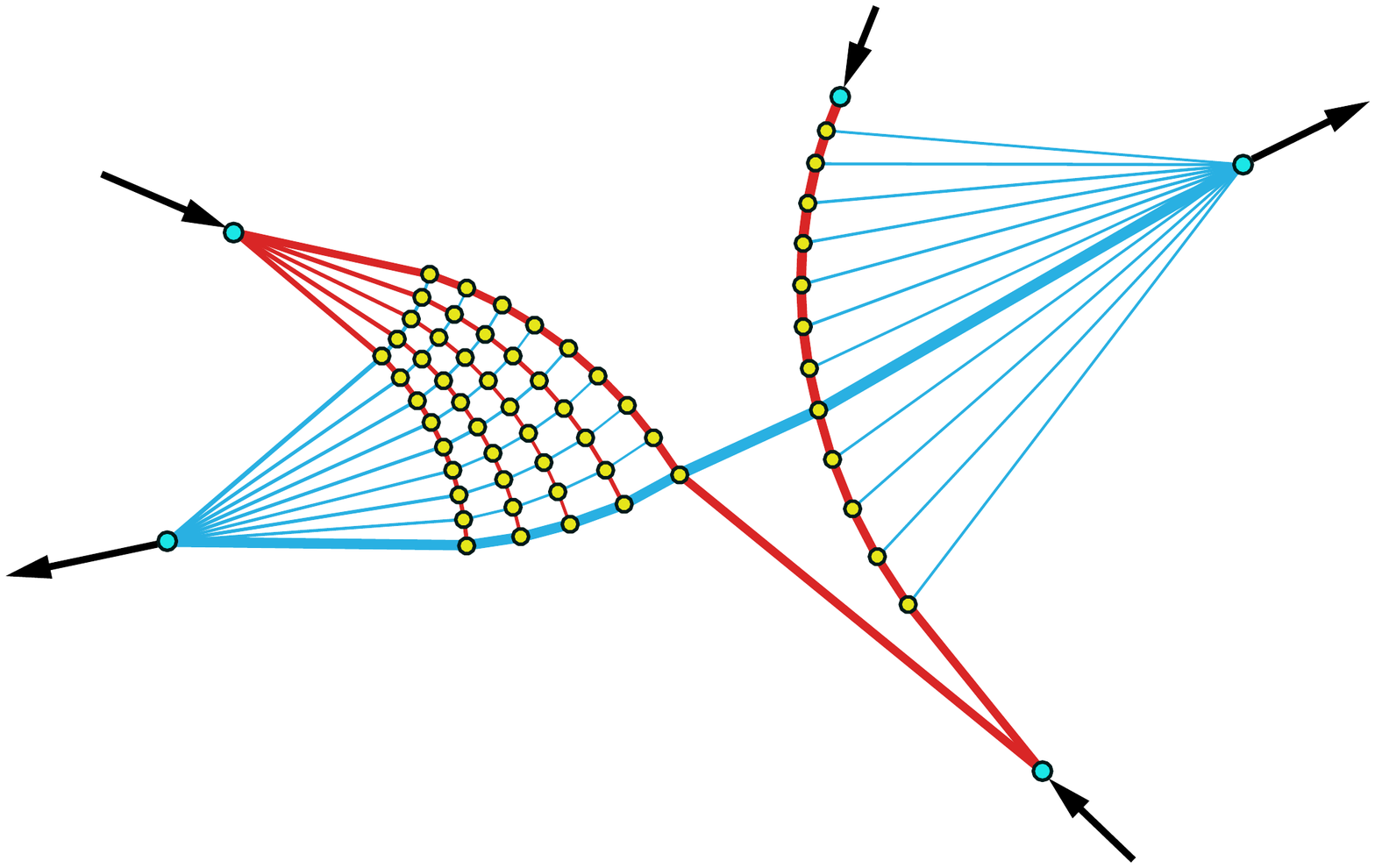}	
			\cput(30,0){\contour{white}{$(c)$}}		
		\end{overpic}
	}	
	\caption{
		\label{fig:concurrentforce}
A truss design with the input being a 2D concurrent force system. Note that the concave quadrilateral in (a) is not subdivided.  }
\end{figure*}

\begin{figure*}[tb]	\centering{
		\begin{overpic}			
			[height=.45\columnwidth,trim={9cm 9cm 7cm 8cm},clip,angle =90]{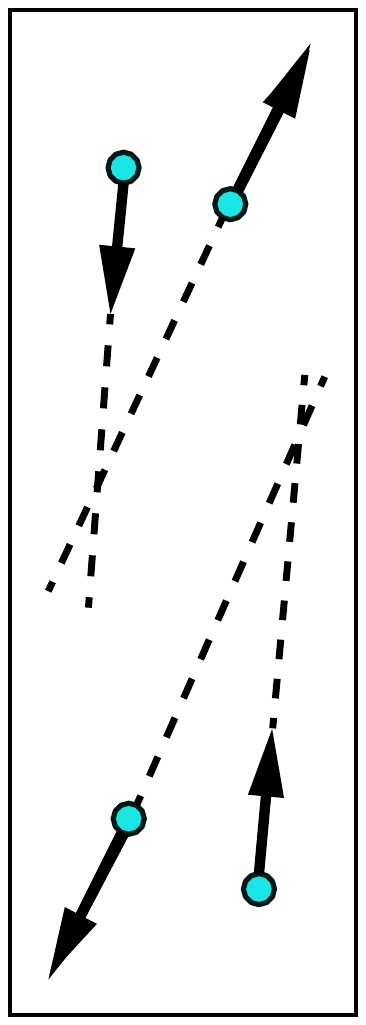}	
		\end{overpic}
		\begin{overpic}			
			[height=.45\columnwidth,trim={5cm 2cm 5cm 1cm},clip,angle =90]{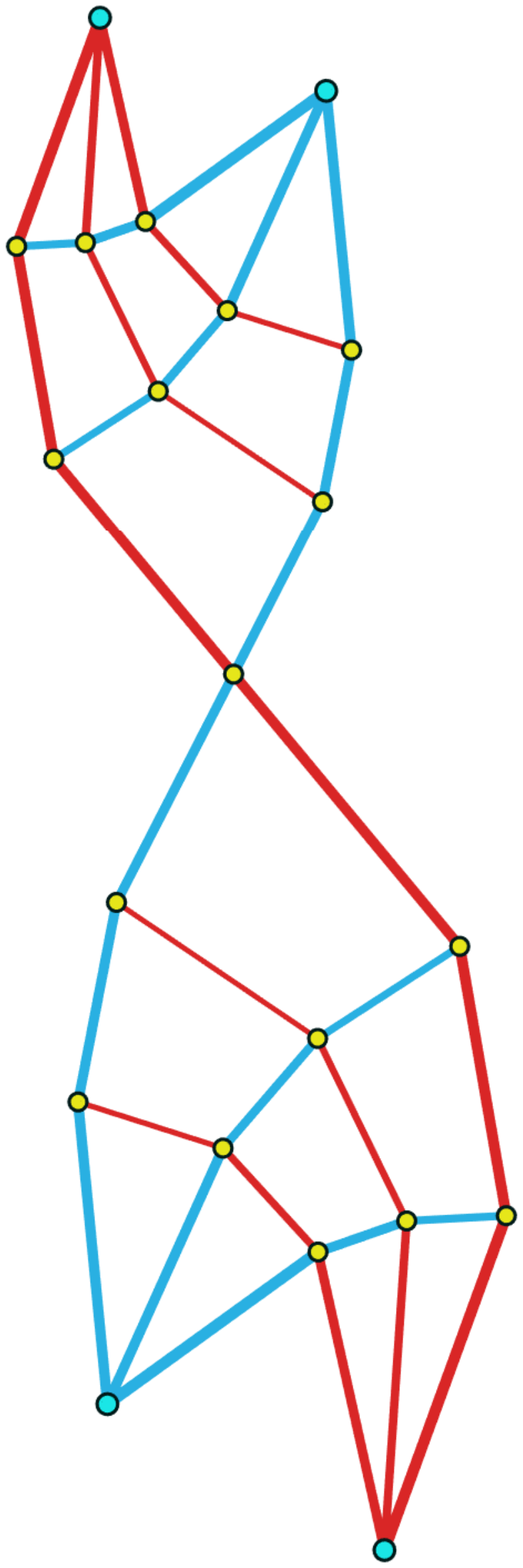}	
			\cput(40,2){\contour{white}{$(a)$}}	
		\end{overpic}
		\begin{overpic}			
			[height=.45\columnwidth,trim={5cm 2cm 5cm 1cm},clip,angle =90]{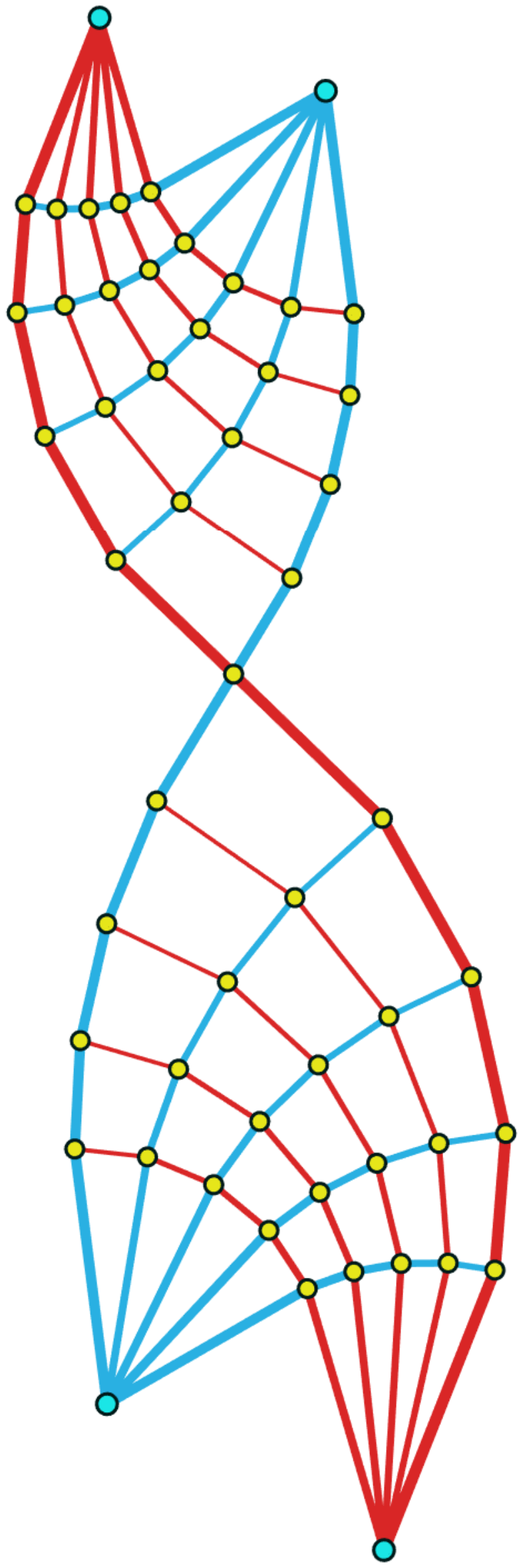}	
			\cput(40,0){\contour{white}{$(b)$}}		
		\end{overpic}
				\begin{overpic}			
					[height=.45\columnwidth,trim={5cm 2cm 5cm 1cm},clip,angle =90]{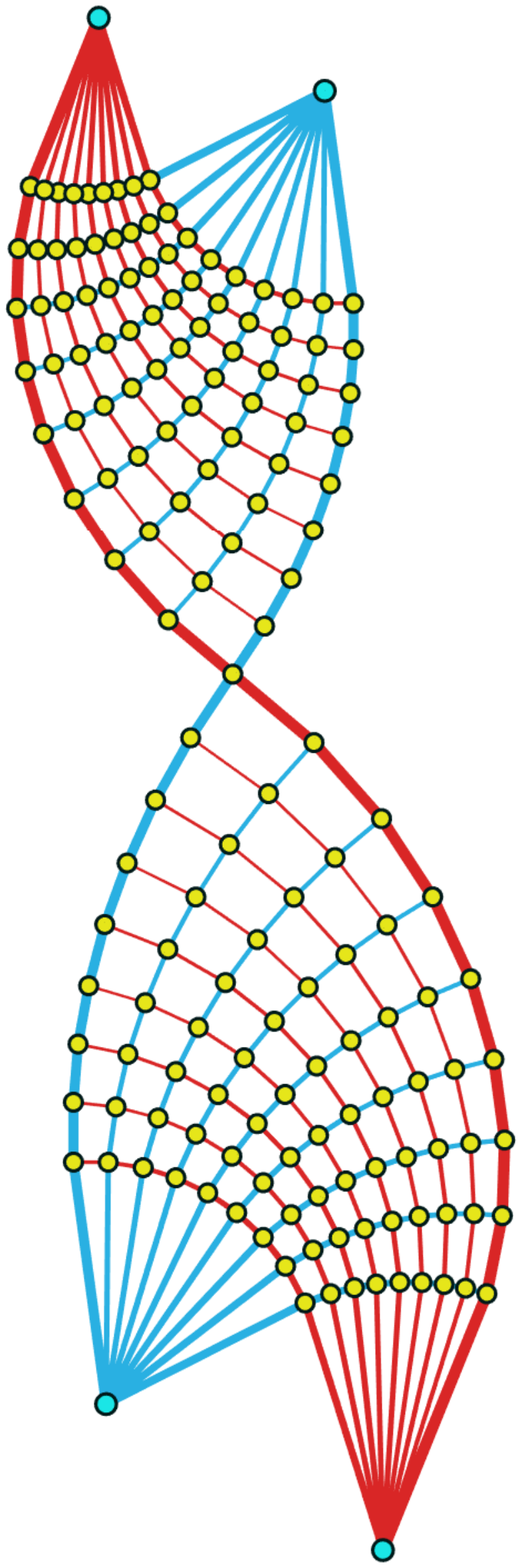}	
					\cput(40,0){\contour{white}{$(c)$}}		
				\end{overpic}
	}	
	\caption{
		\label{fig:nonconcurrentforce}
A truss design with the input being a 2D non-concurrent force system. Note that the orthogonality of the two families of trusses, in compression and tension, is improved with subdivision.}
\end{figure*}

\begin{figure*}[tb]\centering{
		\begin{overpic}			
			[width=0.45\columnwidth,trim={4cm 10cm 4cm 10cm},clip]{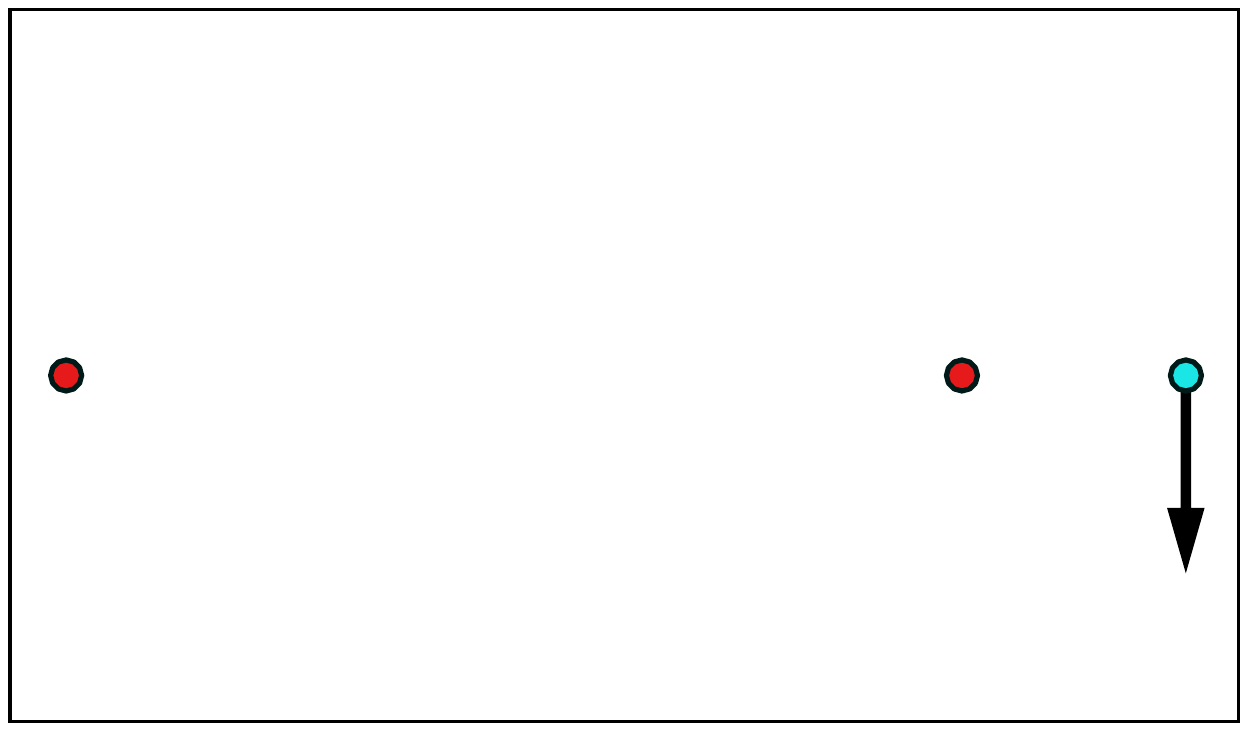}	
			\cput(50,10){\contour{white}{$(a1)$}}		
		\end{overpic}
		\begin{overpic}			
			[width=0.45\columnwidth,trim={4cm 10cm 4cm 10cm},clip]{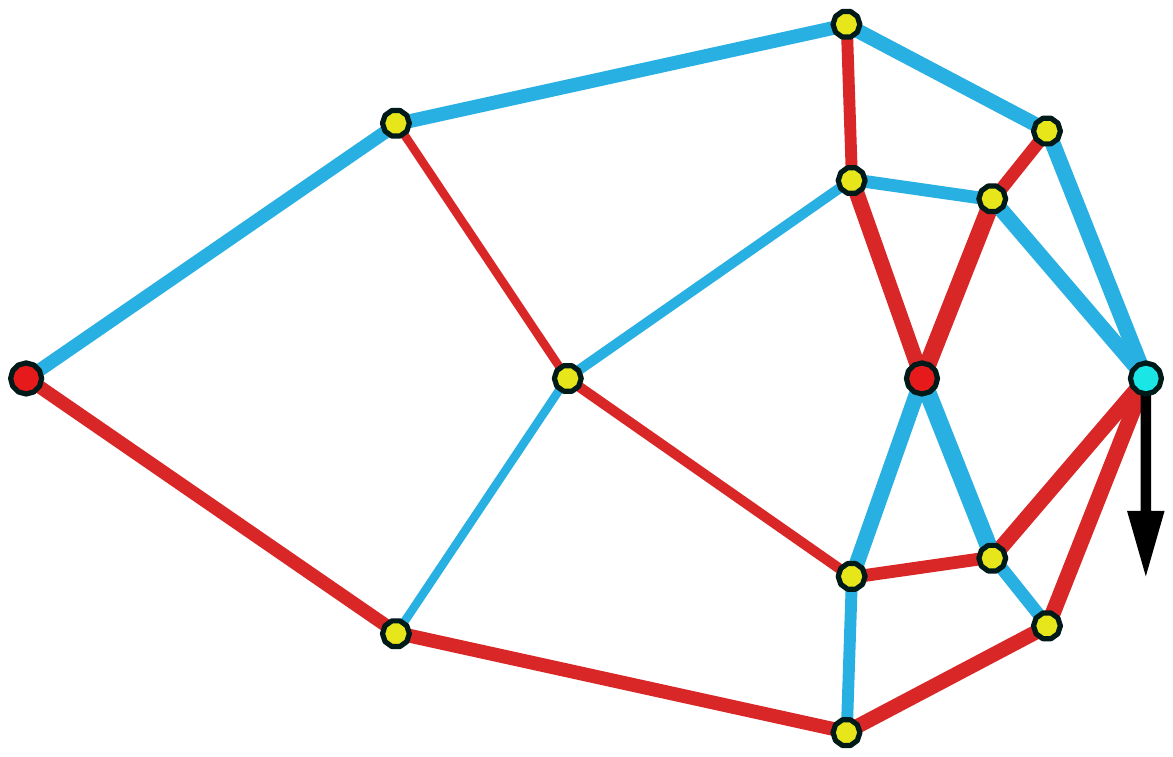}	
			\cput(40,2){\contour{white}{$(a2)$}}	
		\end{overpic}
		\begin{overpic}			
			[width=0.45\columnwidth,trim={4cm 10cm 4cm 10cm},clip]{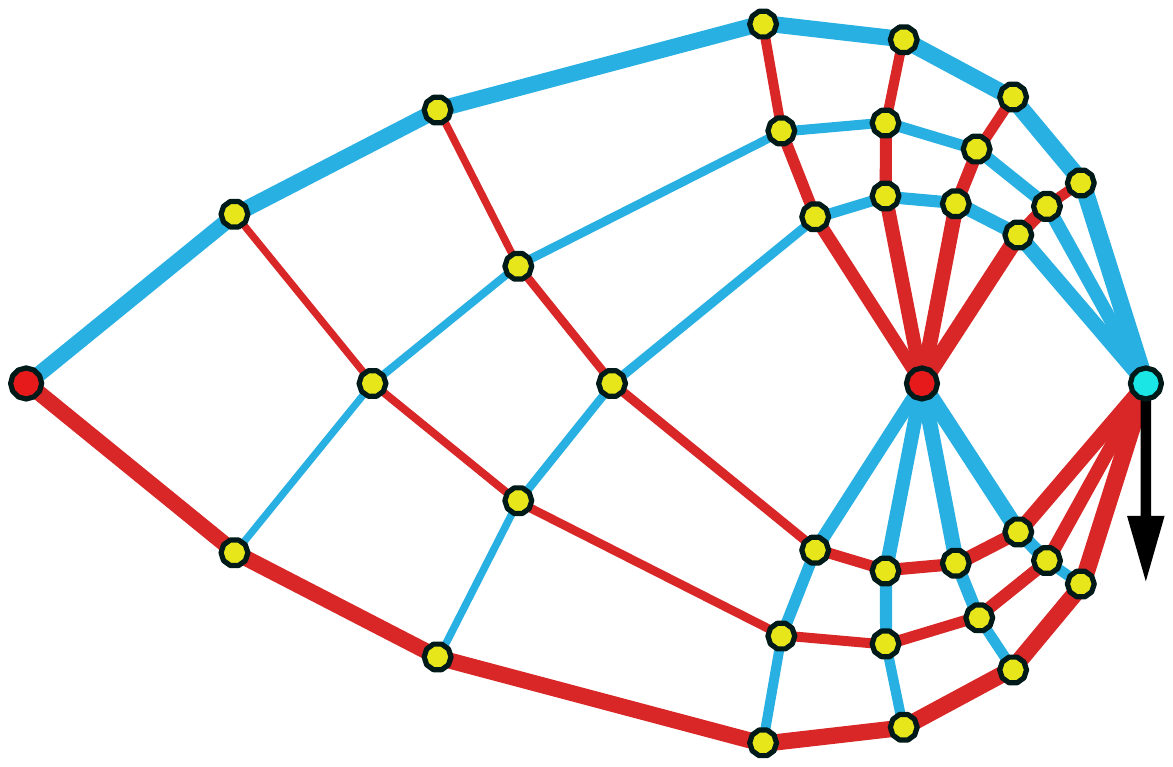}	
			\cput(40,2){\contour{white}{$(a3)$}}	
		\end{overpic}
		\begin{overpic}			
			[width=0.45\columnwidth,trim={4cm 10cm 4cm 10cm},clip]{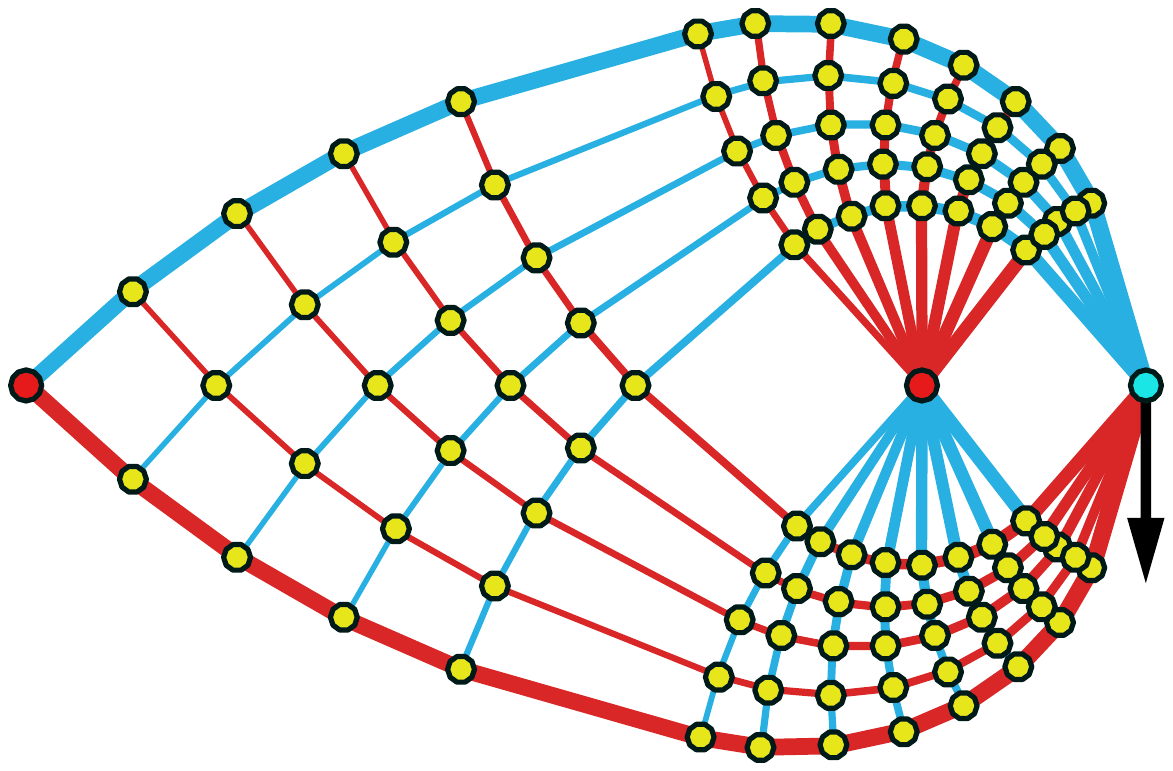}	
			\cput(40,2){\contour{white}{$(a4)$}}	
		\end{overpic}					
		\begin{overpic}			
			[width=0.45\columnwidth,trim={4cm 10cm 4cm 10cm},clip]{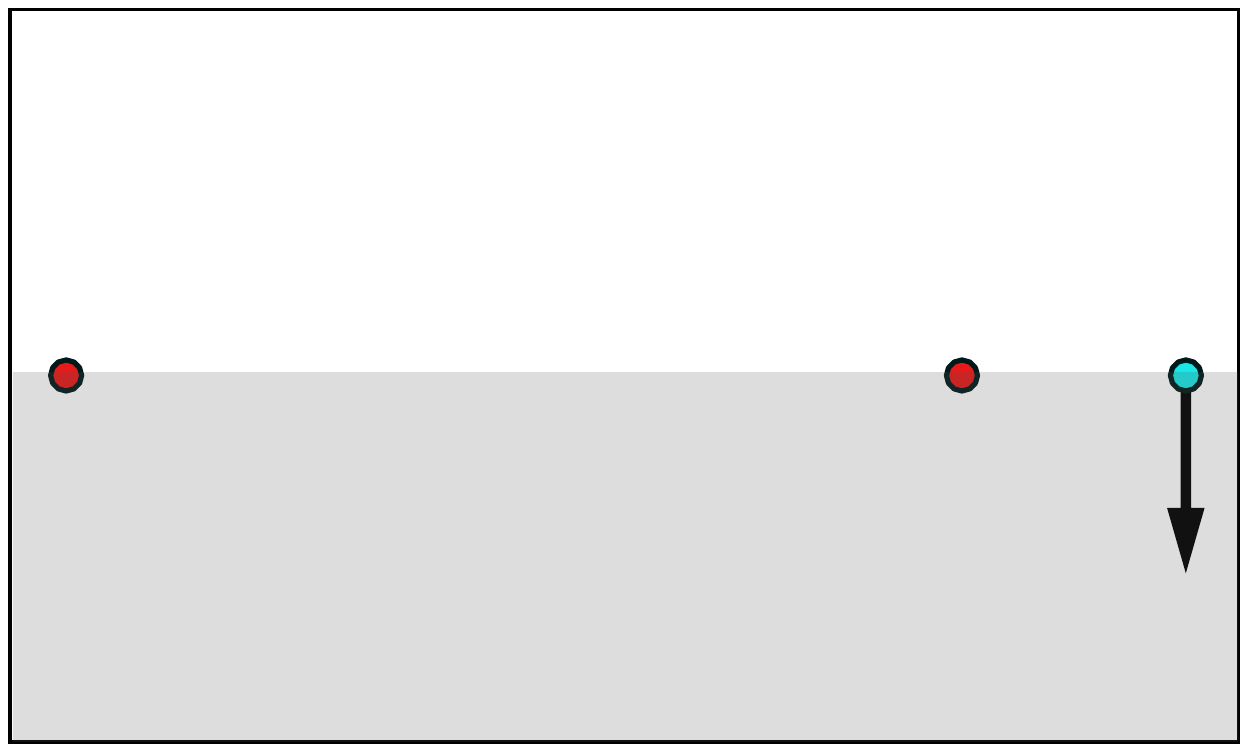}	
			\cput(50,10){\contour{white}{$(b1)$}}		
		\end{overpic}
		\begin{overpic}			
			[width=0.45\columnwidth,trim={4cm 10cm 4cm 10cm},clip]{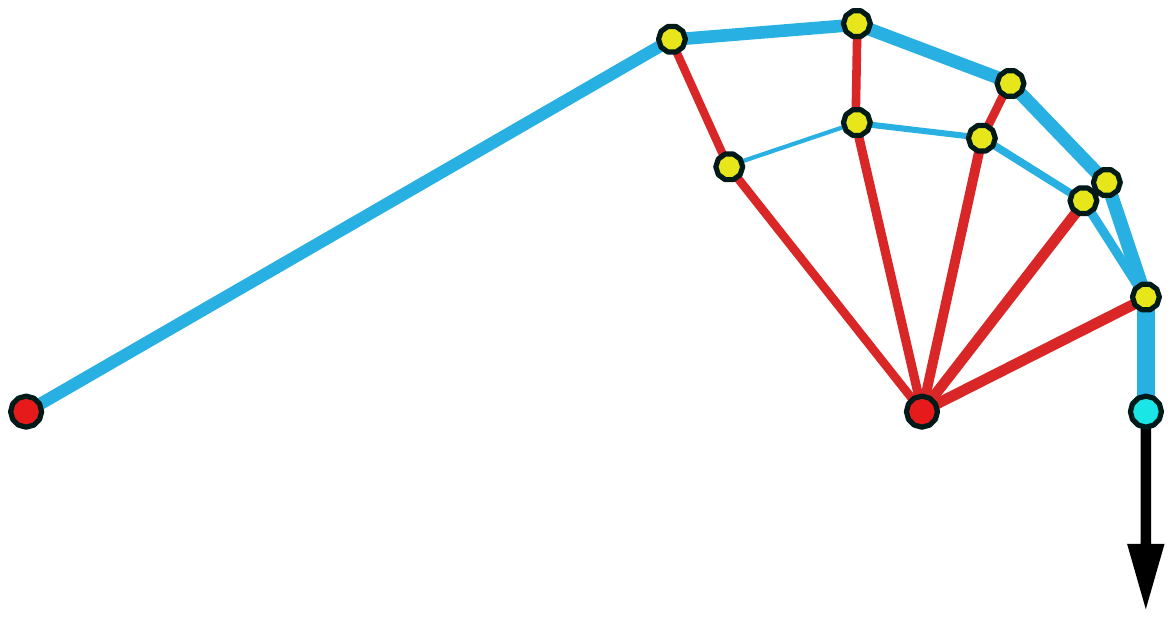}	
			\cput(40,2){\contour{white}{$(b2)$}}	
		\end{overpic}
		\begin{overpic}			
			[width=0.45\columnwidth,trim={4cm 10cm 4cm 10cm},clip]{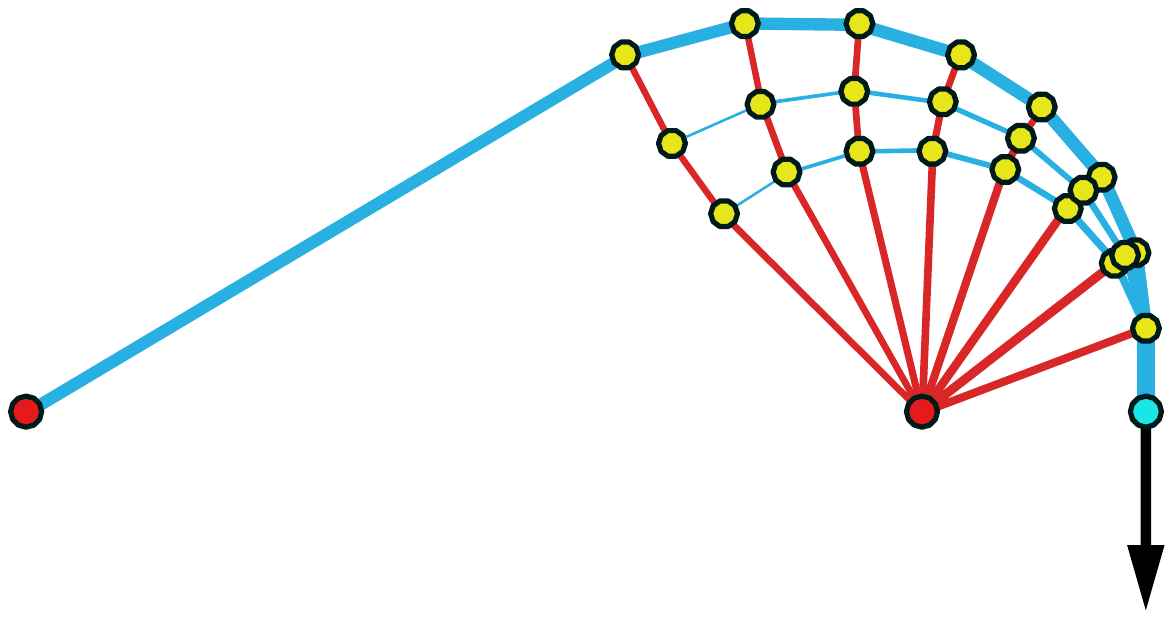}	
			\cput(40,2){\contour{white}{$(b3)$}}	
		\end{overpic}
		\begin{overpic}			
			[width=0.45\columnwidth,trim={4cm 10cm 4cm 10cm},clip]{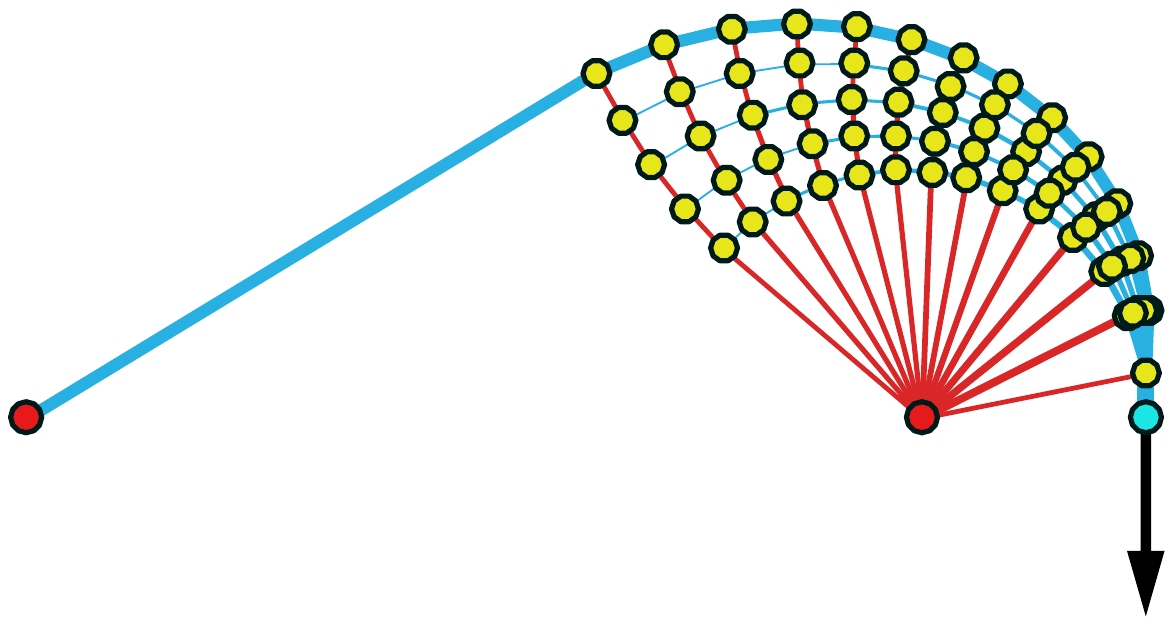}	
			\cput(40,2){\contour{white}{$(b4)$}}	
		\end{overpic}
	}	
	\caption{
		\label{fig:2D005} Truss designs for different design regions.
		Top: truss designs for a functional specification that the design region is the entire plane. Bottom: truss designs for the same functional specification except that the design region is the upper half plane, with the lower half plane being an obstacle to be avoided.
		 }
\end{figure*}

\begin{figure*}[tb]\centering{
		\begin{overpic}			
			[width=0.4\columnwidth,trim={4cm 6cm 4cm 7cm},clip]{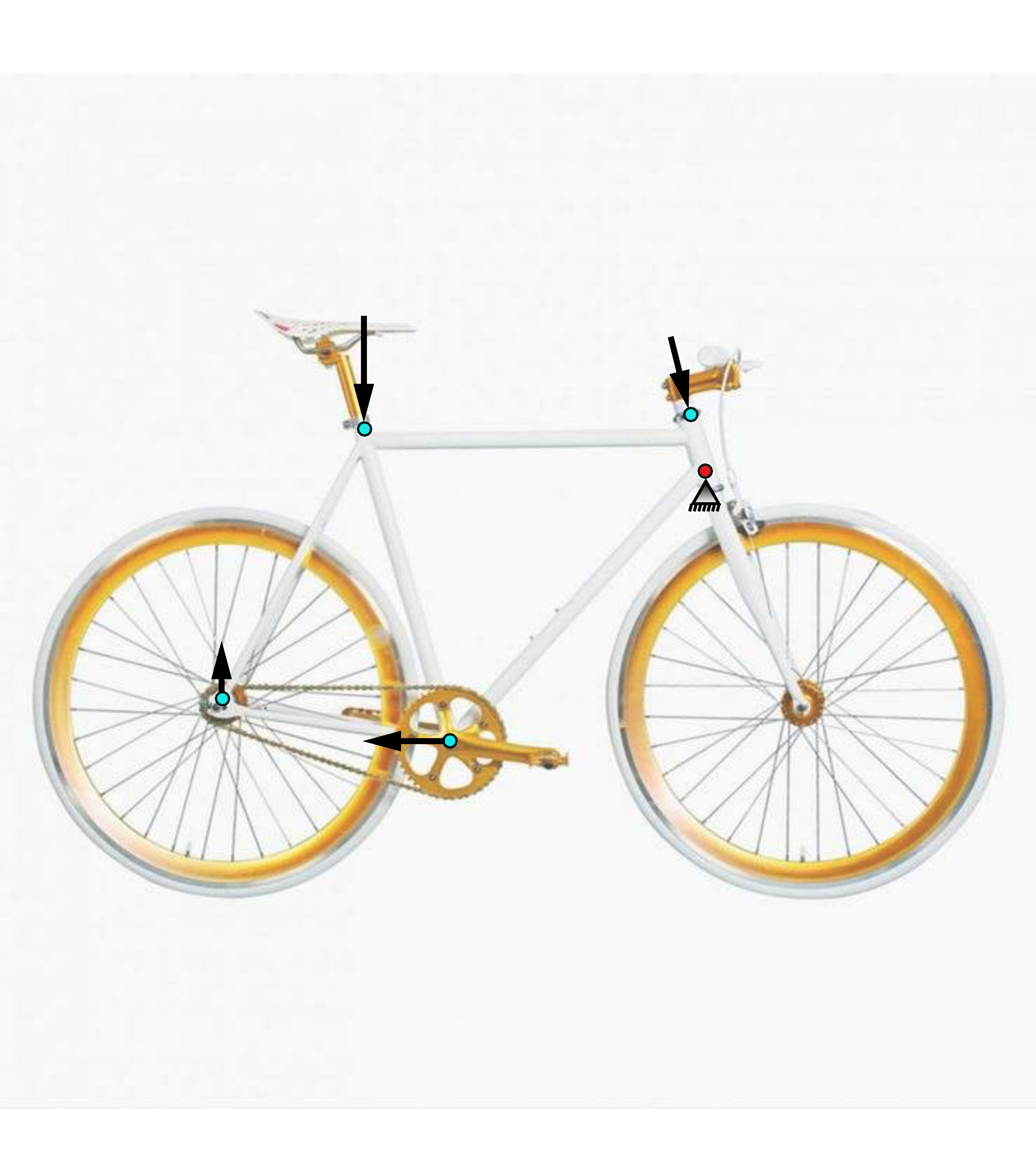}	
			\cput(50,0){\contour{white}{$(a)$}}		
		\end{overpic}
		\begin{overpic}			
			[width=0.4\columnwidth,trim={4cm 6cm 4cm 7cm},clip]{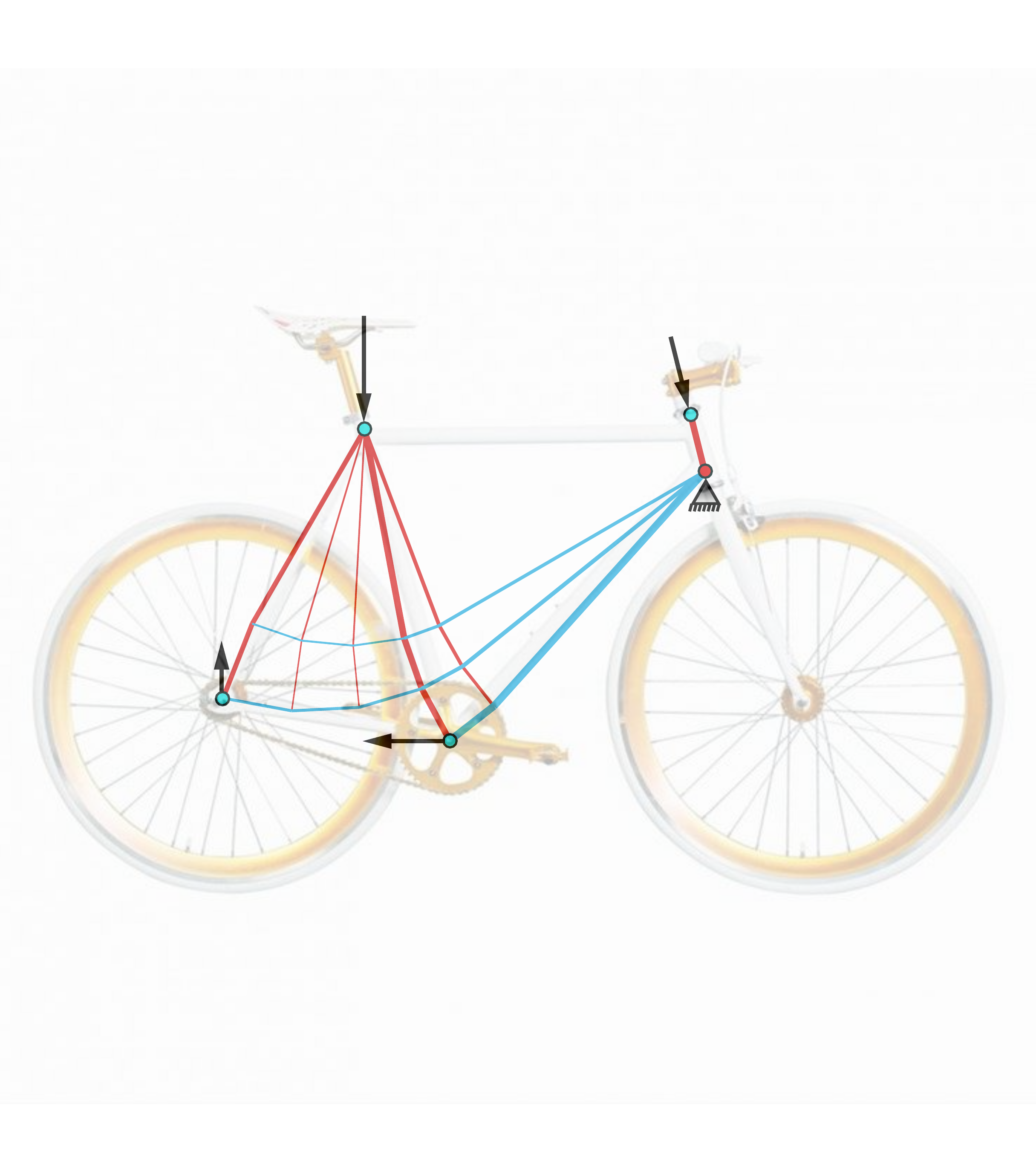}	
			\cput(50,2){\contour{white}{$(b)$}}	
		\end{overpic}
		\begin{overpic}			
			[width=0.4\columnwidth,trim={4cm 6cm 4cm 7cm},clip]{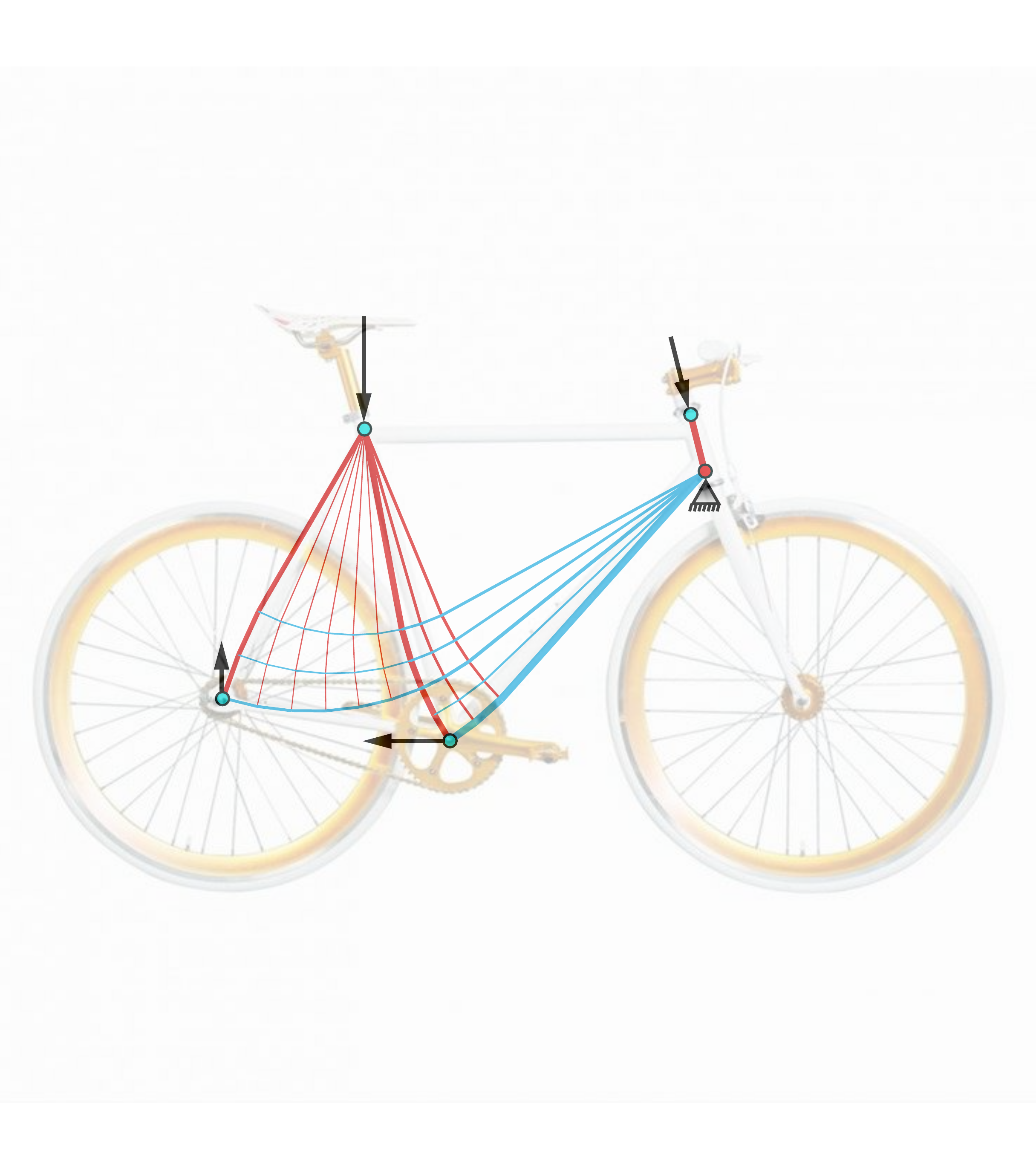}	
			\cput(50,2){\contour{white}{$(c)$}}	
		\end{overpic}
		\begin{overpic}			
			[width=0.4\columnwidth,trim={4cm 6cm 4cm 7cm},clip]{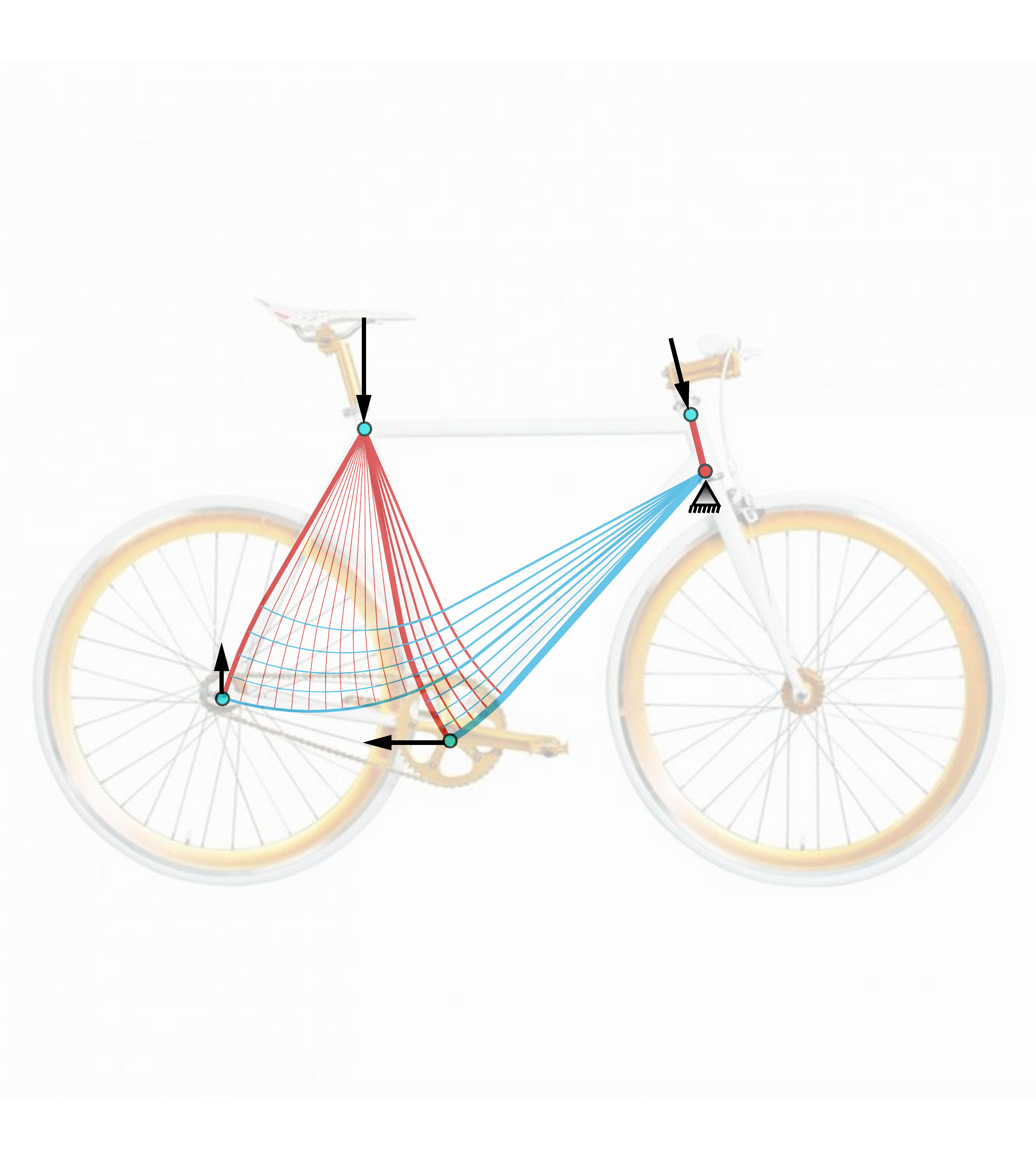}	
			\cput(50,2){\contour{white}{$(d)$}}	
		\end{overpic}
		\begin{overpic}			
			[width=0.4\columnwidth,trim={4cm 6cm 4cm 7cm},clip]{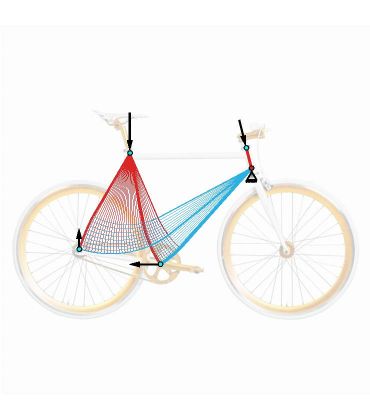}	
			\cput(50,2){\contour{white}{$(e)$}}	
		\end{overpic}
	}	
	\caption{
		\label{fig:2D0042}
	A bike frame design.}
\end{figure*}

\begin{figure*}[tb]\centering{
		\begin{overpic}			
			[width=0.41\columnwidth]{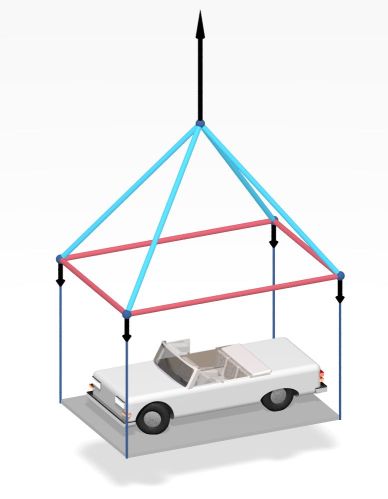}	
			\cput(50,2){\contour{white}{$(a)$}}		
		\end{overpic}
		\begin{overpic}			
			[width=0.41\columnwidth]{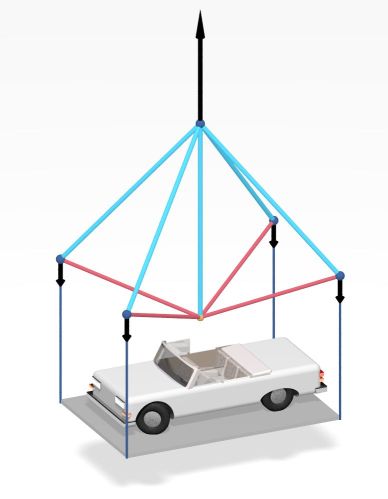}	
			\cput(50,2){\contour{white}{$(b)$}}	
		\end{overpic}
		\begin{overpic}			
			[width=0.41\columnwidth]{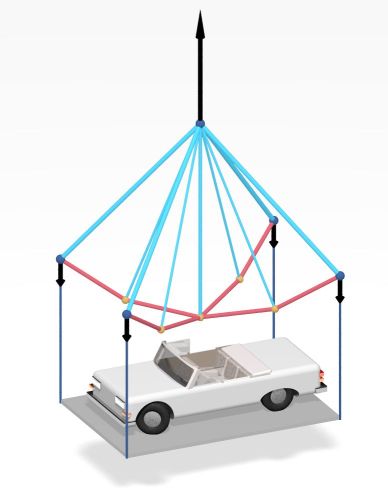}	
			\cput(50,2){\contour{white}{$(c)$}}	
		\end{overpic}
		\begin{overpic}			
			[width=0.41\columnwidth]{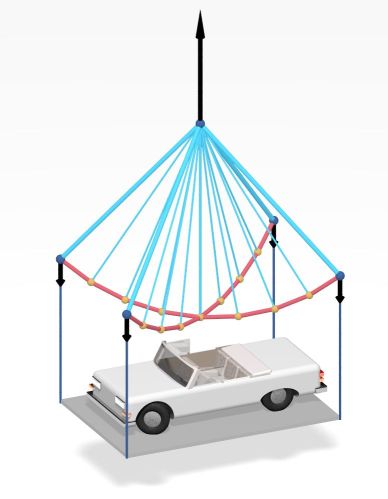}	
			\cput(50,2){\contour{white}{$(d)$}}	
		\end{overpic}
		\begin{overpic}			
			[width=0.41\columnwidth]{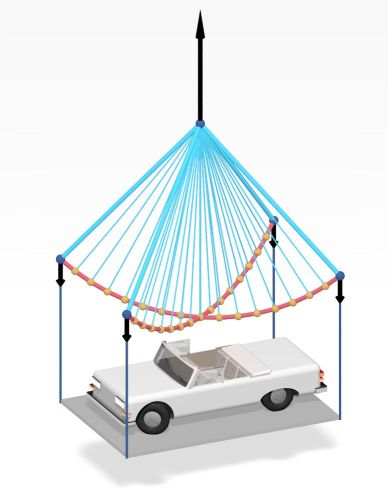}	
			\cput(50,2){\contour{white}{$(e)$}}	
		\end{overpic}
	}	
	\caption{
		\label{fig:3D003}
	A truss design with the input being a 3D parallel force system.}
\end{figure*}

\begin{figure*}[tb]	\centering{
		\begin{overpic}			
			[height=.3\columnwidth]{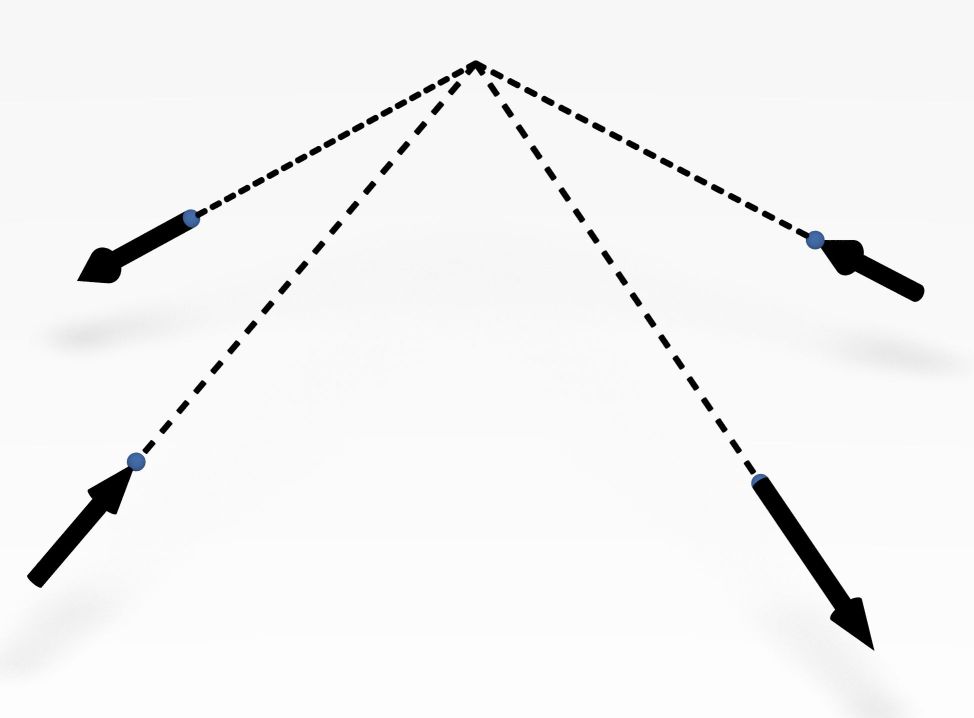}	
		\end{overpic}
		\begin{overpic}			
			[height=.3\columnwidth]{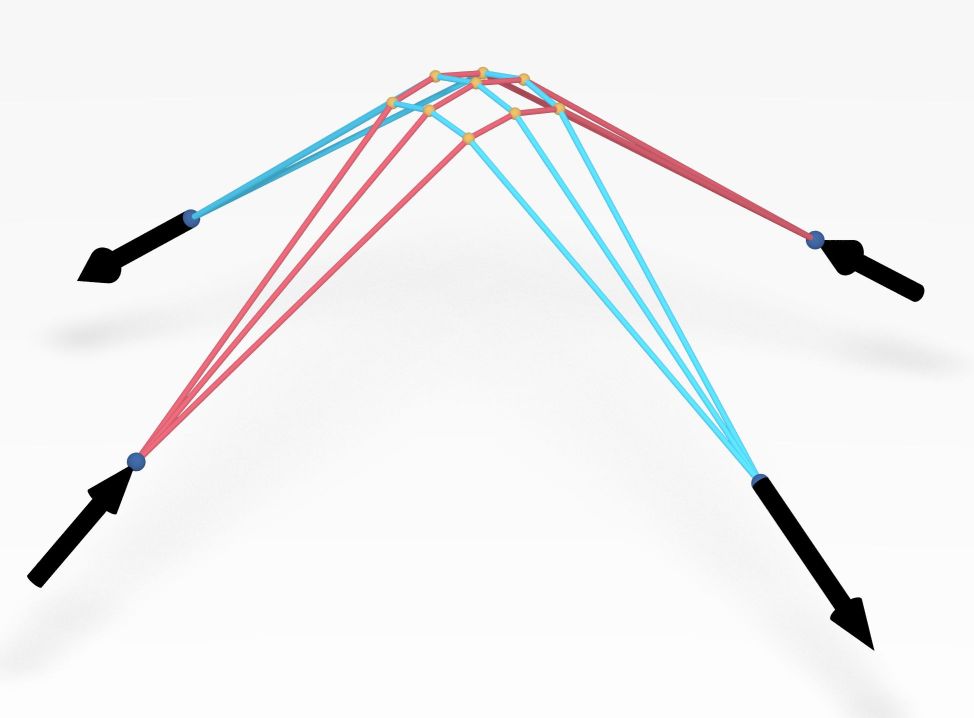}	
			\cput(20,2){\contour{white}{$(a)$}}		
		\end{overpic}
		\begin{overpic}			
			[height=.3\columnwidth]{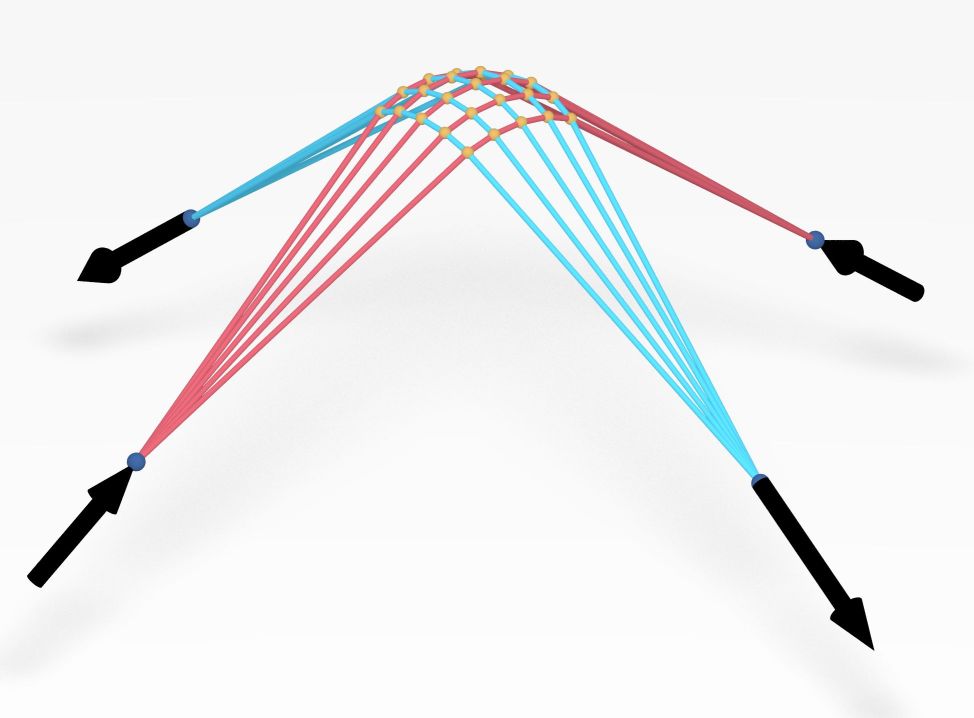}	
			\cput(20,2){\contour{white}{$(b)$}}	
		\end{overpic}
		\begin{overpic}			
			[height=.3\columnwidth]{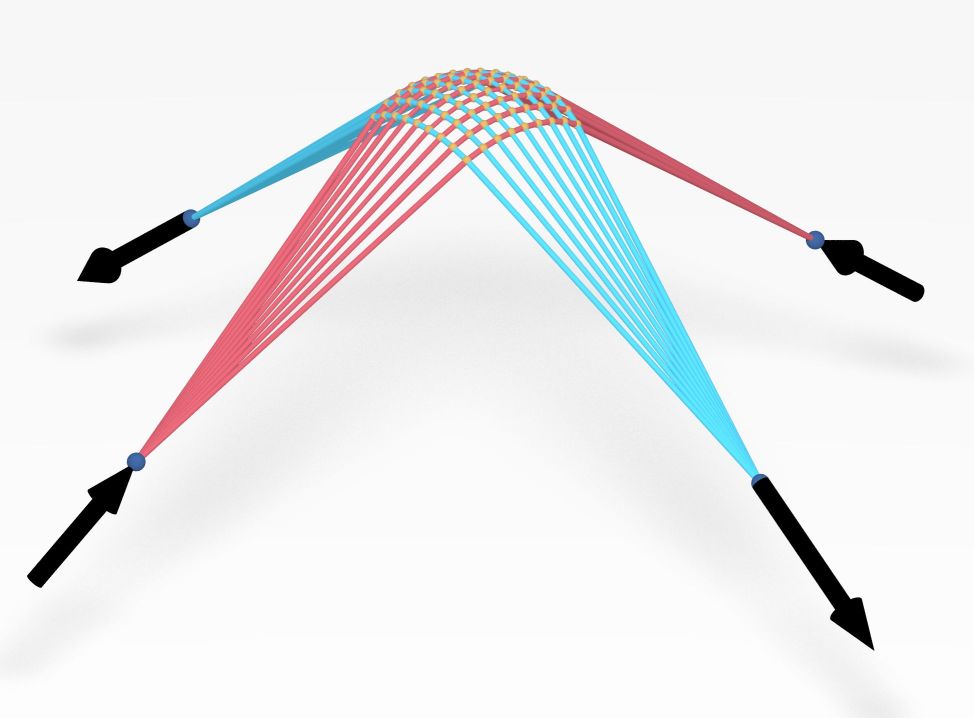}
			\cput(20,0){\contour{white}{$(c)$}}		
		\end{overpic}
		\begin{overpic}			
			[height=.3\columnwidth]{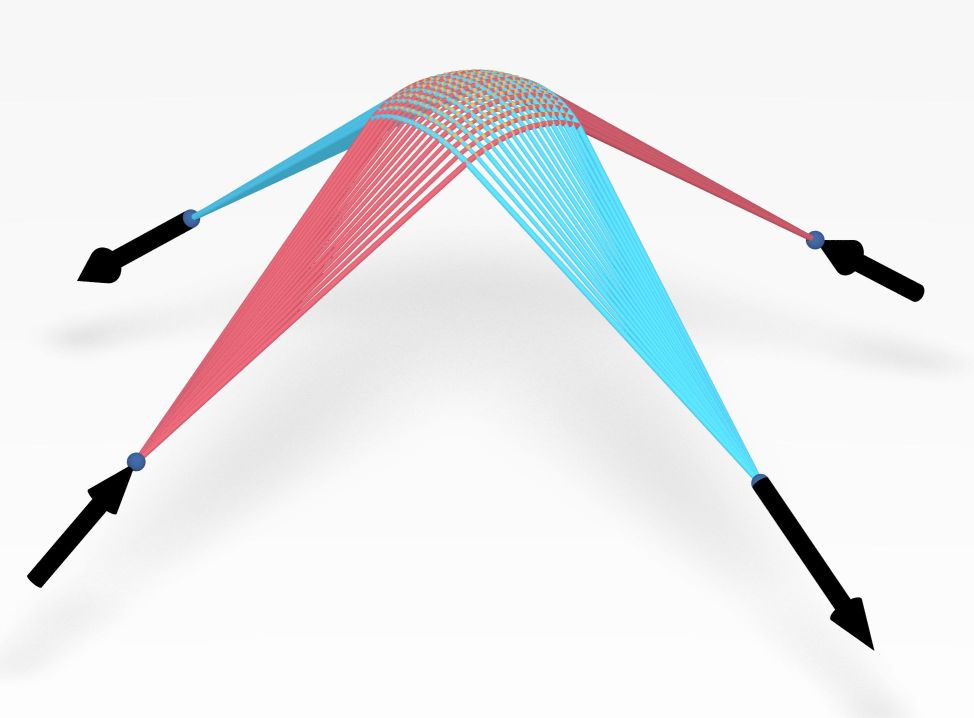}	
			\cput(20,0){\contour{white}{$(d)$}}		
		\end{overpic}
	}	
	\caption{
		\label{fig:concurrentforce3D}
		A truss design with the input being a 3D concurrent force system.}
\end{figure*}

\begin{figure*}[tb]	\centering{
		\begin{overpic}			
			[height=.3\columnwidth]{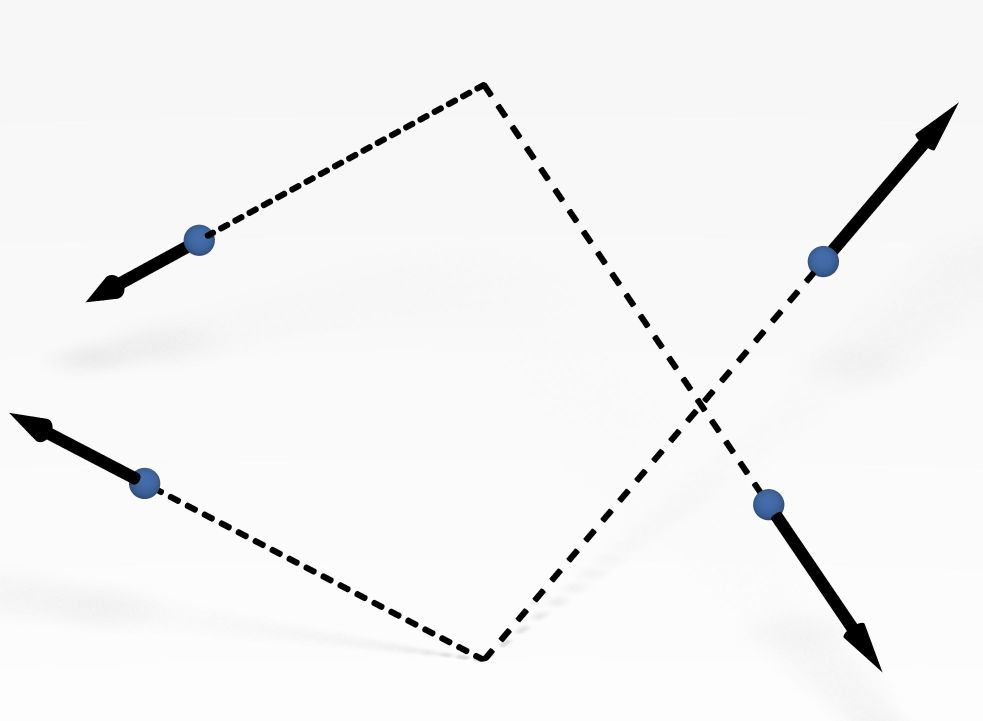}	
		\end{overpic}
		\begin{overpic}			
			[height=.3\columnwidth]{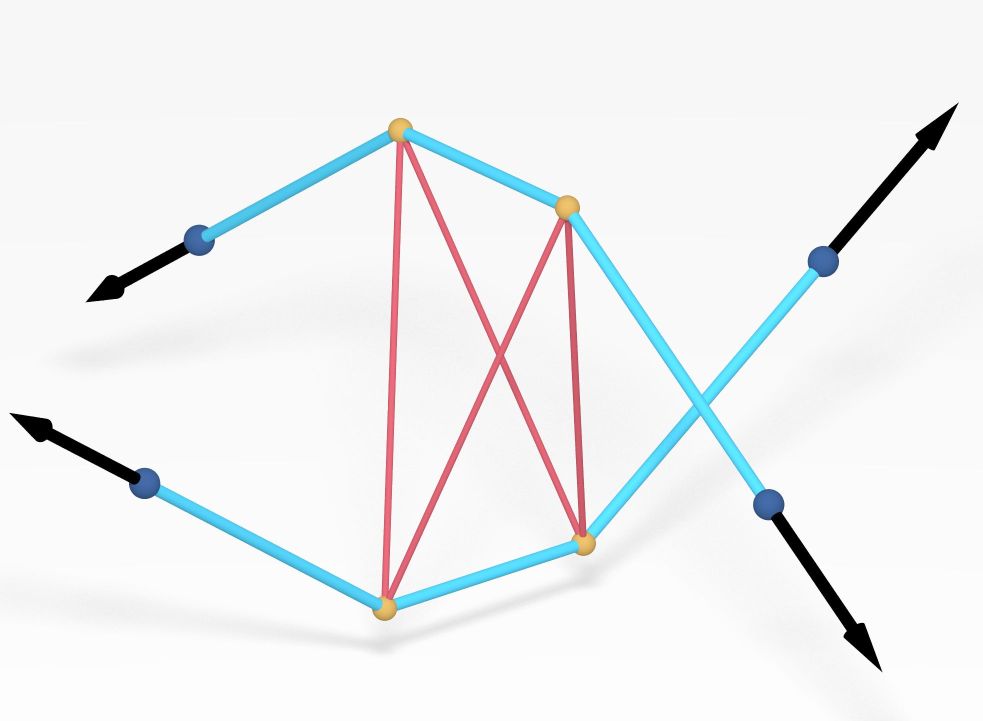}	
			\cput(20,2){\contour{white}{$(a)$}}		
		\end{overpic}
		\begin{overpic}			
			[height=.3\columnwidth]{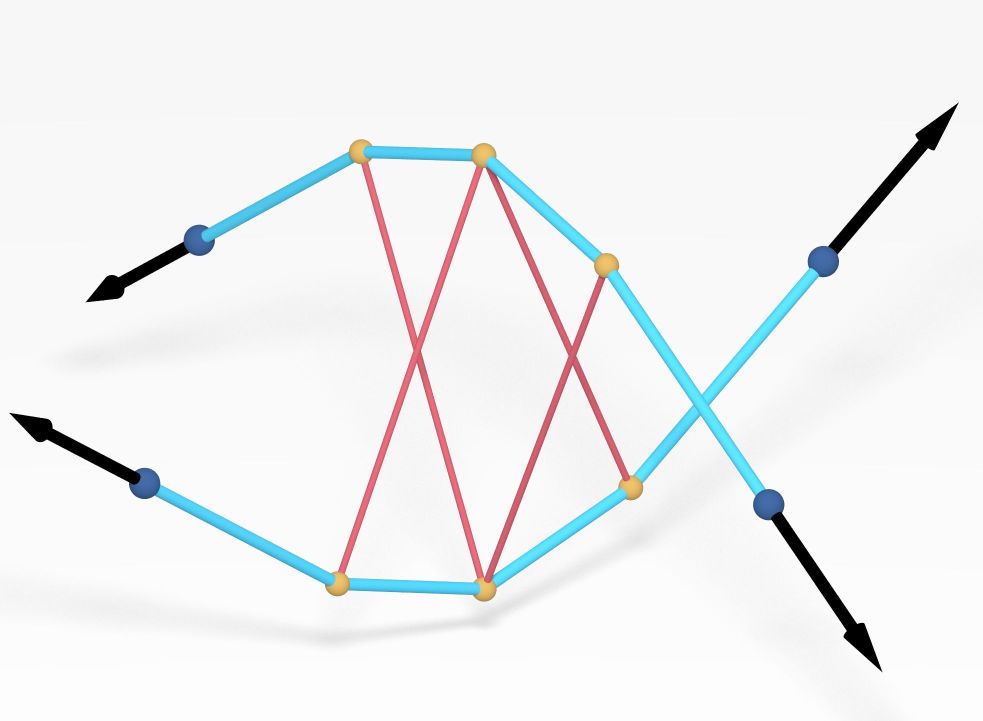}	
			\cput(20,2){\contour{white}{$(b)$}}	
		\end{overpic}
		\begin{overpic}			
			[height=.3\columnwidth]{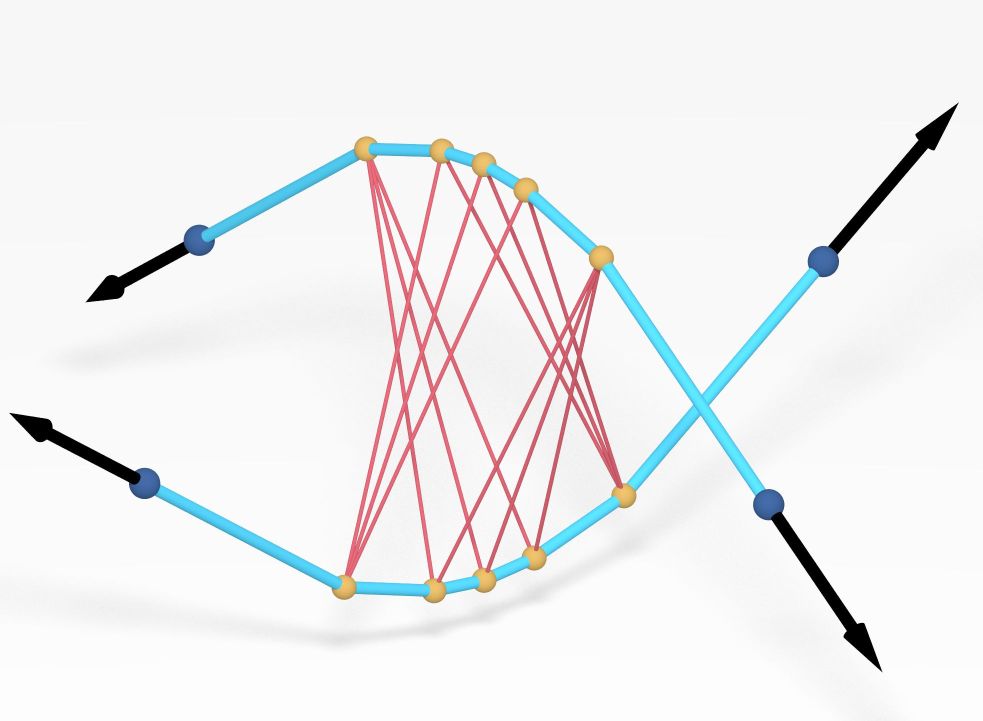}
			\cput(20,0){\contour{white}{$(c)$}}		
		\end{overpic}
		\begin{overpic}			
			[height=.3\columnwidth]{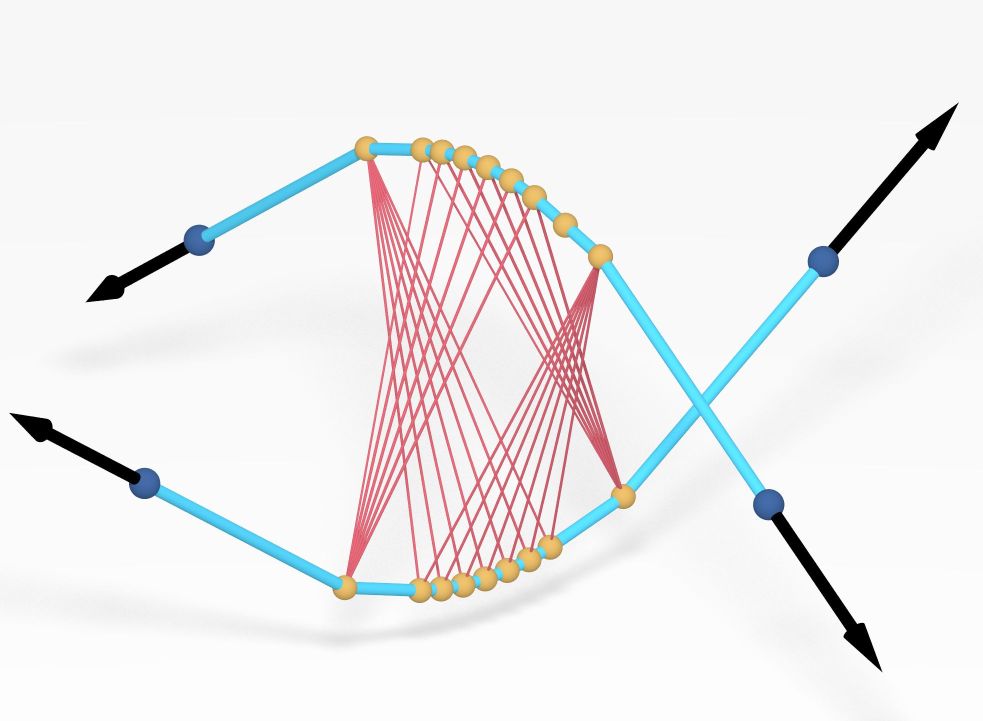}	
			\cput(20,0){\contour{white}{$(d)$}}		
		\end{overpic}
	}	
	\caption{
		\label{fig:nonconcurrentforce3D}
		A truss design with the input being a 3D non-concurrent force system. }
\end{figure*}

\begin{figure*}[tb]\centering{
		\begin{overpic}			
			[width=0.4\columnwidth]{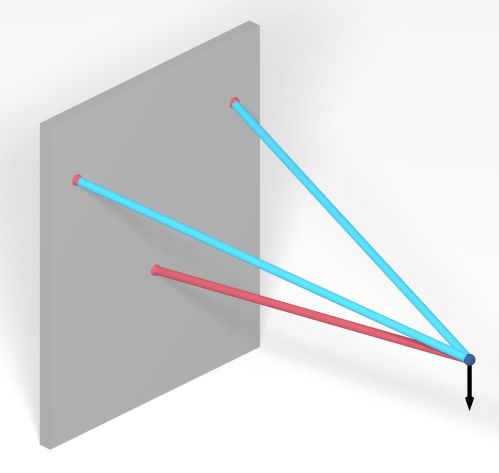}	
			\cput(50,2){\contour{white}{$(a)$}}		
		\end{overpic}
		\begin{overpic}			
			[width=0.4\columnwidth]{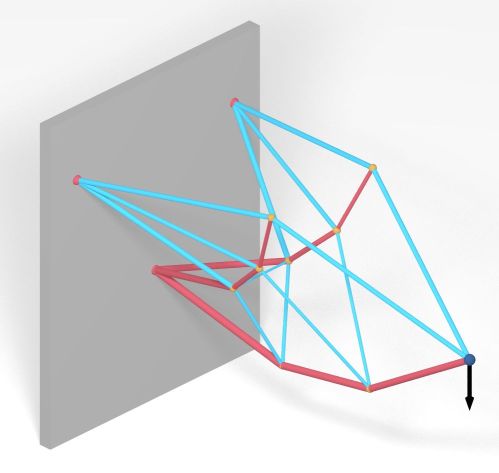}	
			\cput(50,2){\contour{white}{$(b)$}}	
		\end{overpic}
		\begin{overpic}			
			[width=0.4\columnwidth]{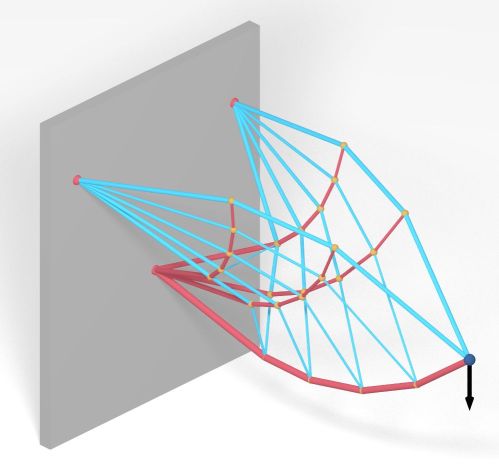}	
			\cput(50,2){\contour{white}{$(c)$}}	
		\end{overpic}
		\begin{overpic}			
			[width=0.4\columnwidth]{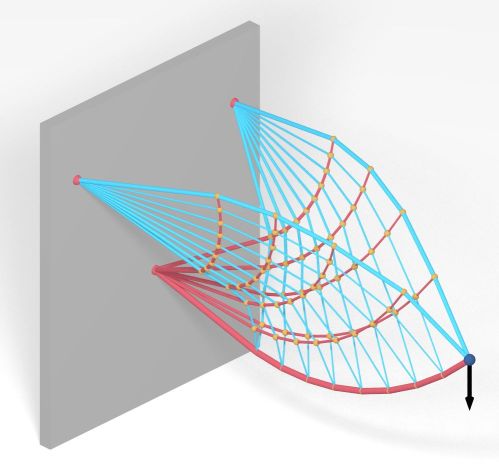}	
			\cput(50,2){\contour{white}{$(d)$}}	
		\end{overpic}
		\begin{overpic}			
			[width=0.4\columnwidth]{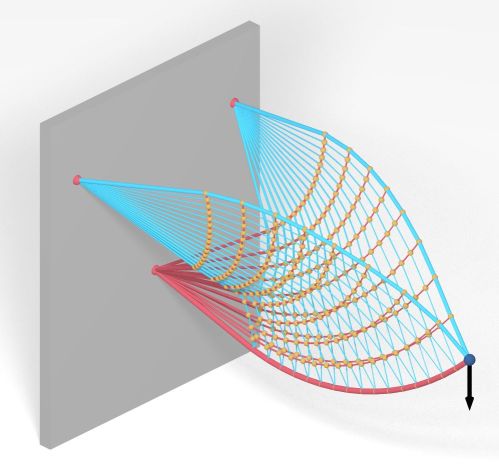}	
			\cput(50,2){\contour{white}{$(e)$}}	
		\end{overpic}
	}	
	\caption{
		\label{fig:3D004}
		A 3D cantilever design.}
\end{figure*}

We show the results of our method for different types of functional specifications. We present 2D truss designs with a parallel equilibrium force system in Figure \ref{fig:parallelforce}, a concurrent equilibrium force system in Figure \ref{fig:concurrentforce}, and a non-concurrent equilibrium force system in Figure \ref{fig:nonconcurrentforce}. We illustrate examples of designs for the same input external forces and supporting joints but different design regions in Figure \ref{fig:2D005}.  For 3D trusses, we demonstrate the results for input of a parallel equilibrium force system in Figure \ref{fig:3D003}, a concurrent equilibrium force system in Figure \ref{fig:concurrentforce3D}, and a non-concurrent equilibrium force system in Figure \ref{fig:nonconcurrentforce3D}. In addition, we show examples based on real functional requirements such as a 2D bike frame in Figure \ref{fig:2D0042}, a 3D cantilever in Figure \ref{fig:3D004}, and a 3D bridge in Figure \ref{fig:Teaser}.

\subsection{Quantitative Evaluation}
Our framework is implemented in Matlab R2016 on a workstation with an Intel Xeon X5550 2.67 GHz processor. 
We use Mosek~\cite{mosek} as the solver for linear programming. For the ALP algorithm, we set the maximum iteration number $N_\mathrm{max}$=500 and the maximum line search step $S_\mathrm{max}$=10. For the number of grid points in the initialization, $n$, we use the default value, the number of joints specified in the functional specification. 
We set the number of max iteration of Phase 1, $P_\mathrm{max1}=5$. The number of subdivisions in Phase 2, $P_\mathrm{max2}$, controls the trade off between number of bars and truss weight and is set depending on the context. We set the thresholds $\epsilon_1$ and $\epsilon_2$ in Section  \ref{sec:GeoOperations} as $\epsilon_1=0.002\bar{a}$, $\epsilon_2=0.01\bar{d}$, where $\bar{a}$ is the average cross-section, and $\bar{d}$ is the average value of distances between joints specified in the functional specification.
 In Table \ref{table1}, for each optimized truss, we report parameters of trusses such as the number of bars, the total volume of material, and the computation time of coarse truss optimization and different levels of structure refinement. 

\subsection{Evaluation of the Initialization}

\begin{figure}[t!]	\centering{
\begin{overpic}			
	[width=0.99\columnwidth]{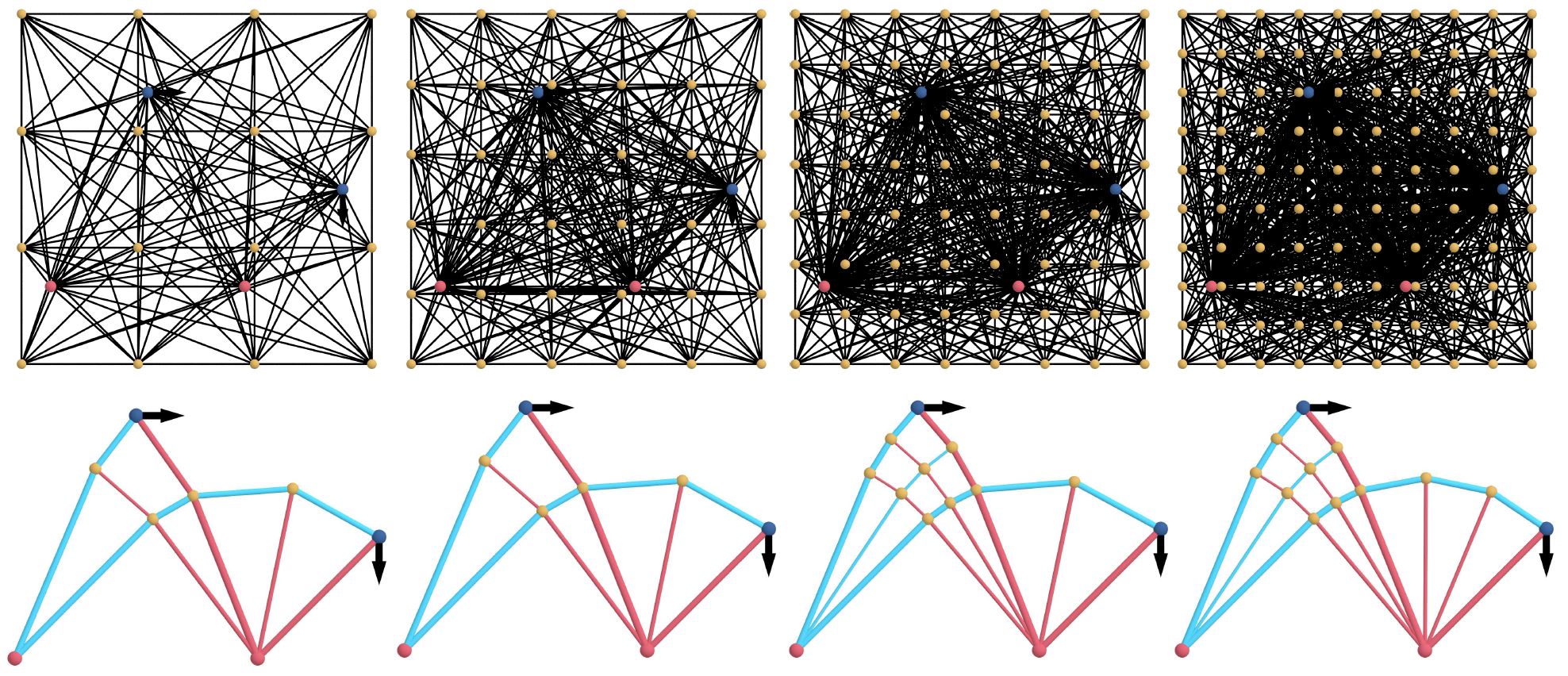}	
	\cput(10,0){\contour{white}{$(a)$}}	
		\cput(35,0){\contour{white}{$(b)$}}
			\cput(60,0){\contour{white}{$(c)$}}
				\cput(85,0){\contour{white}{$(d)$}}
\end{overpic}
	}	
	\caption{
		\label{fig:ini}
		Comparing different initializations. From left to right we initialize with 4$\times$4, 6$\times$6, 8$\times$8, and 10$\times$10 intermediate joints and 124, 364, 732, and 1228 intermediate bars.}
\end{figure}

To test the sensitivity of our framework to the initialization, we show optimized trusses starting from initializations with different numbers of intermediate joints and bars (before subdivision operations are conducted). Figure \ref{fig:ini} shows that adding different intermediate joints and bars results in different trusses. However, these trusses have similar structure and they all function as robust discrete approximations of the optimal truss for further subdivision.

\subsection{Comparisons}
\input{ALP_table}

\subsubsection{Comparing ALP to alternative solvers}

\begin{figure*}[h]\centering{
				\begin{overpic}			
			[width=0.29\columnwidth]{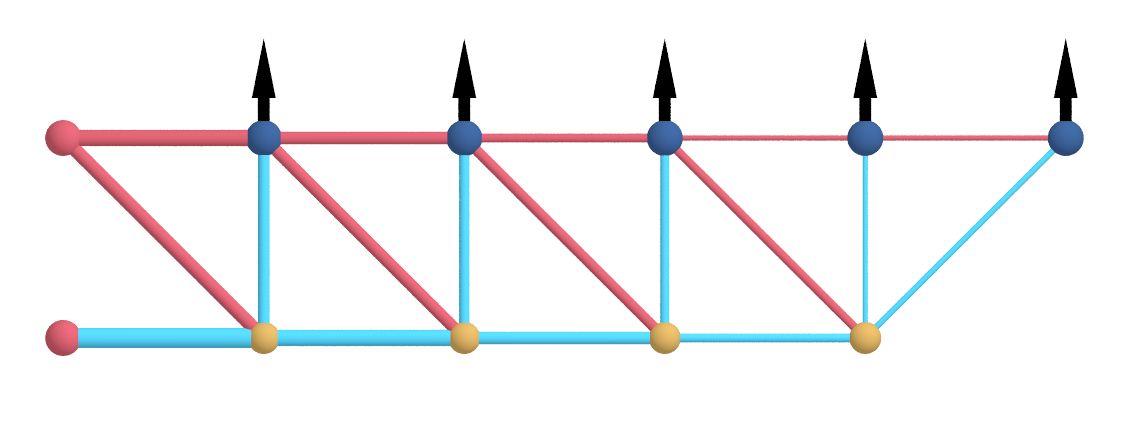}	
			\cput(90,10){\contour{white}{$(a1)$}}		
		\end{overpic}
		\begin{overpic}			
			[width=0.29\columnwidth]{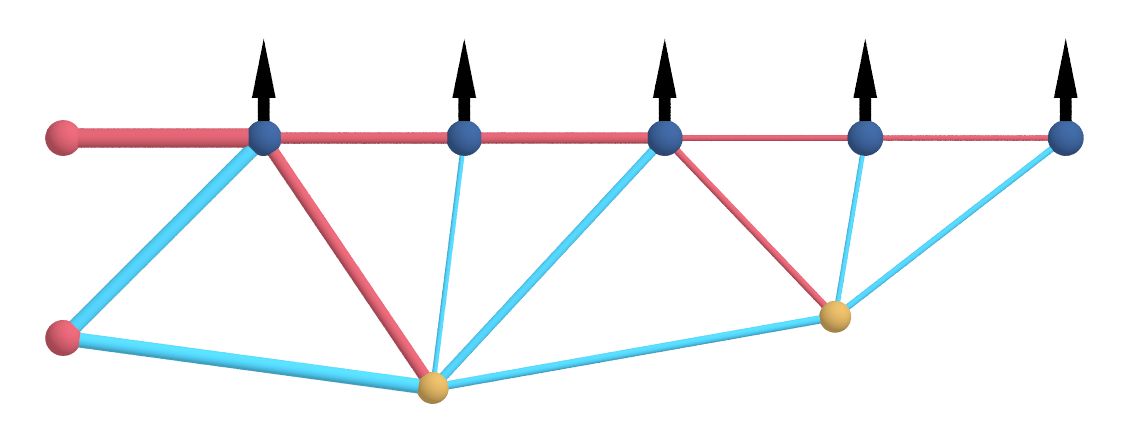}	
			\cput(90,10){\contour{white}{$(b1)$}}		
		\end{overpic}
		\begin{overpic}			
			[width=0.29\columnwidth]{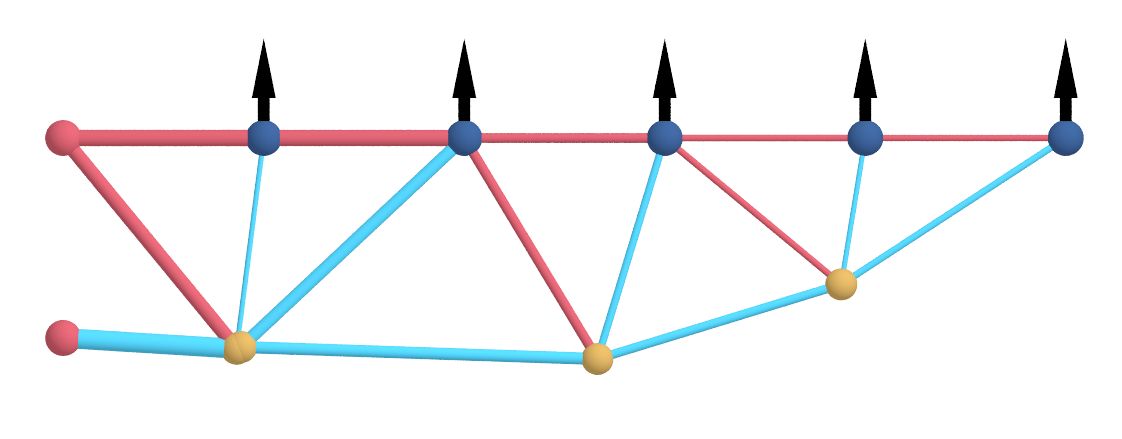}	
			\cput(90,10){\contour{white}{$(c1)$}}		
		\end{overpic}
		\begin{overpic}			
			[width=0.29\columnwidth]{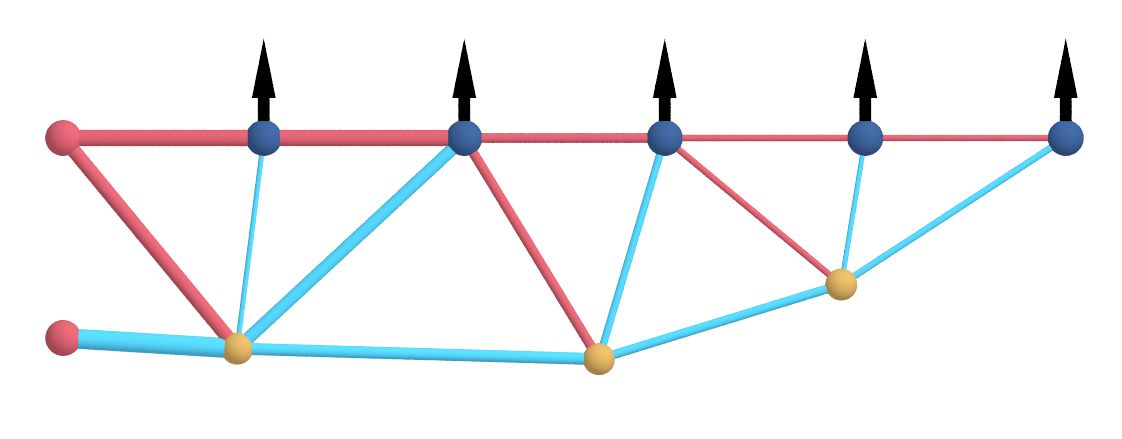}	
			\cput(90,10){\contour{white}{$(d1)$}}		
		\end{overpic}
		\begin{overpic}			
			[width=0.29\columnwidth]{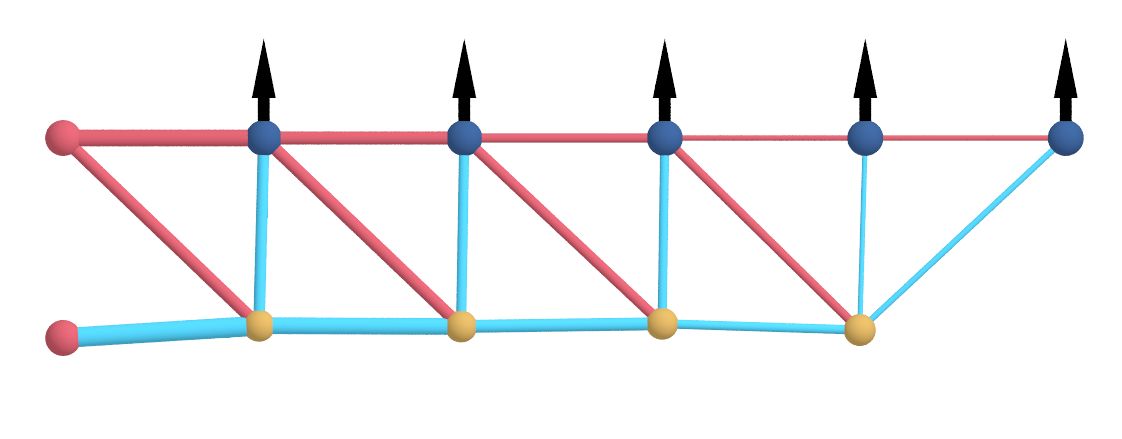}	
			\cput(90,10){\contour{white}{$(e1)$}}		
		\end{overpic}
		\begin{overpic}			
			[width=0.29\columnwidth]{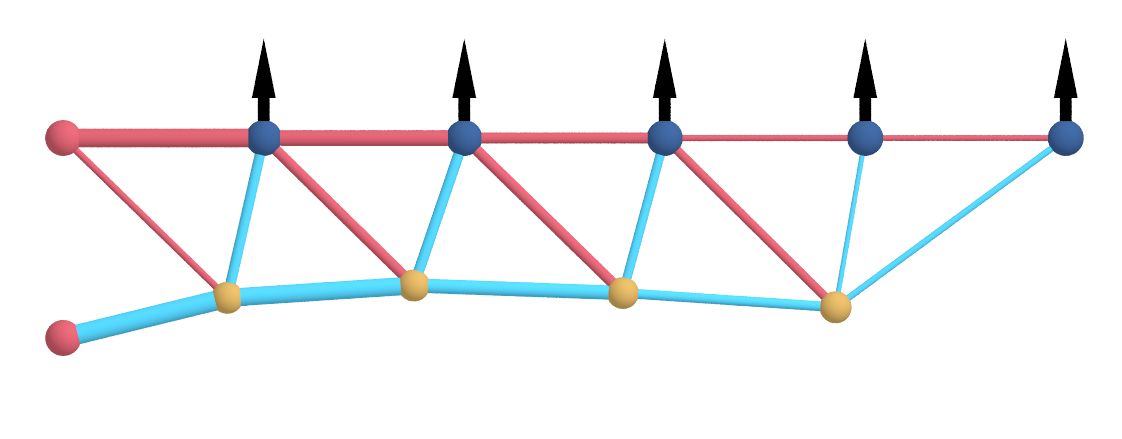}	
			\cput(90,10){\contour{white}{$(f1)$}}		
		\end{overpic}
			\begin{overpic}			
		[width=0.29\columnwidth]{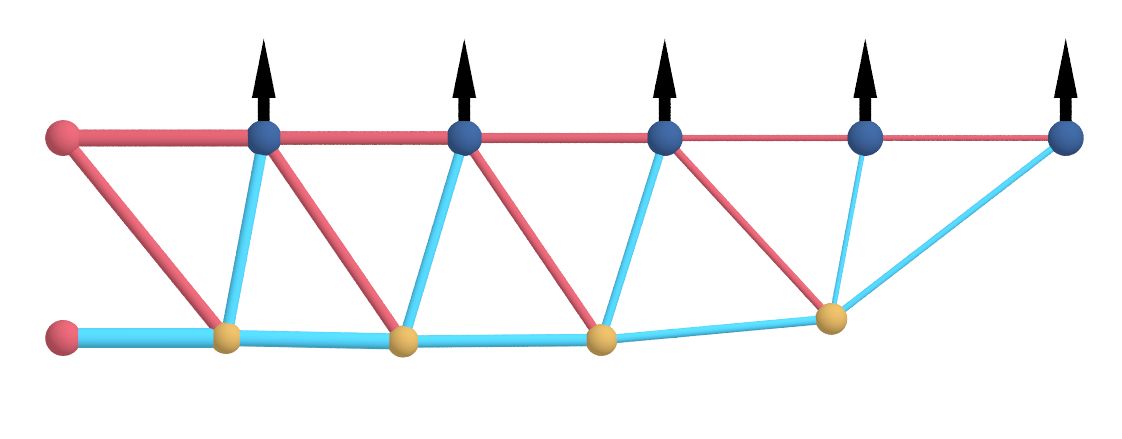}	
		\cput(90,10){\contour{white}{$(g1)$}}		
	\end{overpic}
		\begin{overpic}			
			[width=0.29\columnwidth]{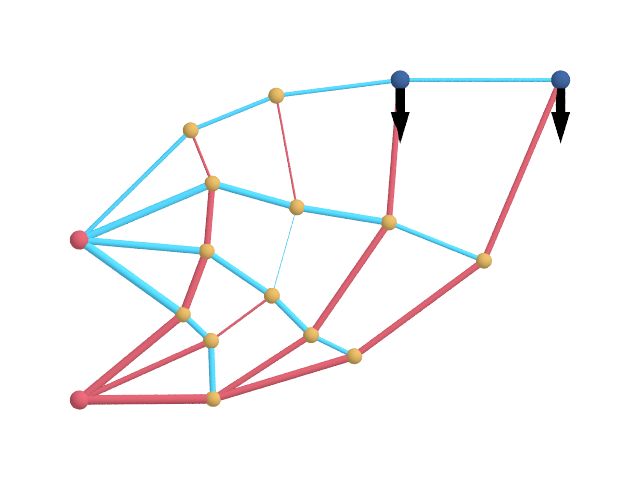}	
			\cput(70,10){\contour{white}{$(a2)$}}		
		\end{overpic}
		\begin{overpic}			
			[width=0.29\columnwidth]{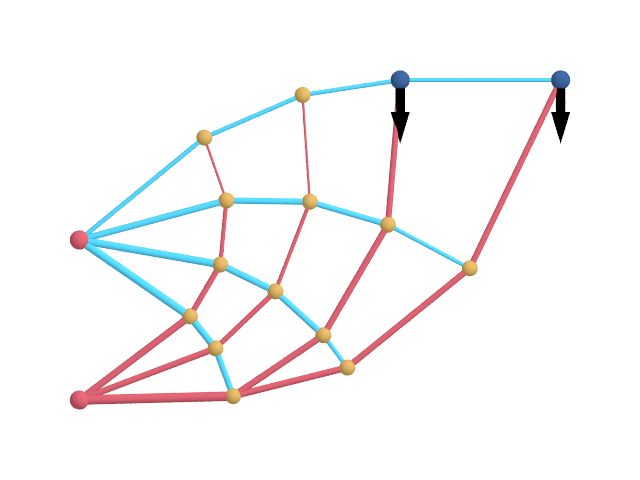}	
			\cput(70,10){\contour{white}{$(b2)$}}		
	    \end{overpic}
		\begin{overpic}			
			[width=0.29\columnwidth]{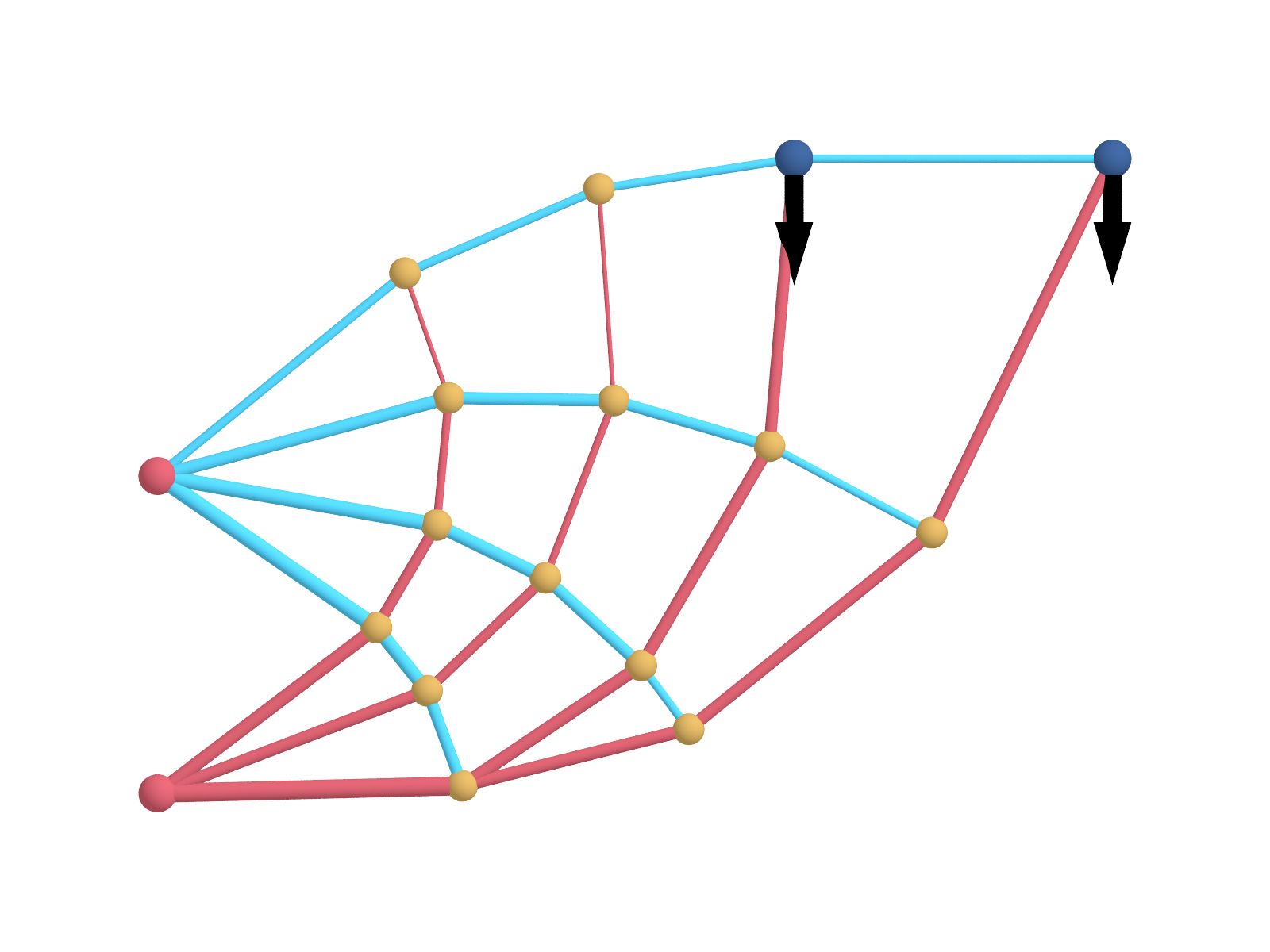}	
			\cput(70,10){\contour{white}{$(c2)$}}		
		\end{overpic}
		\begin{overpic}			
			[width=0.29\columnwidth]{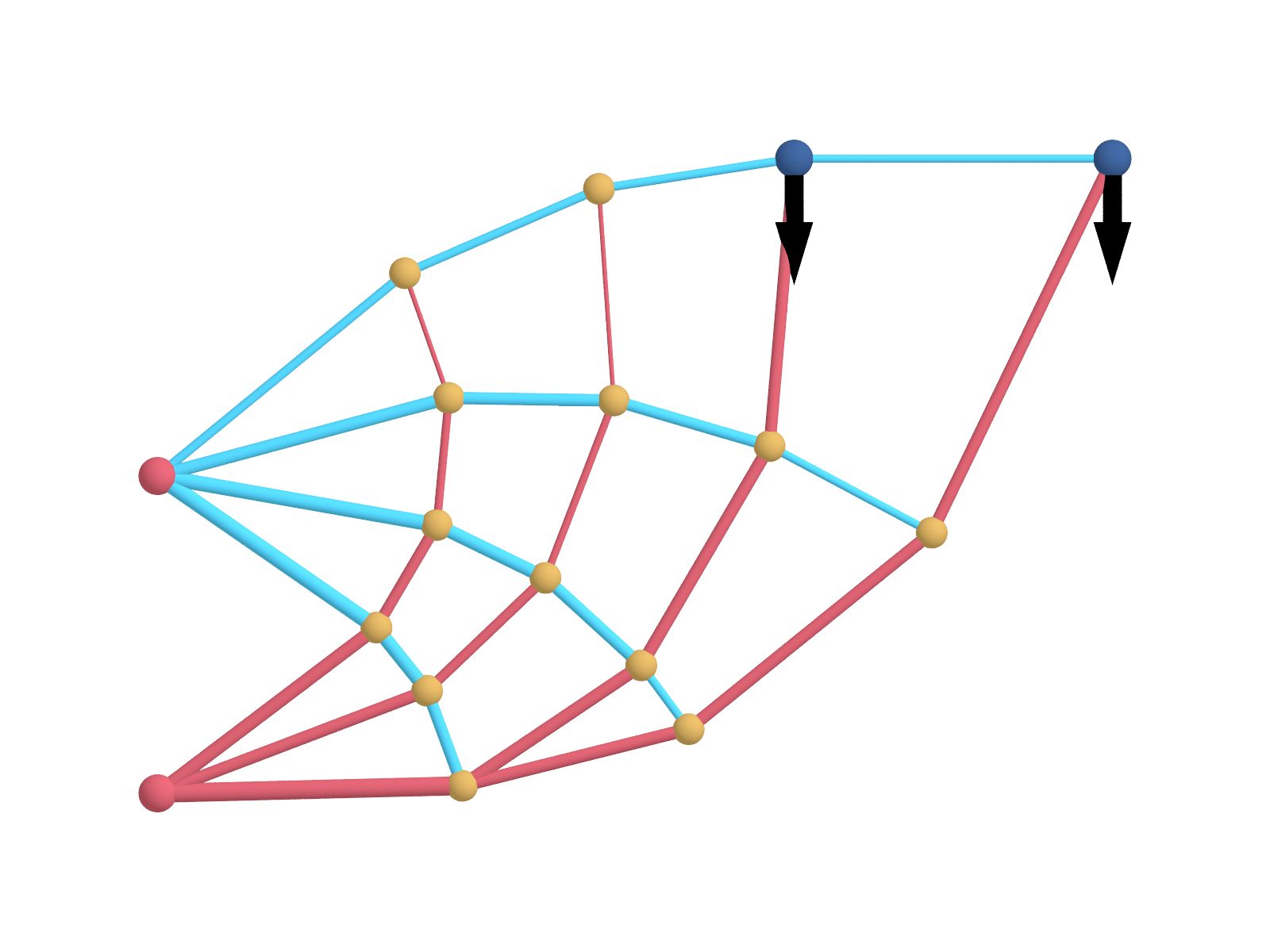}	
			\cput(70,10){\contour{white}{$(d2)$}}		
		\end{overpic}
		\begin{overpic}			
			[width=0.29\columnwidth]{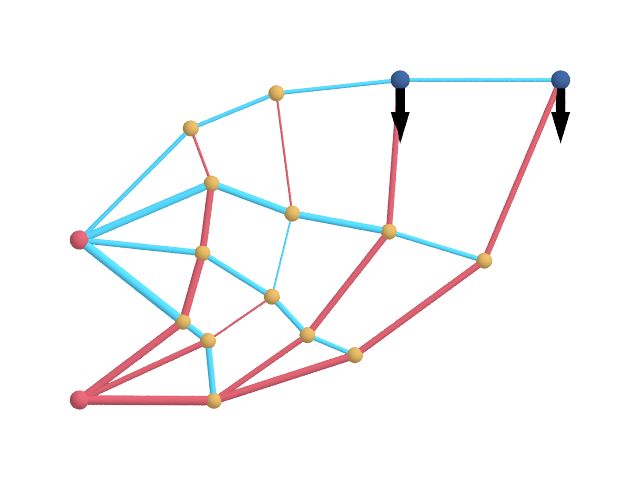}	
			\cput(70,10){\contour{white}{$(e2)$}}		
		\end{overpic}
		\begin{overpic}			
			[width=0.29\columnwidth]{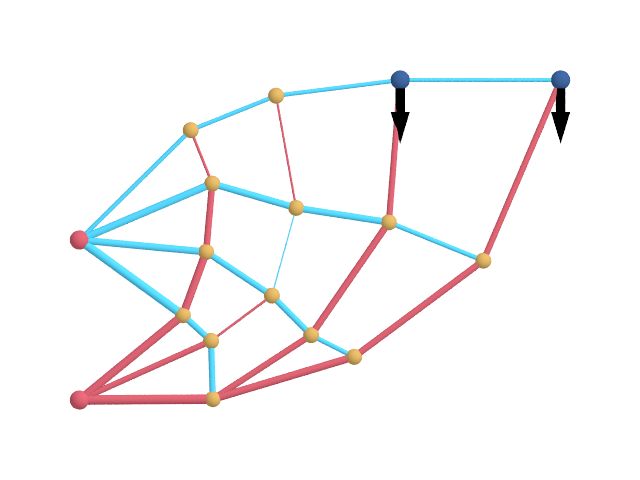}	
			\cput(70,10){\contour{white}{$(f2)$}}		
		\end{overpic}
			\begin{overpic}			
		[width=0.29\columnwidth]{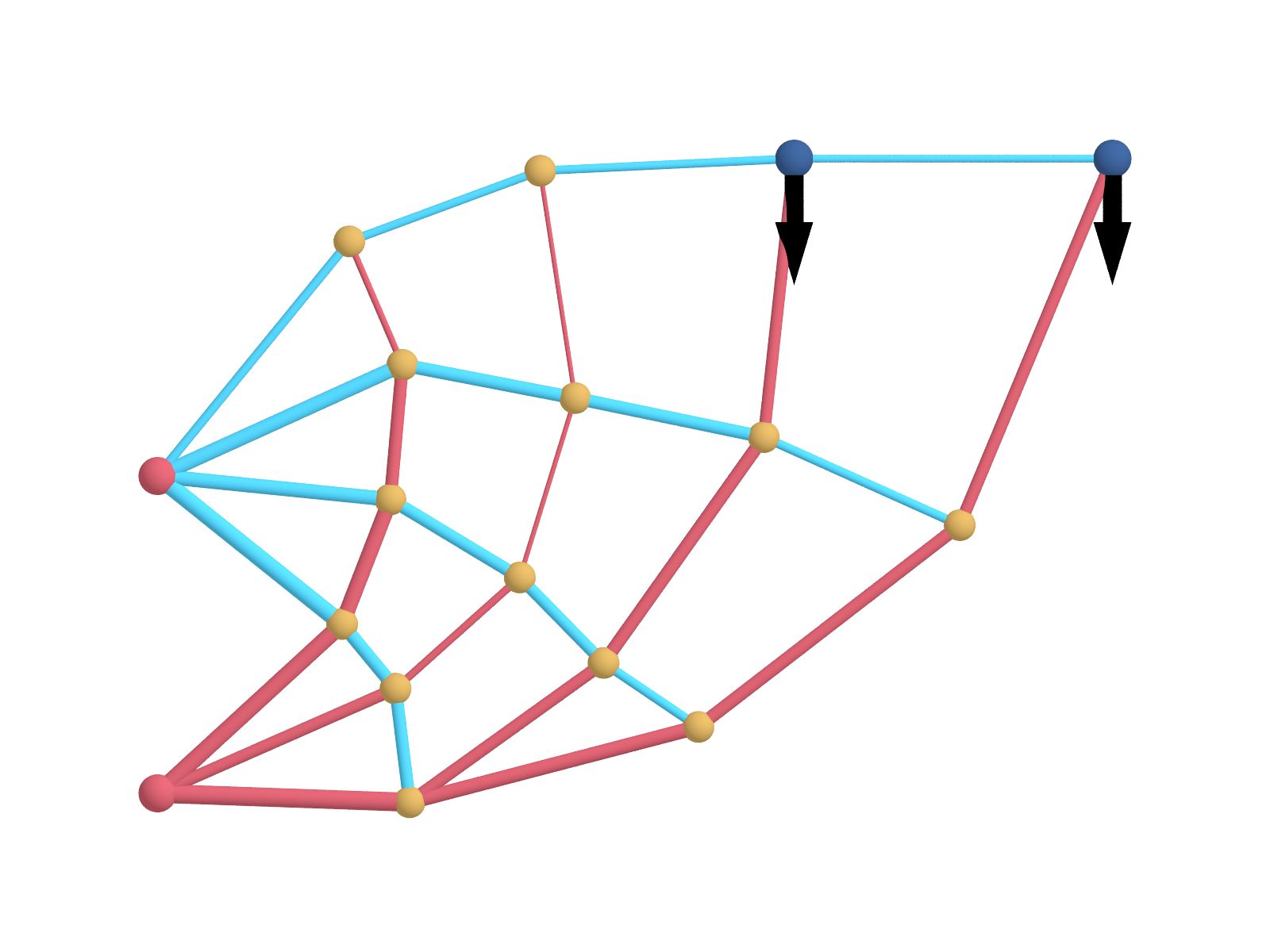}	
		\cput(70,10){\contour{white}{$(g2)$}}		
		\end{overpic}
	}	
	\caption{
		\label{fig:methodcomparison}
		Volume optimization comparison with different methods. From left to right: (a) the ground structure method (GSM), (b) ALP (ours), (c) SQP with initialization of cross-section of constant values, (d) SQP with initialization of cross-section from (a), (e) the gradient descent method (GDM) with initialization of cross-section of constant values, (f) uses the gradient descent method with initialization of cross-section from (a). (g) the method used in ~\cite{jiang2017design}. 
        Table \ref{table_ALP} shows our ALP method has better performance.}
\end{figure*}

We compare the proposed ALP algorithm with three alternative optimization methods: sequential quadratic programming (SQP), gradient descent (GD), and the method of Jiang et al.~\cite{jiang2017design}. The details of the implementation of SQP and GDM are given in the appendix. Jiang et al.'s framework contains multiple parts and we only compare to the solver that is comparable to our proposed ALP algorithm (The objective function in Jiang et al. is different though). For various input specifications, we compare the quality (volume) of the output. Fig. \ref{fig:methodcomparison} and Table \ref{table_ALP} show the results. The results demonstrate that our proposed optimization converges to a better solution than competing approaches, especially GD struggles to find meaningful solutions. We noticed that competing algorithms can be improved when using the ground structure method (GSM) for initialization. In the table, we use a * to indicate the version of SQP and GD that has been initialized using GSM. In our comparison, our method generates the best results for all input specifications. SQP* can match our result in some cases. We can also observe that our method is fast and better scales to larger inputs than SQP. The reason for the 
poor performance of Jiang et al.~\cite{jiang2017design} is that they use soft constraints instead of hard constraints in the formulation.

\subsubsection{Comparisons with Previous Numerical Methods}
\begin{figure}[h]	\centering{
		\begin{overpic}	
			[width=.34\columnwidth,trim={4cm 3cm 3cm 13.5cm},clip]{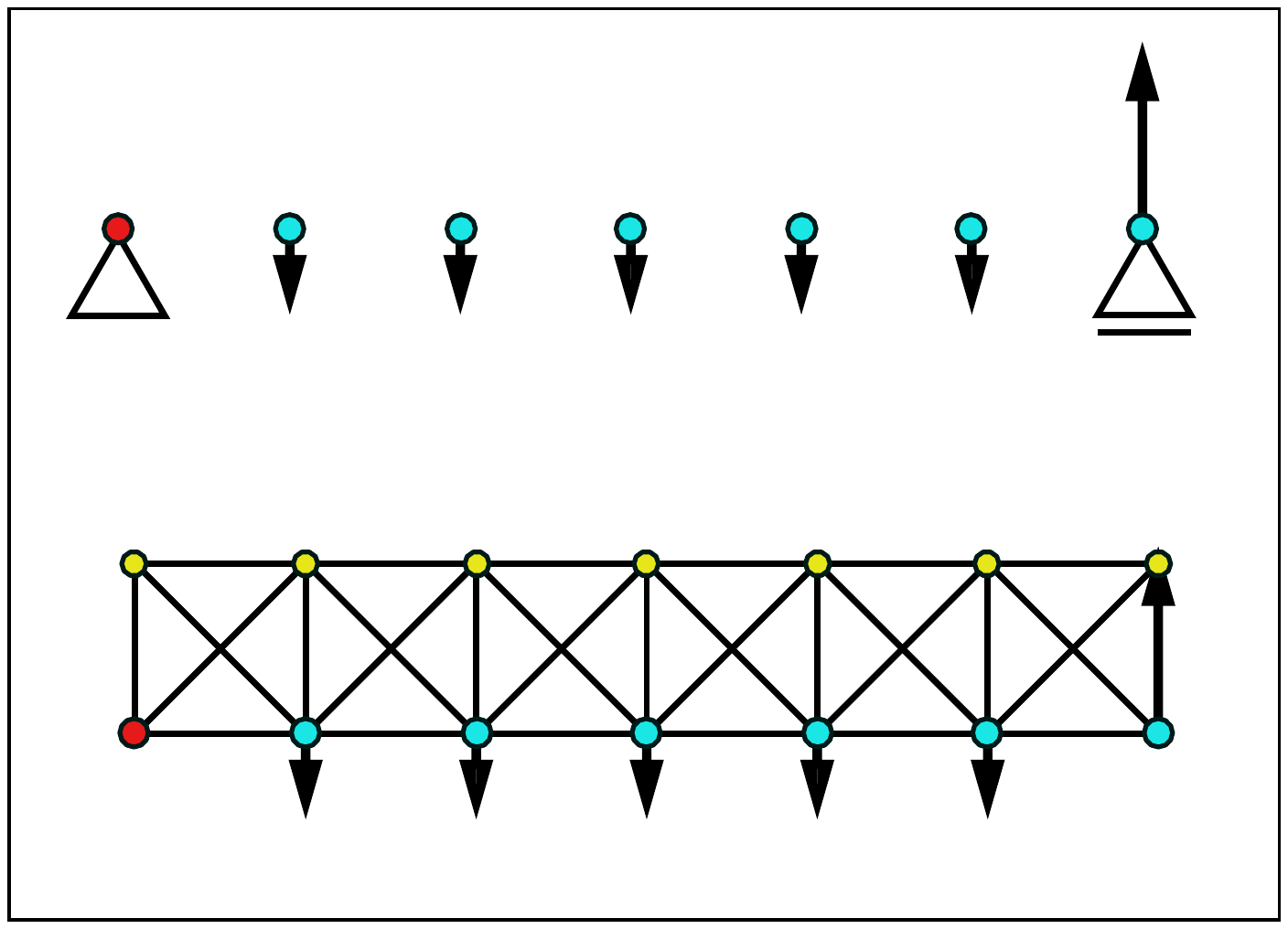}	
			\cput(50,0){\contour{white}{$(a)$}}					
		\end{overpic}
		\begin{overpic}	
			[width=.34\columnwidth,trim={4cm 8cm 4cm 8cm},clip]{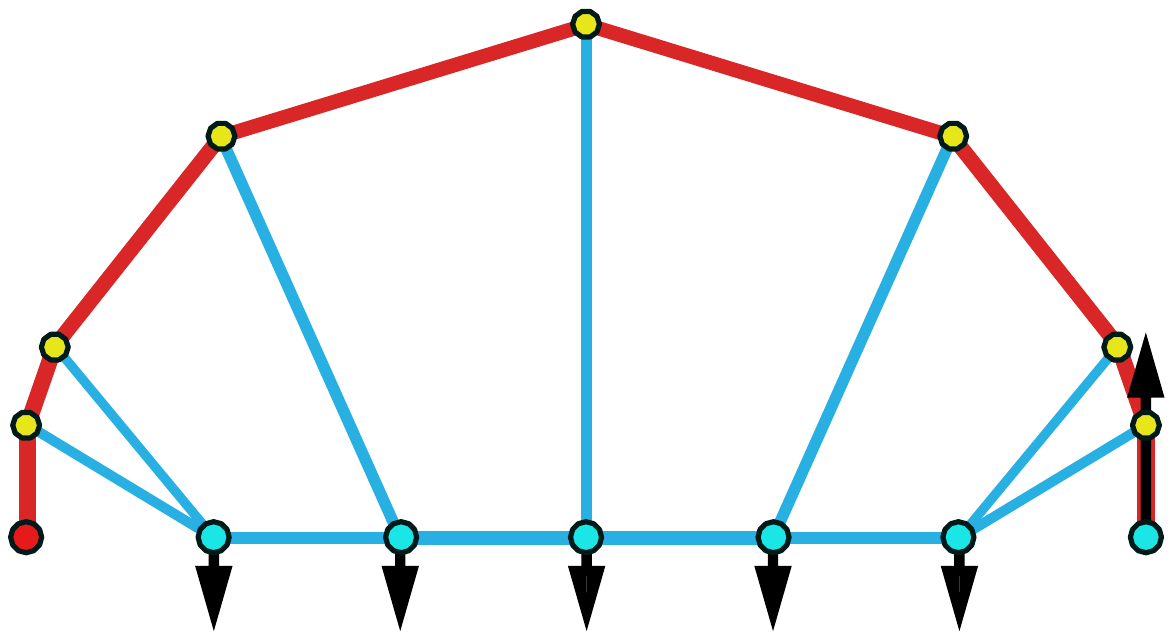}	
			\cput(50,0){\contour{white}{$(b)$}}					
		\end{overpic}
		\begin{overpic}	
			[width=.32\columnwidth,trim={4cm 8cm 4cm 8cm},clip]{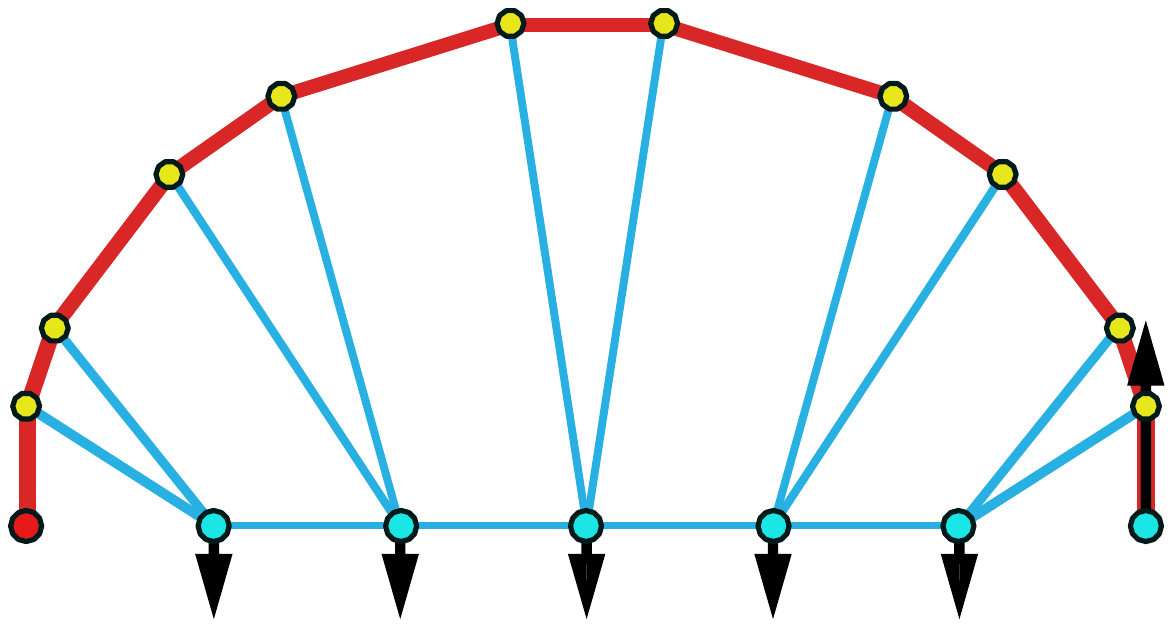}	
			\cput(50,0){\contour{white}{$(c)$}}					
		\end{overpic}
		\begin{overpic}	
			[width=.32\columnwidth,trim={4cm 8cm 4cm 8cm},clip]{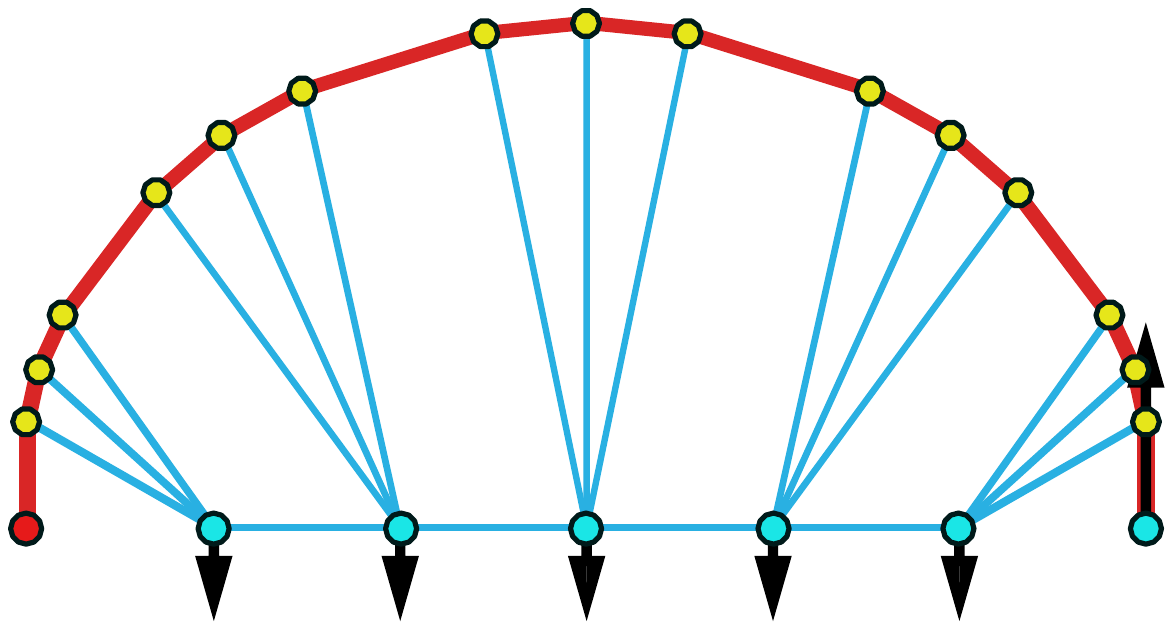}	
			\cput(50,0){\contour{white}{$(d)$}}					
		\end{overpic}
		\begin{overpic}	
			[width=.32\columnwidth,trim={4cm 8cm 4cm 8cm},clip]{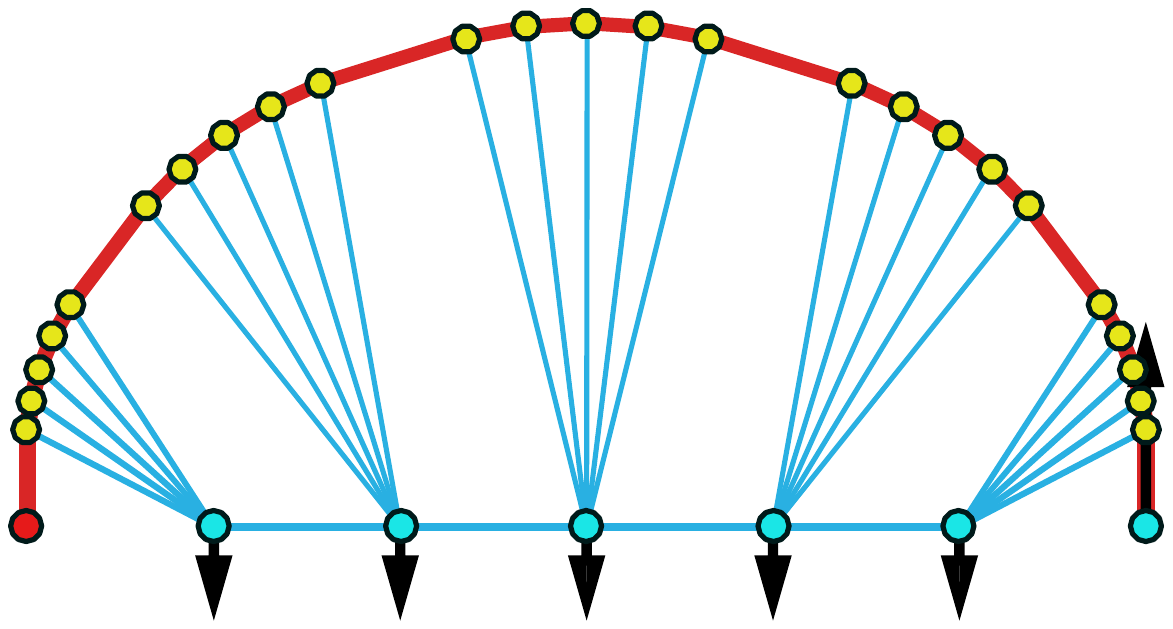}	
			\cput(50,0){\contour{white}{$(e)$}}					
		\end{overpic}
	}	
	\caption{A comparison with the method in \cite{descamps2013lower} on a 2D bridge model. (a) The input functional specification and initial truss for the method in \cite{descamps2013lower}. (b) Optimal truss in \cite{descamps2013lower} (number of bars: 19, volume: 34.977). (c) Our optimal coarse truss (number of bars: 25, volume: 34.593). Refined structures after 1 and 2 rounds of subdivisions are shown in (d) and (e).
		\label{fig:comparison3}
	}
\end{figure}

\begin{figure}[h]	\centering{
		\begin{overpic}			
			[height=.25\columnwidth]{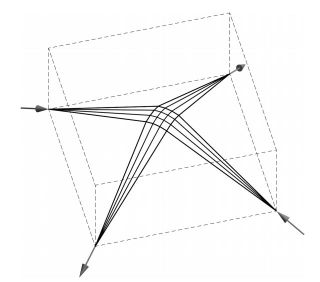}	
			\cput(50,2){\contour{white}{$(a)$}}		
		\end{overpic}
		\begin{overpic}			
			[height=.25\columnwidth]{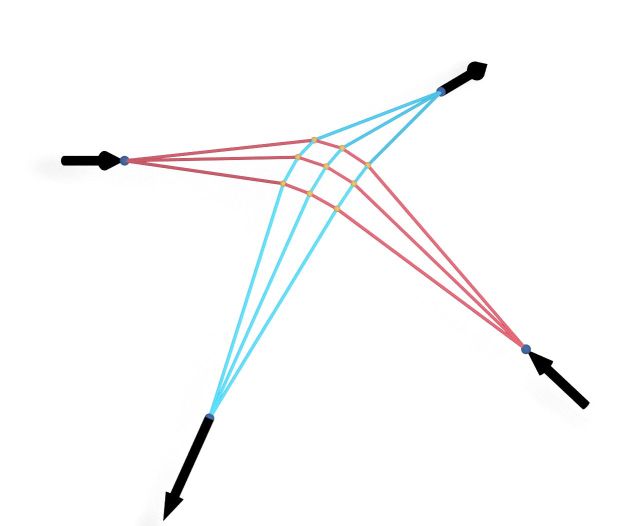}	
			\cput(50,2){\contour{white}{$(b)$}}		
		\end{overpic}
		\begin{overpic}			
			[height=.25\columnwidth]{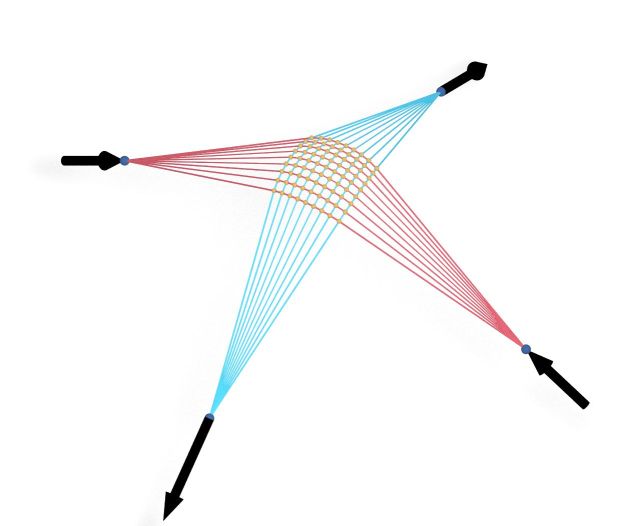}	
			\cput(50,2){\contour{white}{$(c)$}}		
		\end{overpic}
	}	
	\caption{A comparison with the method in \cite{sokol2017numerical} on a 3D model. (a) is the optimal structure in~\cite{sokol2017numerical}. To get this topology, they used a 3D grid of 50X50X50 joints and 7,318,049,198 bars, the running time is close to 2 hours. (b) and (c) are results of our method. To get the initial structure in (b), we use 1118 bars and running time is less than 10s. To get the optimization result in (c) our running time is less than 20s.
		\label{fig:comparison2}
	}
\end{figure}

\input{1_table}
\input{2_table}
We compare the performance of our method with several previous methods using the functional specifications, solutions, and running times provided in their papers. We denote the methods in \cite{descamps2013lower}, \cite{gilbert2003layout}, \cite{he2015rationalization}, and \cite{sokol2017numerical} as D2013, G2003, H2015 and S2017, respectively.
In Figure \ref{fig:Teaser} and Figure \ref{fig:comparison3}, we compare our method with \cite{descamps2013lower} for a 2D and a 3D bridge design problem. In Figure \ref{fig:comparison1} and \ref{fig:comparison5}, we compare our method with \cite{he2015rationalization} and \cite{gilbert2003layout} on a benchmark problem---Hemp cantilever design.
For 3D truss optimization, we compare our method with \cite{sokol2017numerical} on a simple 3D model in Figure \ref{fig:comparison2}. The comparison is shown in Table \ref{table2}. We can observe that our method is orders of magnitude faster, even though we can achieve a lower volume than previous work. Unfortunately, these methods do not have code or executables publicly available, so we cannot test them on the same machine. However, it seems unlikely that this significant difference in running time can be overcome by slightly faster hardware.

\subsubsection{Comparisons with Analytical Solutions}
\begin{figure}[h]	\centering{
		\begin{overpic}			
			[height=.38\columnwidth]{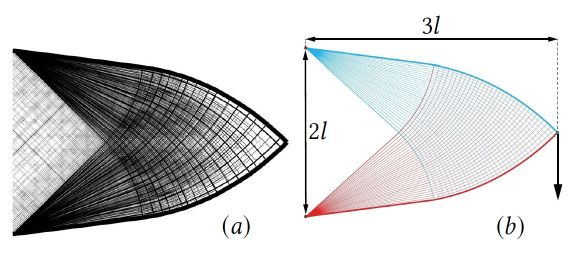}	
	\cput(80,2){\contour{white}{$ $}}		
		\end{overpic}
	}	
	\caption{
		\label{fig:comparison1}
		A comparison with the method in \cite{gilbert2003layout} on a benchmark design problem (Hemp cantilever). Compared with the best result presented in \cite{gilbert2003layout}, obtained with 6h50m of computation, shown in (a), which used an initial truss of 116,288,875 bars and obtain a total volume of 4.499827, we obtain a result, shown in (b), with 2178 bars in 30s, which achieves a total volume of 4.498635. For this problem, an analytical solution with an infinite amount of bars exists, which has a total volume of 4.498115.}
\end{figure}

\begin{figure}[h]	\centering{
		\begin{overpic}			
			[height=.35\columnwidth]{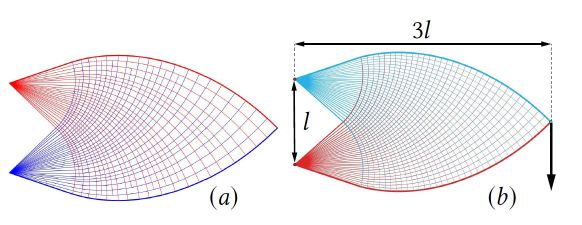}	
			\cput(80,0){\contour{white}{$ $}}		
		\end{overpic}
	}	
	\caption{
		\label{fig:comparison5}
		A comparison with the method in \cite{he2015rationalization} on a benchmark design problem (Hemp cantilever). (a) The optimal truss of \cite{he2015rationalization} (number of bars: 4244, running time: 4875s, volume: 4.3228). (b) Our optimized truss (number of bars: 2178, running time: 30s, volume: 4.3223). The analytical solution of optimal volume is 4.3217.}
\end{figure}

\begin{figure}[h]	\centering{
		\begin{overpic}	
			[width=.40\columnwidth]{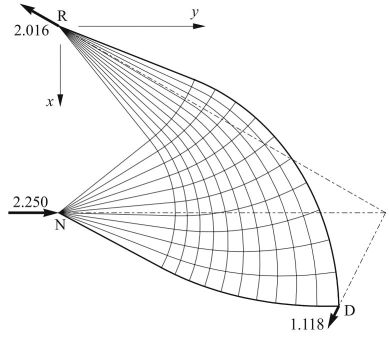}	
			\cput(50,0){\contour{white}{$(a)$}}			
		\end{overpic}
		\begin{overpic}	
			[width=.40\columnwidth,trim={5cm 9cm 5cm 9cm},clip]{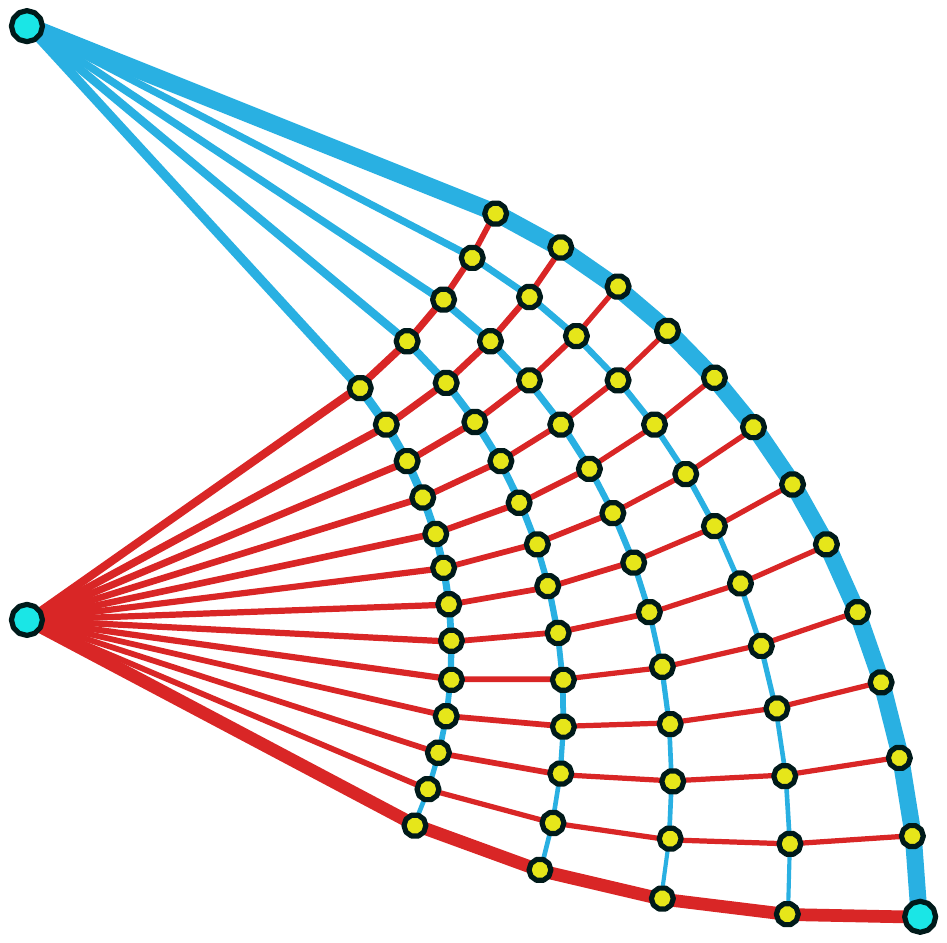}	
			\cput(50,0){\contour{white}{$(b)$}}				
		\end{overpic}
	}	
	\caption{A comparison with an analytic solution in \cite{sokol2010solution}. The volume of the analytic solution with an infinite amount of bars is 6.831 and the volume of our optimized discrete truss (130 bars) is 6.838. 
		\label{fig:comparison4}
	}
\end{figure}

For the cantilever design problems in Figure \ref{fig:comparison1} and \ref{fig:comparison5}, the total volumes of analytical solutions are 4.498115 and 4.232168. 
The volumes of our discrete designs for these two cases are  4.498635 and 4.3223, which are closer to the analytical solutions than the previous work, \cite{gilbert2003layout} and \cite{he2015rationalization}. For the three force problem, we compare our result with the analytical solution presented in ~\cite{sokol2010solution}. In Figure \ref{fig:comparison4}, we show the computed discrete truss (130 bars, volume: 6.838) on the right which is visually similar to the analytical solution on the left (volume: 6.831).


\begin{figure*}[tb]\centering{
		\begin{overpic}			
			[width=0.4\columnwidth]{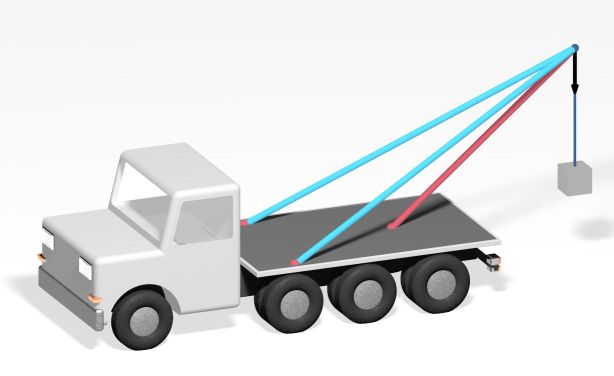}	
			\cput(50,2){\contour{white}{$(a)$}}		
		\end{overpic}
		\begin{overpic}			
			[width=0.4\columnwidth]{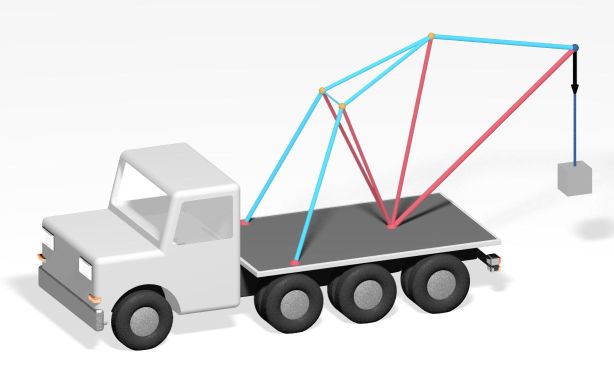}	
			\cput(50,2){\contour{white}{$(b)$}}	
		\end{overpic}
		\begin{overpic}			
			[width=0.4\columnwidth]{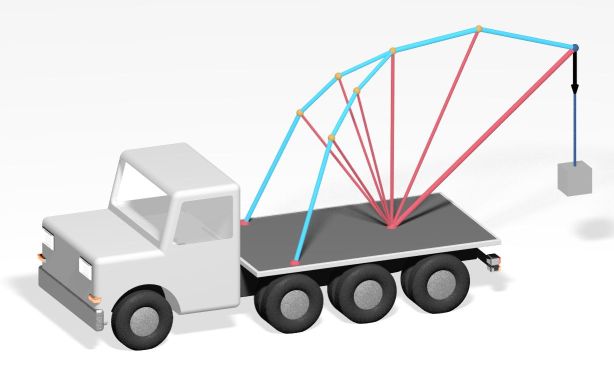}	
			\cput(50,2){\contour{white}{$(c)$}}	
		\end{overpic}
		\begin{overpic}			
			[width=0.4\columnwidth]{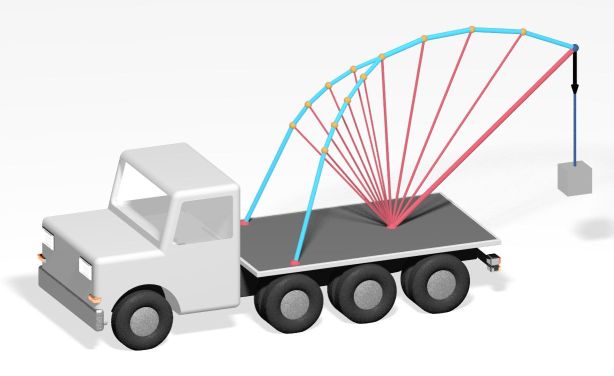}	
			\cput(50,2){\contour{white}{$(d)$}}	
		\end{overpic}
		\begin{overpic}			
			[width=0.4\columnwidth]{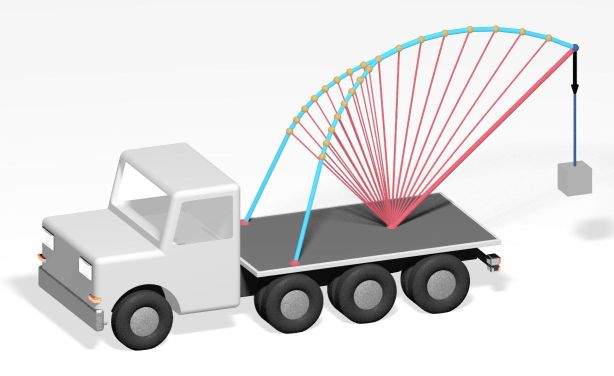}	
			\cput(50,2){\contour{white}{$(e)$}}	
		\end{overpic}
	}	
	\caption{
		\label{fig:3D002}
		A 3D example.}
\end{figure*}

\paragraph{Discussion and Limitations.}
Compared with previous work, the geometry optimization in terms of both the axial forces and joint positions through two linear programming problems alternatively applied in ALP provides more degrees of freedom compared with the original formulation of the ground structure method which solves a single linear programming problem. Splitting a highly nonlinear programming problem into two linear ones also attains better efficiency compared with the original nonlinear formulation. Moreover, the two categories of topological operations, local and global, complementarily allow both flexible yet stable topology changes and manipulation. Most importantly, the subdivision approach, which has been overlooked by previous work, is a natural choice for topology refinement from coarse to fine, which creates valid topologies at different levels both efficiently and robustly. Despite the efficiency and efficacy, our optimization framework cannot guarantee a global optimum. However, in simple special cases where the analytical optimum is known, we observe that the method almost reaches the known global optimum. \AD{ As our method is based on subdivision of edge-networks on surfaces, the optimal trusses in 3D also need to constitute sheets of surfaces. For general 3D specifications, the subdivision approach requires an initial surface-like structure, or a structure that consists of multiple sheets, which is a challenging problem for future work. The proposed ALP algorithm will work for general 3D structures, however.
}

%% file: ALP_table.tex
\begin{table*}[]
\centering
\begin{tabular}{|p{1cm}|p{0.8cm}|p{0.8cm}|p{0.8cm}|p{0.8cm}|p{0.8cm}|p{0.8cm}|p{0.8cm}|p{0.8cm}|p{0.8cm}|p{0.8cm}|p{0.8cm}|p{0.8cm}|p{0.8cm}|p{0.8cm}|}
\hline
{Fig.} & \multicolumn{2}{c|}{GSM} & \multicolumn{2}{c|}{ALP} & \multicolumn{2}{c|}{SQP} & \multicolumn{2}{c|}{SQP*} & \multicolumn{2}{c|}{GD} & \multicolumn{2}{c|}{GD*} & \multicolumn{2}{c|}{{[}Jiang et al. 2017{]}} \\ \cline{2-15} 
                     & Volume      & Time(s)    & Volume      & Time(s)    & Volume      & Time(s)    & Volume      & Time(s)     & Volume     & Time(s)    & Volume     & Time(s)     & Volume            & Time(s)           \\ \hline
\ref{fig:methodcomparison}(1)                  & 98.000      & 0.25       & \textbf{85.416}      & 2.35       & 87.420      & 6.10       & 87.331     & 6.24        & 97.731     & 1.93       & 96.993      & 1.88        & 92.992             & 40.80             \\ \hline
\ref{fig:methodcomparison}(2)                   & 23.868      & 0.38       & \textbf{22.318}      & 4.01       & \textbf{22.318}      & 11.47      & \textbf{22.318}     & 9.17        & 22.662     & 2.53       & 22.638     & 2.33        & 22.790            & 42.50             \\ \hline
\ref{fig:parallelforce}                 & 6.516       & 0.23       & \textbf{6.183}       & 2.99       & 6.482       & 1.68       & 6.399       & 2.16        & 6.656      & 1.13       & 6.516      & 1.07        & 6.396             & 29.56             \\ \hline
\ref{fig:concurrentforce}                  & 3.125       & 0.23       & \textbf{3.041}       & 6.52       & 3.569       & 2.23       & 3.099       & 0.77        & 3.101      & 1.39       & 3.100      & 1.15        & 3.124             & 33.13             \\ \hline
\ref{fig:nonconcurrentforce}                 & 3.872       & 0.23       & \textbf{3.555}       & 5.76       & 4.064       & 3.68       & 3.609       & 1.15        & 4.197      & 1.38       & 3.872      & 1.17        & 3.783             & 31.62             \\ \hline
\ref{fig:2D005}(a1)                  & 3.375       & 0.48       & \textbf{3.238}       & 1.50       & 3.536       & 5.89       & \textbf{3.238}       & 0.72        & 3.371      & 2.03       & 3.356      & 1.22        & 3.343             & 33.26             \\ \hline
\ref{fig:3D003}                & 19.510      & 0.22       & \textbf{19.081}      & 0.79       & 21.478      & 2.14       & 19.141      & 0.50        & 19.338     & 1.86       & 19.358     & 1.37        & 19.268            & 48.11             \\ \hline
\ref{fig:concurrentforce3D}                  & 13.891      & 0.36       & \textbf{11.885}      & 0.54       & 12.000      & 6.57       & 11.901      & 17.40       & 12.767     & 3.70       & 13.217     & 3.48        & 13.653            & 57.30             \\ \hline
\ref{fig:nonconcurrentforce3D}                 & 15.429      & 0.35       & \textbf{14.627}      & 1.11       & 17.282      & 20.36      & \textbf{14.627}      & 17.37       & 15.135     & 3.43       & 15.173     & 2.79        & 16.722            & 70.10             \\ \hline
\end{tabular}
\caption{Comparison of the achieved volume and running time for different numerical optimization methods. We only compare the result of the geometry optimization part without performing topology operations or subdivision.}
\label{table_ALP}
\end{table*}

%% file: 1_table.tex
\begin{table*}[t]
\centering

\begin{tabular}{|l|l|l|l|l|l|l|l|l|l|l|l|l|l|l|}
\hline
{Fig.} & \multicolumn{2}{l|}{Initial Truss}                          & \multicolumn{3}{l|}{Optimal Coarse Truss} & \multicolumn{3}{l|}{Truss Subdivision 1} & \multicolumn{3}{l|}{Truss Subdivision 2} & \multicolumn{3}{l|}{Truss Subdivision 3} \\ \cline{2-15} 
                     & Bars & \begin{tabular}[c]{@{}l@{}}Volume\\ GSM\end{tabular} & Bars       & Volume       & Time(s)       & Bars       & Volume       & Time(s)      & Bars       & Volume       & Time(s)      & Bars       & Volume       & Time(s)      \\ \hline
\ref{fig:parallelforce}                   & 800  & 6.516                                                & 14         & 5.830        & 3.76          & 26         & 5.732        & 0.31         & 50         & 5.709        & 1.41         & 98         & 5.703        & 2.97         \\ \hline
\ref{fig:concurrentforce}                   & 404  & 3.125                                                & 22         & 2.967        & 5.86          & 46         & 2.913        & 5.85         & 118        & 2.909        & 8.86         & 358        & 2.907        & 15.16        \\ \hline
\ref{fig:nonconcurrentforce}                  & 124  & 3.872                                                & 16         & 3.365        & 0.64          & 36         & 3.248        & 0.88         & 100        & 3.203        & 5.46         & 324        & 3.188        & 17.28        \\ \hline
\ref{fig:2D005}(a1)                & 835  & 3.375                                                & 19         & 3.210        & 5.02          & 45         & 3.183        & 1.43         & 133        & 3.170        & 8.79         & 453        & 3.163        & 29.56        \\ \hline
\ref{fig:2D005}(b1)                & 835  & 3.335                                                & 25         & 3.114        & 5.42          & 67         & 3.041        & 1.81         & 211        & 3.021        & 7.77         & 739        & 3.016        & 17.84        \\ \hline
\ref{fig:2D0042}                   & 466  & 4.340                                                & 25         & 4.294        & 4.92          & 67         & 4.289        & 3.22         & 211        & 4.288        & 9.68         & 739        & 4.287        & 35.63        \\ \hline
\ref{fig:3D003}                   & 2061 & 19.510                                               & 9          & 19.081       & 4.28          & 18         & 18.700       & 0.46         & 37         & 18.610       & 2.39         & 76         & 18.587       & 5.56         \\ \hline
\ref{fig:concurrentforce3D}                   & 1118 & 13.891                                               & 24         & 11.742       & 5.87          & 60         & 11.724       & 7.73         & 180        & 11.718       & 19.09        & 612        & 11.713       & 57.41        \\ \hline
\ref{fig:nonconcurrentforce3D}                   & 3674 & 15.429                                               & 10         & 14.628       & 1.61          & 16         & 14.416       & 1.53         & 28         & 14.325       & 1.97         & 52         & 14.316       & 2.11         \\ \hline
\ref{fig:3D004}                   & 3674 & 30.850                                               & 24         & 29.049       & 8.70          & 58         & 28.507       & 1.96         & 174        & 28.285       & 14.75        & 598        & 28.217       & 64.01        \\ \hline
\ref{fig:3D002}                     & 3674 & 20.545                                               & 10         & 18.900       & 3.41          & 16         & 18.542       & 0.4          & 30         & 18.412       & 6.22         & 59         & 18.375       & 14.10        \\ \hline
\end{tabular}
\caption{Statistics of results presented in this paper: For each truss design, we report the number of bars, the total volume of material consumption, as well as the computational time for different stages. Please note that the results from optimal coarse trusses are already better than the results from the ground structure method without introducing additional joints and bars. Topological refinement through subdivision provides additional reduction in material consumption generally in less than one minute in our current implementation which still has a significant room for further improvement.}
\label{table1}
\end{table*}

%% file: 2_table.tex
\begin{table}[]
\centering
\begin{tabular}{|c|c|c|c|c|c|}
\hline
Fig.                   & Method & \begin{tabular}[c]{@{}c@{}}Bars\\ (initial)\end{tabular} & \begin{tabular}[c]{@{}c@{}}Bars\\ (final)\end{tabular} & Time           & Volume           \\ \hline
{\ref{fig:Teaser}}   & D2013  & 258                                                      & 96                                                    & 1376s          & 408.807          \\ \cline{2-6} 
                         & Ours   & 258                                                      & 201                                                    & \textbf{7.7s}  & \textbf{333.395} \\ \hline
{\ref{fig:comparison3}}  & D2013  & 31                                                       & 19                                                     & n/a            & 34.977           \\ \cline{2-6} 
                         & Ours   & 804                                                       & 25                                                     & \textbf{3.4s}  & \textbf{34.593}  \\ \hline
{\ref{fig:comparison1}}  & G2003  & \textgreater1 billion                                     & n/a                                                    & \textgreater6h & 4.4998           \\ \cline{2-6} 
                         & Ours   & 105                                                      & 2178                                                   & \textbf{30s}   & \textbf{4.4986}  \\ \hline
{\ref{fig:comparison5}}  & H2015  & 12,456,601                                                 & 4244                                                   & 4875s          & 4.3228           \\ \cline{2-6} 
                         & Ours   & 105                                                      & 2178                                                   & \textbf{30s}   & \textbf{4.3223}  \\ \hline
{\ref{fig:comparison2}} & S2017  & \textgreater7 billion                                     & 40                                                     & \textgreater1h & n/a              \\ \cline{2-6} 
                         & Ours   & 1118                                                     & 24                                                     & \textbf{13.6s} & \textbf{11.742}  \\ \hline
\end{tabular}
\caption{Comparisons with previous work. Compared with previous approaches, our framework consistently creates truss designs with smaller volumes with significantly shorter computational times.}
\label{table2}
\end{table}

%% file: 7_Conclusion.tex
\section{Conclusions and Future Work}
We present a method for the design of optimal trusses satisfying functional specifications with minimized material consumption. The core components of the proposed approach include an alternating linear programming formulation for geometry optimization and two sets of topological operations. The subdivision scheme inspired by Michell's theoretical studies utilized in the global topology refinement step plays a crucial role for the efficiency and efficacy of the proposed approach. The performance of our framework is validated by comparisons with multiple previous studies in different scenarios\AD{, which indicate that our method creates trusses with smaller volumes and is over two orders of magnitude faster in terms of computational speed}.
For future work, it would be exciting to study dynamic structures created by trusses with movable parts that have multiple configurations, e.g., robotic and mechatronic systems with trusses. Moreover, a better theoretical understanding of optimal trusses in 3D would be inspiring for the entire field.

%% file: 9_Appendix2.tex
\section*{Appendix}

\subsection{Equilibrium Force Systems}

\begin{figure}[H]	\centering{
		\begin{overpic}
			[width=.89\columnwidth]{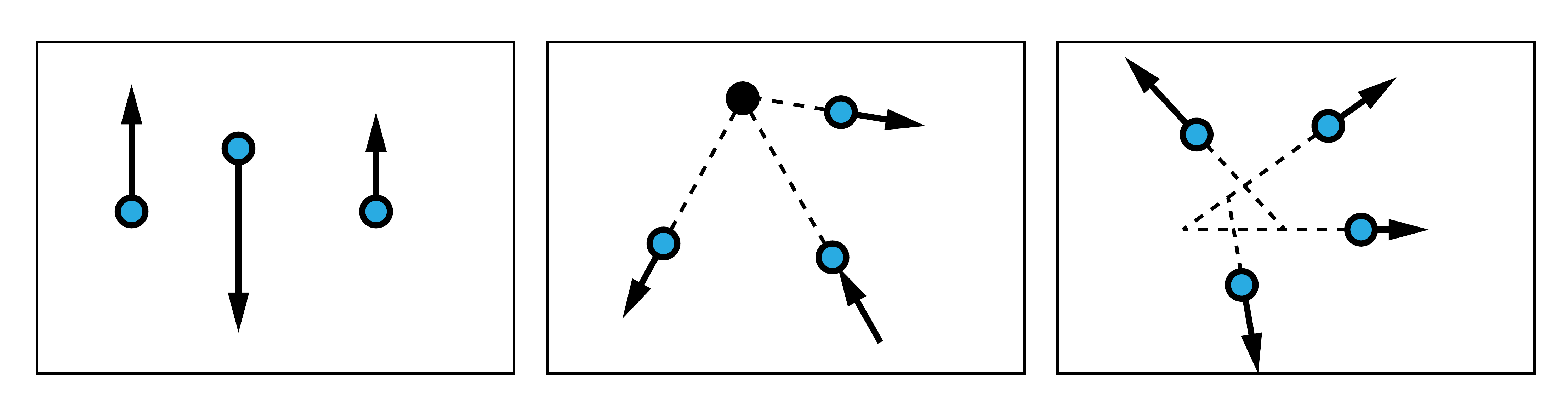}
			\cput(18,-2){\contour{white}{$(a)$}}	
			\cput(50,-2){\contour{white}{$(b)$}}
			\cput(82,-2){\contour{white}{$(c)$}}				
		\end{overpic}
	}	
	\caption{
		\label{fig:forcesys}
		Three equilibrium force systems. (a) a parallel force system. (b) a concurrent force system. (c) a non-concurrent force system.}
\end{figure}

The input functional specification is essentially an equilibrium force system. For a system with forces $F_1$,$F_2$,...,$F_n$ acting at points $P_1$,$P_2$,...,$P_n$, there are two constraints of equilibrium: the balance of forces, $\sum_{i=1}^{n}F_i=\vec{0}$, and the balance of torques, $\sum_{i=1}^{n}{F_i\times P_i} =\vec{0}$. In mechanical engineering, the equilibrium force systems are usually classified into three types: parallel force systems, concurrent force systems, and non-concurrent force systems as shown in Figure \ref{fig:forcesys}. 

\subsection{SQP}
First, lets recall the formulation in Eq.\ref{eqn:lp2}.
In this formulation, joint positions are considered as known values. The objective function and constraints in the above formulation are both linear. If the joint positions $\uw_j, j=1,...,|V|$ are considered as design variables, the formulation is a nonlinear programming with nonlinear constraints. To solve such a problem by SQP, we reformulate the problem similarly as in ~\cite{descamps2013lower}. We introduce slack variables, $w_{i}^+$ and $w_{i}^-$, {which represent the magnitudes of compressive and tensile forces respectively:} $w_{i}^+ = \max (0, w_{i})$, $w_{i}^- = \max (0, -w_{i})$. Then, we have $w_i=w_{i}^+ - w_{i}^-$, and $|w_i|=w_{i}^+ + w_{i}^-$. Here we denote the vectors consisted of all $w_{i}^+$ and $w_{i}^+$ as $\ww^+$ and $\ww^-$. We assume the $i$th bar is connected by joints $i1$ and $i2$, then $l^{2}_i=(\uw_{i1}-\uw_{i2})^2$. The elements in matrix $\mathbf{C}$ are linear combinations of joint positions. 
The  whole problem is reformulated as
 
\begin{align}
& \underset{\uw_j,\ww_i^+,\ww_i^-}{\text{minimize}}
& & \sum_{i=1}^{|E|} (\uw_{i1}-\uw_{i2})^2(w_{i}^+ + w_{i}^-), \label{eqn:lp8} \\
& \text{subject to}
& & \mathbf{C}^T (\ww^+ -\ww^-) = -\fw, \tag{\theequation a} \label{eqn:lp8_a} \\
& & & w_{i}^+ \geq 0 ,  \tag{\theequation b} \label{eqn:lp8_b} \\
& & & w_{i}^- \geq 0 ,  \tag{\theequation c} \label{eqn:lp8_c} 
\end{align}

The algorithm in ~\cite{descamps2013lower} used the above formulation by SQP method.
We implement the algorithm using the $fmincon$ routine in Matlab.

\subsection{The Gradient Decent Method}
To use the GDM, we first transfer the constrained optimization problem in Eq. \ref{eqn:lp8} into an unconstrained formulation by adding an extra penalty term into the objective function as 
\begin{equation}
f=\lambda_1E_{volume}+\lambda_2E_{static},
\end{equation}
where
\begin{equation*}
E_{volume}=\sum_{i=1} ^{|E|}(\uw_{i1}-\uw_{i2})^2(w_{i}^+ + w_{i}^-),
\end{equation*}
and
\begin{equation*}
E_{static}=(\mathbf{C}^T (\ww^+ -\ww^-) + \fw)^2.
\end{equation*}
This objective function is defined similarly in ~\cite{jiang2017design}. The constraint of $w_{i}^+ \geq 0$ and $w_{i}^- \geq 0$ are avoided by replacing $w_{i}^+$ and $w_{i}^-$ with $(\varrho_{i}^+)^2$ and $(\varrho_{i}^+)^2$, where $\varrho_{i}^+$ and $\varrho_{i}^-$ are new variables without any constraints. Usually, the weight of penalty term $\lambda_2$ should be much larger than $\lambda_1$. The above object function is differentiable. We do the comparison using standard gradient decent method.

\clearpage
\newpage
\section*{Additional Materials}

\subsection{Special Cases}

\begin{figure}[H]	\centering{
		\begin{overpic}			
			[width=.45\columnwidth,trim={2cm 11.5cm 2cm 11.5cm},clip]{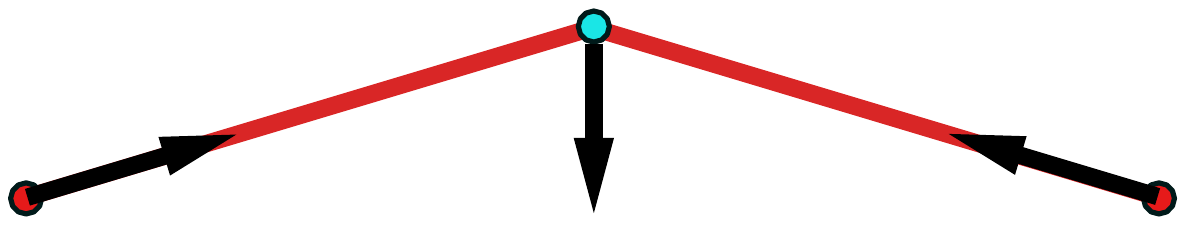}	
			\cput(50,2){\contour{white}{$(a)$}}		
		\end{overpic}
		\begin{overpic}			
			[width=.45\columnwidth,trim={2cm 10.5cm 2cm 11.5cm},clip]{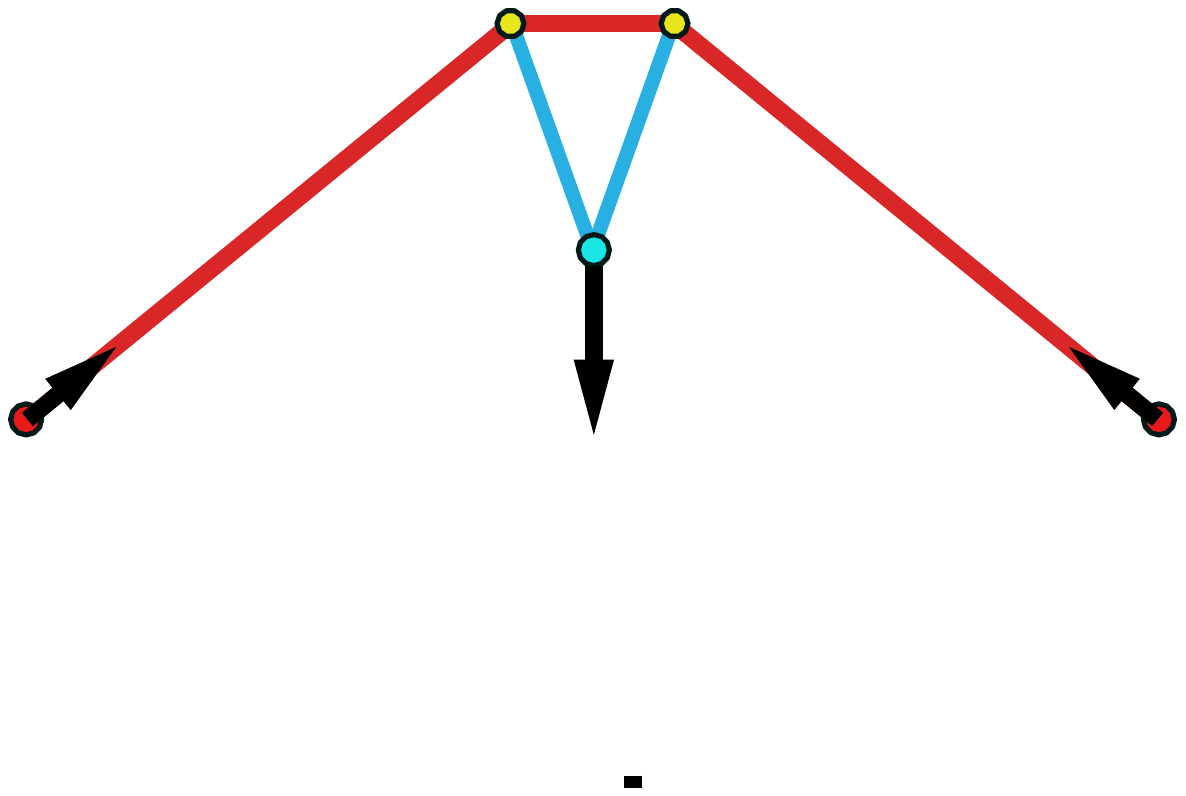}	
			\cput(50,2){\contour{white}{$(b)$}}	
		\end{overpic}
		
	}	
	\caption{
		\label{fig:2bars}
		The inputs are two supporting points and one external load. (a) a truss of two bars in compression. (b) a better design as reaction forces at supporting points are changed to obtain a less material structure.}
\end{figure}

The theorem of Maxwell(1872) states that $\sum{f_il_i}=C$, where $C$ is the same constant for any statically admissible truss layout considering a given set of external forces. As a consequence, if it is found possible to design a structure all of whose members are in tension, or alternatively compression, then the optimum design has been achieved, because $\sum{\left | f_i \right |l_i}=\left | C \right |$. The above statement assumes the input force system is precisely given. For example, a self-equilibrium force system with no supporting points, such as the 3-force problem in 2D and 4-force problem in 3D. However, it is not correct when some supporting points are given without prescribing their reaction forces. For example, as shown in Figure \ref{fig:2bars}, we can find a structure of two bars with compression on the left, but it is not an optimal structure because reaction forces at the boundary could be changed to achieve a better results as show on the right.

\begin{figure}[H]	\centering{
		\begin{overpic}			
			[width=.32\columnwidth,trim={5cm 9cm 4.5cm 9cm},clip]{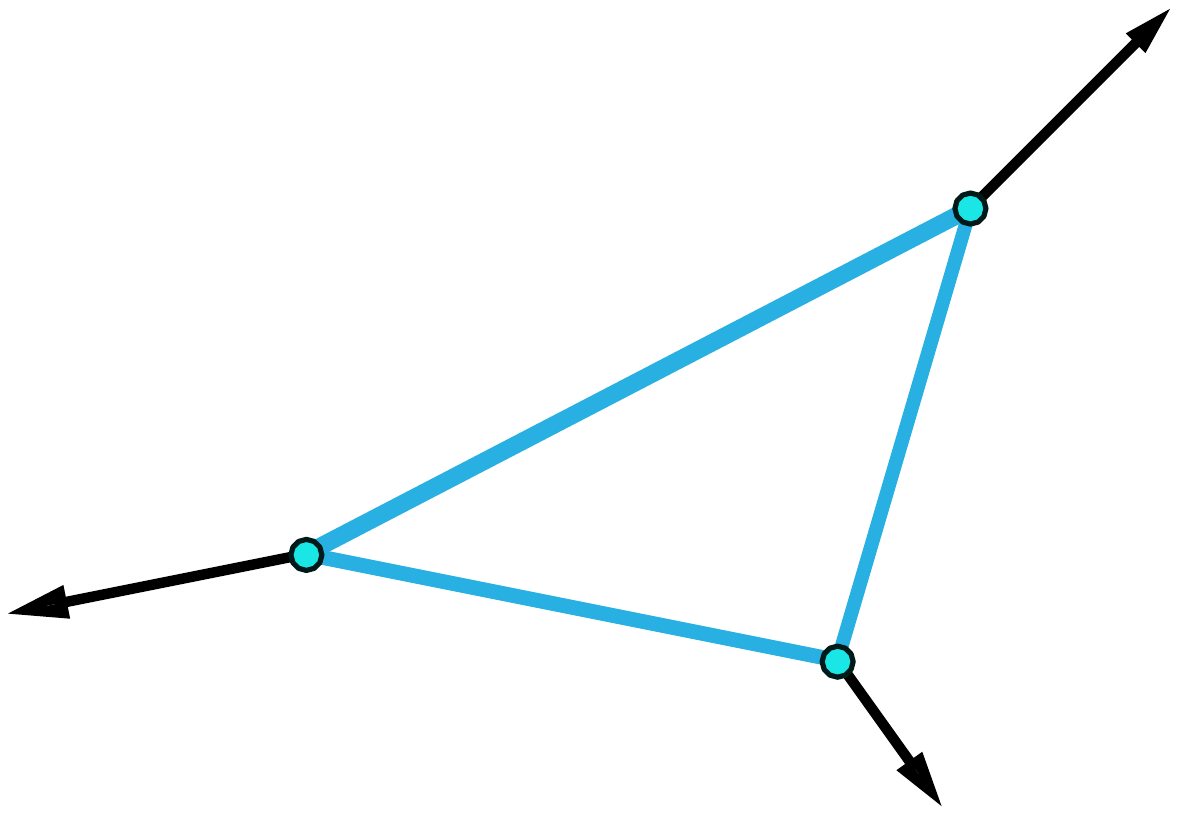}	
			\cput(50,2){\contour{white}{$(a)$}}		
		\end{overpic}
		\begin{overpic}			
			[width=.32\columnwidth,trim={5cm 9cm 4.5cm 9cm},clip]{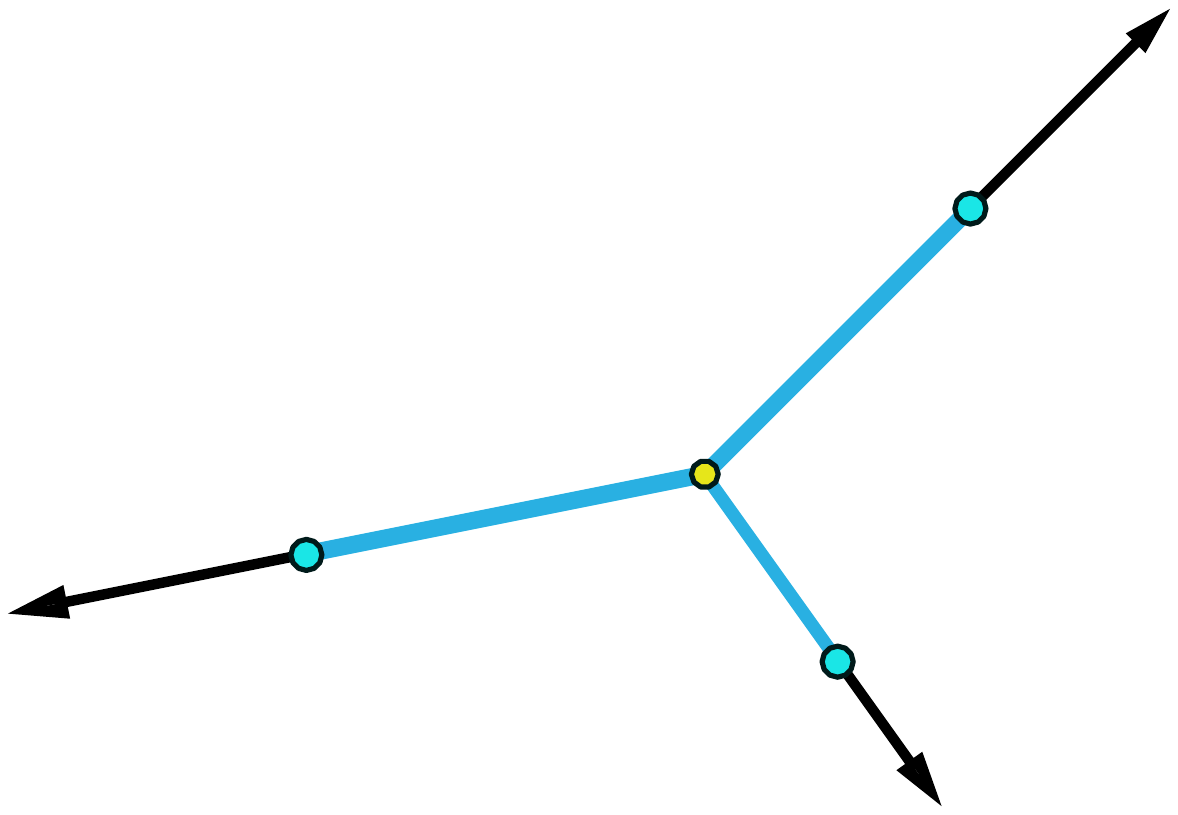}	
			\cput(50,2){\contour{white}{$(b)$}}	
		\end{overpic}
		\begin{overpic}			
			[width=.32\columnwidth,trim={5cm 9cm 4.5cm 9cm},clip]{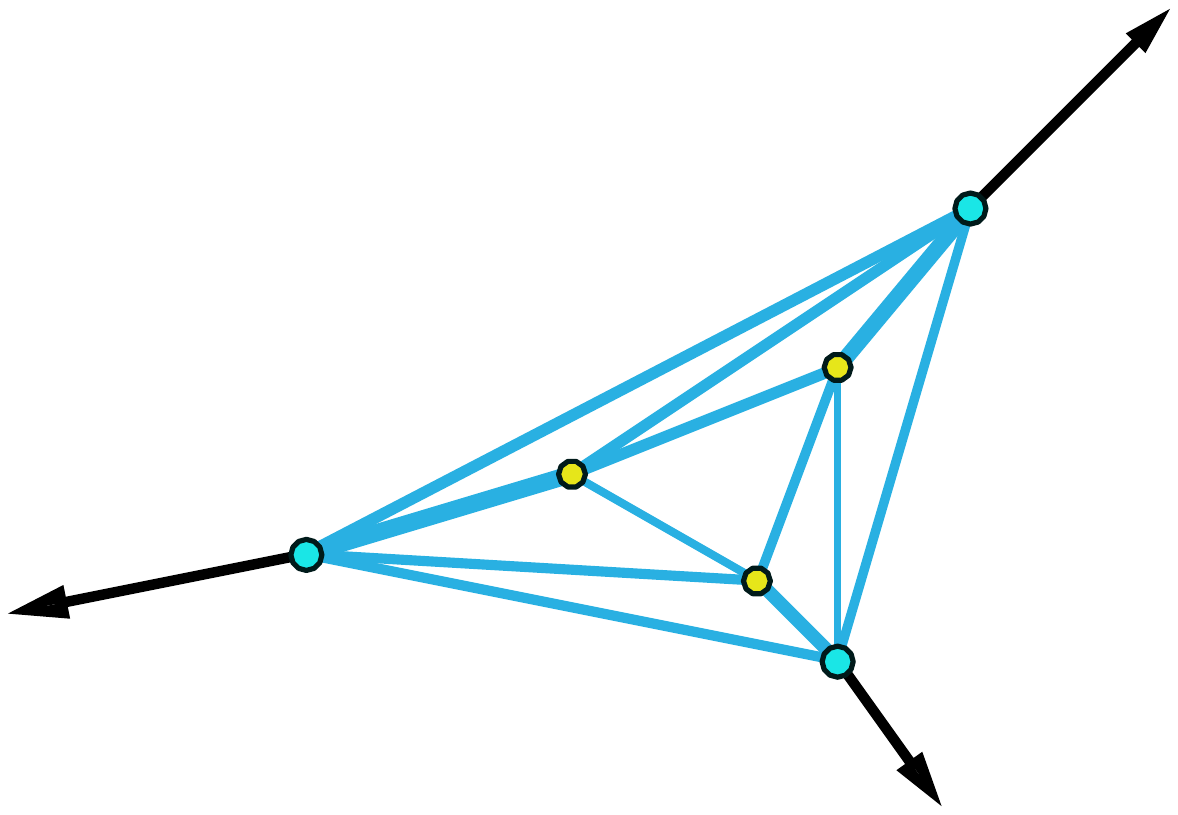}	
			\cput(50,0){\contour{white}{$(c)$}}		
		\end{overpic}
	}	
	\caption{
		\label{fig:nonunique}
	Three different truss designs for the same special boundary condition. These truss use the same amount of material. There are infinite number of such optimal trusses for this special boundary condition. }
\end{figure}

According to the theorem of Maxwell, we can easily find some special cases. For example, given a concurrent force systems with forces $F_1$,$F_2$,...,$F_n$ through a common point $O$ and their corresponding action points $P_1$,$P_2$,...,$P_n$, we have $F_i=\lambda_i\overrightarrow{OP_i}$. If all the $\lambda_i$ have the same sign, either positive or negative, then the truss connect all $P_i$ and $O$ is an optimum design. Note that the optimum for these special boundary conditions is usually not unique. Figure \ref{fig:nonunique} shows multiple optimum designs for the same boundary condition where 3 forces are self-equilibrium in 2D plane. From a design standpoint, the non-uniqueness provide a flexibility of topology and geometry design for these special boundary conditions. 

To systematically investigate the boundary condition under which a design with all bars in compression or tension can be found is interesting but out of the scope of this paper. Our numerical computational method can be use to detect such kind of boundary condition in the initial truss generation stage. Under such kind of special boundary conditions, further subdivision is not necessary.

\CG{

\subsection{Nodal Equilibrium Matrix}
The nodal equilibrium matrix, $\mathbf{B}$, transforms the magnitudes of internal forces, $\sw = (s_1, s_2, \dots, s_{|E|}$), at each bar, into forces along the $x$, $y$ and $z$ axes at each joint. If a truss is in equilibrium, then each joint must be in equilibrium. 
 The equilibrium equation is written as

\begin{equation}
\mathbf{B}^T \sw = -\fw.  \label{eq:nodal}
\end{equation}

Let's take a look at the equilibrium equations of a particular $i$-th joint with a valence of $n_i$, connected to nodes $p_{i,j}, j=1,\dots,n_i$ through bars $b_{i,j}, j=1,\dots,n_i$ respectively. Then the equilibrium equations at the joint $i$ are
\begin{align*} 
\sum_{j=1}^{n_i} \frac{x_{p_{i,j}}-x_i}{l_{b_{i,j}}}  s_{b_{i,j}} & = -f_{ix}, \\
\sum_{j=1}^{n_i} \frac{y_{p_{i,j}}-y_i}{l_{b_{i,j}}}  s_{b_{i,j}} & = -f_{iy}, \\
\sum_{j=1}^{n_i} \frac{z_{p_{i,j}}-z_i}{l_{b_{i,j}}}  s_{b_{i,j}} & = -f_{iz},
\end{align*}
where $ l_{b_{i,j}} =\sqrt{(x_{p_{i,j}}-x_i)^2+(y_{p_{i,j}}-y_i)^2+(z_{p_{i,j}}-z_i)^2}$ is the length of bar $b_{i,j}$. By writing all the equilibrium equations in a matrix form, we get the Equation \ref{eq:nodal}, where the element in the matrix $B$ are linear combinations of the directional cosines of the bars.

If we use force densities $w_i=s_i/l_i$ as variables, the above equations are transformed to simpler forms:
\begin{align*} 
\sum_{j=1}^{n_i} (x_{p_{i,j}}-x_i) w_{b_{i,j}} & = -f_{ix}, \\
\sum_{j=1}^{n_i} (y_{p_{i,j}}-y_i) w_{b_{i,j}} & = -f_{iy}, \\
\sum_{j=1}^{n_i} (z_{p_{i,j}}-z_i) w_{b_{i,j}}  & = -f_{iz}.
\end{align*}

The above equilibrium equations could be written in a similar matrix form as
\begin{equation}
\mathbf{C}^T \ww = -\fw.  
\end{equation} 
All the elements in the matrix $\mathbf{C}$ are linear combinations of the joint coordinates. Similarly, the elements in the matrix $\Delta \mathbf{C}$ in Equation \ref{eqn:lp5_a} are linear combinations of nodal variations.
}